\documentclass{article}

\usepackage{iclr2027_conference,times}

\usepackage{amsmath,amsfonts,bm}

\def\eqref#1{equation~\ref{#1}}

\def\1{\bm{1}}

\DeclareMathAlphabet{\mathsfit}{\encodingdefault}{\sfdefault}{m}{sl}
\SetMathAlphabet{\mathsfit}{bold}{\encodingdefault}{\sfdefault}{bx}{n}

\usepackage{graphicx}
\usepackage{float}
\usepackage{needspace}
\usepackage{caption}
\usepackage{booktabs}
\usepackage{hyperref}
\usepackage{url}

\usepackage{xcolor}
\usepackage{colortbl} %
\usepackage{amssymb} %
\definecolor{bobapearl}{HTML}{6B3E26}

\hypersetup{pdftitle={Built on Broken Assumptions (BOBA): Why Coding Agents Cannot Be Left Alone}}
\definecolor{bobalink}{HTML}{1E7A46}
\hypersetup{colorlinks=true,linkcolor=bobalink,citecolor=bobalink,urlcolor=bobalink,filecolor=bobalink}

\title{Built on Broken Assumptions \mbox{(\BobaTitleMark BOBA)}: Why Coding Agents Cannot Be Left Alone}

\author{Alejandro Cuadron \\ Amazon \And Arpit Gupta \\ Amazon}
\iclrfinalcopy

\usepackage{listings}
\definecolor{bobapromptframe}{HTML}{477DA5}
\definecolor{bobapromptback}{HTML}{F7F5F2}
\lstdefinestyle{bobaprompt}{
  basicstyle=\footnotesize\ttfamily,
  columns=fullflexible, keepspaces=true,
  breaklines=true, breakatwhitespace=true, breakindent=1em,
  showstringspaces=false, showspaces=false, showtabs=false,
  numbers=none, frame=single, framerule=0.4pt,
  rulecolor=\color{bobapromptframe},
  backgroundcolor=\color{bobapromptback},
  framesep=6pt, xleftmargin=7pt, xrightmargin=7pt,
  aboveskip=0.7\baselineskip, belowskip=0.7\baselineskip,
  abovecaptionskip=0pt, belowcaptionskip=0pt,
  language={}, keywordstyle={}, commentstyle={}, stringstyle={}
}

\usepackage{tabularx}

\input{boba_icons}

\begin{document}

\maketitle
\lhead{Preprint}  %

\enlargethispage{\baselineskip}%
\begin{abstract}
Coding agents are autonomous developers. Given a request, they change the code and report back
when done. As such, they can handle not only independent requests but whole chains of them, each
request built on the last. Agents are known to report wrong work as done, yet benchmarks still test
them one request at a time, on human-written code, checked once. We build \textsc{BOBA-Bench}, 280
chains of 20 consecutive requests drawn from 140 real open-source projects, and evaluate ten coding
agents on both. All ten do far worse on chains. On average, they complete 95\% of requests taken
one at a time, but when requests are chained together, they complete 51\%. Worse code is not the
cause. Once checked, agent code is nearly as good a base for later requests as human code. The
cause is the mistakes agents miss. Coding agents declare a request done once their own tests pass,
even when those tests miss a mistake. The next request then builds on broken code and fails about
twice as often. The mistake is rarely found later. Nor can agents catch such mistakes by reviewing
code. Asked whether an implementation is correct, they are right 53\% of the time, barely better
than a coin flip. Coding agents are therefore heavily dependent on access to verification. A single
external ``that's wrong'' after a failed request and one repair round lift chain completion from
51\% to 59\%, and a repaired mistake no longer hurts the requests that follow. In simpler terms, coding
agents are only as reliable as the checks they can reach: \emph{without verification, every request
that follows is built on broken assumptions}.
\end{abstract}

\begin{figure}[!h]
\vspace{-8pt}%
\centering
\includegraphics[width=\textwidth]{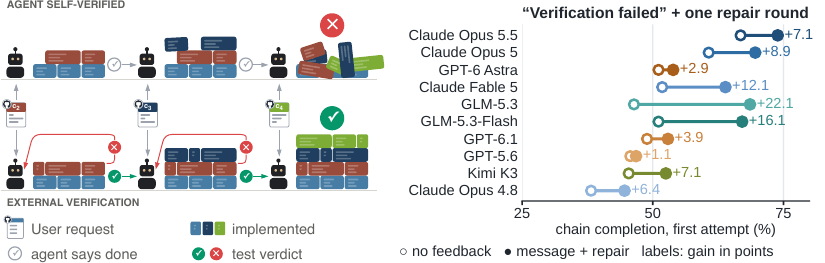}%
\caption{\textbf{An outside check lifts chain completion.} Left: unchecked work collapses, verified work holds (schematic). Right: chains completed on the first attempt without feedback (hollow) and with one ``verification failed'' message and one repair after each failed request (filled), ten models; labels give the gains in percentage points.}
\label{fig:illusion}
\vspace{-8pt}
\end{figure}

\section{Introduction}
\label{sec:intro}

Today's coding agents (\emph{e.g.}, Codex \citep{openai2025codex}, Claude Code
\citep{anthropic2025claudecode}, OpenHands \citep{wang2024openhands}) let developers hand over a
request and receive a finished change.
Delegation promises to help not only when developers review each change before asking for the
next, but also when they cannot, and each request then builds on code the agent wrote and reported
complete \citep{liu2026copilottraces,duma2026reviewsaiprs}. Though trajectory studies confirm that
agents rarely catch a mistake nothing flags for them \citep{traverse2026} and report wrong work as
complete \citep{confidentwrong2026,advani2026falsesuccess,smyth2026overclaim}, they are typically
evaluated on a single request, from human-written code, verified once
\citep{jimenez2024swebench,deng2025swebenchpro}.

Even though a growing body of work proposes to evaluate agents over successive requests
\citep{swemilestone2026,swechain2026,slopcodebench2026,sweci2026,shastry2026swesteps,jin2026chainswe,shen2026evocodebench,sobal2026staminabench},
we identify in our review (\hyperref[sec:related]{Section~\ref*{sec:related}}) that prior work either carries an agent's
verified and unverified work forward as one, even against an oracle control
\citep{jin2026chainswe,swemilestone2026}, so a later failure cannot be told apart as building on
agent-written code or on a mistake nobody checked \citep{peng2026residual}, or stops at the first
failure \citep{shen2026evocodebench,sobal2026staminabench} and never continues from unverified work.
We argue that unattended development, where mistakes go unnoticed
\citep{traverse2026,confidentwrong2026,advani2026falsesuccess,smyth2026overclaim} and get built on
\citep{traverse2026,cuadron2025saber,zhu2025agenterrortaxonomy,laban2025lostmultiturn}, can only be
measured by controlling what it lacks: verification.

In this work, we close this gap by creating a benchmark for continuous, unattended development --
\textsc{BOBA-Bench} -- built from the commit histories of 140 real repositories. At a high level, the
chaining process we propose transforms a stretch of a repository's history into a \emph{chain} of
20 requests, specified from the tests the developers added, that jointly rebuild their changes.
Autonomous evaluation then hands each request to the agent in a fresh conversation, on the state the
previous request left behind, with hidden tests after every request telling verified work from a
\emph{silent miss}: the chain runs through all 20 requests, and the agent, told nothing and remembering
nothing of the earlier requests, builds each one on whatever they left, verified or not.

On the 280 chains that we experimented on, we observed that agents left alone completed an
average of 51\% of chains (pass@1, estimated from eight attempts per chain), and only 39\% without
a single request missed or earlier test broken along the way, although the same agents complete
66--84\% of the individual requests those chains are made of (\hyperref[fig:illusion]{Figure~\ref*{fig:illusion}}). Notably, we
observe silent misses even within the first five requests, and across all ten agents we test, from
open-weights (GLM-5.3-Flash) to state-of-the-art (Claude Opus~5.5).

We then decompose the degradation into two components: (1) who wrote the code beneath a
request, and (2) whether it was verified. On verified code, agents leave a foundation
nearly as buildable as the developers' own, within 4 points for every model. By contrast,
all agents fail far more often on an unverified state, regardless of aptitude. We call this
the \emph{built on broken assumptions} phenomenon: what limits coding agents left alone is not the
code they write but the mistakes they cannot see: without verification, every request that
follows is built on broken assumptions.

We investigate several explanations for this effect and show that the agents tend to (1) submit
work as complete on the strength of their own tests, which barely lower the chance of a miss,
leading them to (2) proceed onto a broken state unaware,
(3) fail more as such mistakes accumulate, and (4) accept or reject finished work almost
indiscriminately when asked to review it.

Finally, we test what the missing verification buys. Told only that verification failed, and
given one repair round, agents fix a tenth to a third of their misses, a quarter to a half when the
failing test is named. Delivered after every failed request, the bare message raises chain
completion by 1 to 22 points, at the price of 3 to 9 points more chains with a broken test
(\hyperref[sec:verification]{\S\ref*{sec:verification}}).

Our findings highlight a gap between how coding agents are used in practice and how they are being
evaluated. Ubiquitous silent misses are likely a reason why human verification remains the bottleneck
\citep{krentsel2026reality,winninger2026steerability,garousi2026oversightoverload}, particularly
for teams whose review of agent work is already thinning
\citep{duma2026reviewsaiprs,yu2026habituation,ye2026sabotage}.

The rest of the paper is structured as follows: \hyperref[sec:related]{Section~\ref*{sec:related}} reviews prior work.
\hyperref[sec:benchmark]{Section~\ref*{sec:benchmark}} describes the benchmark, its protocol, and the ten agents we evaluate,
and \hyperref[sec:agents]{Section~\ref*{sec:agents}} presents and analyzes our main experiment. Finally,
\hyperref[sec:verification]{Section~\ref*{sec:verification}} tests external verification and draws recommendations.

\begin{table}[!t]
\centering
\fontsize{7.5}{9}\selectfont %
\makeatletter\@ifpackageloaded{colortbl}{}{\providecommand{\rowcolor}[2][]{}}\makeatother
\setlength{\tabcolsep}{2.45pt}
\renewcommand{\arraystretch}{1.35}%
\caption{\textsc{BOBA-Bench} next to representative coding benchmarks. Bold: best among the repository-evolution benchmarks; underlined: second.}
\label{tab:benchmarks}
\begin{tabular}{@{}llrrrrrrrl@{}}
\toprule
 & & & & \multicolumn{2}{c}{Graded patches} & 95\% CI & \multicolumn{2}{c}{Per task} & Task \\
\cmidrule(lr){5-6}\cmidrule(lr){7-7}\cmidrule(lr){8-9}
Benchmark & Granularity & Tasks & Repos & per task & total & ($\pm$pp) & spec.\ words & patch lines & collection \\
\midrule
\multicolumn{10}{@{}l}{\emph{Issue resolution}} \\
SWE-bench Verified \citep{openai2024swebenchverified} & Commit & 500 & 12 & 1 & 500 & 4.4 & 189 & 14 & Rule-based \\
SWE-bench Pro \citep{deng2025swebenchpro} & Commit & 1.9k & 41 & 1 & 1.9k & 2.3 & 462 & 170 & Rule-based \\
CodeThread \citep{patel2026codethread} & Commit & 1.4k & 78 & 1 & 1.4k & 2.6 & n/r & n/r & Rule-based \\
\midrule
\multicolumn{10}{@{}l}{\emph{Codebase generation}} \\
FeatureBench \citep{zhou2026featurebench} & Feature & 200 & 24 & 1 & 200 & 6.9 & 4.8k & 790 & Test-driven \\
SWE-EVO \citep{le2025sweevo} & Release & 48 & 7 & 1 & 48 & 13.6 & 2.4k & 611 & Rule-based \\
Commit0 \citep{zhao2024commit0} & Project & 54 & 54 & 1 & 54 & 12.9 & $>$7k & $>$3k & Rule-based \\
DeepSWE \citep{huang2026deepswe} & Feature & 113 & 91 & 1 & 113 & 9.1 & 302 & 701 & Manual \\
\midrule
\multicolumn{10}{@{}l}{\emph{Repository evolution}} \\
SlopCodeBench \citep{slopcodebench2026} & Checkpoint & \underline{36} & synth. & 5.4 & \underline{196} & \underline{15.5} & 4.3k & n/r & Manual \\
SWE-Chain \citep{swechain2026} & Release & 12 & \underline{9} & 12.9 & 155 & 24.6 & \underline{15.5k} & \textbf{8.9k} & Agentic \\
SWE-Milestone \citep{swemilestone2026} & Milestone & 7 & 7 & \underline{14.0} & 98 & 29.8 & \textbf{18.9k} & 6.9k & Agentic \\
\rowcolor{bobapromptback}
\textbf{\textsc{BOBA-Bench} (ours)} & Request & \textbf{280} & \textbf{140} & \textbf{20} & \textbf{5.6k} & \textbf{5.8} & 12.1k & \underline{8.2k} & Agentic \\
\bottomrule
\end{tabular}

\end{table}

\Needspace{10\baselineskip}%
\section{Background and Related Work}
\label{sec:related}

Early code-generation benchmarks asked models to implement isolated functions, checked against
unit tests \citep{chen2021codex}. As models began to edit real projects, repository-level benchmarks
gave agents an issue and a human-written codebase and graded the resulting patch once
\citep{jimenez2024swebench,deng2025swebenchpro}. Later benchmarks enlarged the task to features,
releases, and repositories \citep{zhou2026featurebench,le2025sweevo,zhao2024commit0,yang2026programbench},
but each task still began from a reference state and ended with a single verdict.

Recent benchmarks instead evaluate agents over successive requests on the same codebase, so an
agent's earlier implementation becomes the foundation for later work
\citep{swemilestone2026,swechain2026,slopcodebench2026,sweci2026,jin2026chainswe,shen2026evocodebench,sobal2026staminabench}.
This exposes what single-request evaluation cannot: regressions, accumulated errors, and declining
performance as development continues.

Crucially, these benchmarks differ in what happens after an agent fails. Some carry the failed
work forward with everything else \citep{swemilestone2026,jin2026chainswe}, so later failures mix
the difficulty of building on agent-written code with the cost of building on a mistake nobody
checked \citep{peng2026residual}. Others stop at the first failure
\citep{shen2026evocodebench,sobal2026staminabench}, so the remaining evaluation concerns only
verified code. Unattended development guarantees neither: agents can report wrong work as
complete \citep{confidentwrong2026} and proceed as though it were right
\citep{traverse2026}. We therefore carry the agent's work forward, verified or not, grade every
request, and make verification, not authorship, the central variable of our evaluation.

Separating the two requires controls that prior work provides only in part. CodeThread compares a
follow-on issue on human-written and agent-written code \citep{patel2026codethread}, and
EvoCode-Bench compares reference states with agent prefixes that passed cumulative verification
\citep{shen2026evocodebench}. We run the same later request on human code, verified agent code,
and unverified agent code, and then add repair: comparisons that separate what an agent leaves
behind when right from when~wrong.%

Measuring these effects requires checking every request as the code evolves. Static measures
describe the implementation \citep{paul2025smells,sun2026nonfunctional}, while completion of
the current request ignores what earlier requests established. Like continuous benchmarks
\citep{swemilestone2026,sweci2026}, we test new behavior alongside earlier behavior, with hidden
tests from real development histories, so that an unresolved error, a later regression, and a
repair stay distinguishable.

\Needspace{8.5\baselineskip}%
Scaling these experiments requires an external check that does not depend on the agent. Agents
can debug from execution feedback \citep{chen2023selfdebug} and write their own tests
\citep{mundler2024swtbench}, but their self-repair depends on the quality of that feedback
\citep{olausson2024selfrepair}, and their tests can share their implementation errors
\citep{execcritic2026}. Model reviewers scale but remain imperfect judges \citep{swereview2026}, and
human review does not scale \citep{krentsel2026reality}. We use hidden tests as the check, ask
whether agents can supply that check themselves, and measure how much external feedback is needed
to recover. \hyperref[app:related-details]{Appendix~\ref*{app:related-details}} provides supporting details.
\Needspace{11\baselineskip}%
\section{BOBA-Bench: Built on Broken Assumptions}
\label{sec:benchmark}

\begin{figure}[t]
\centering
\includegraphics[width=\columnwidth]{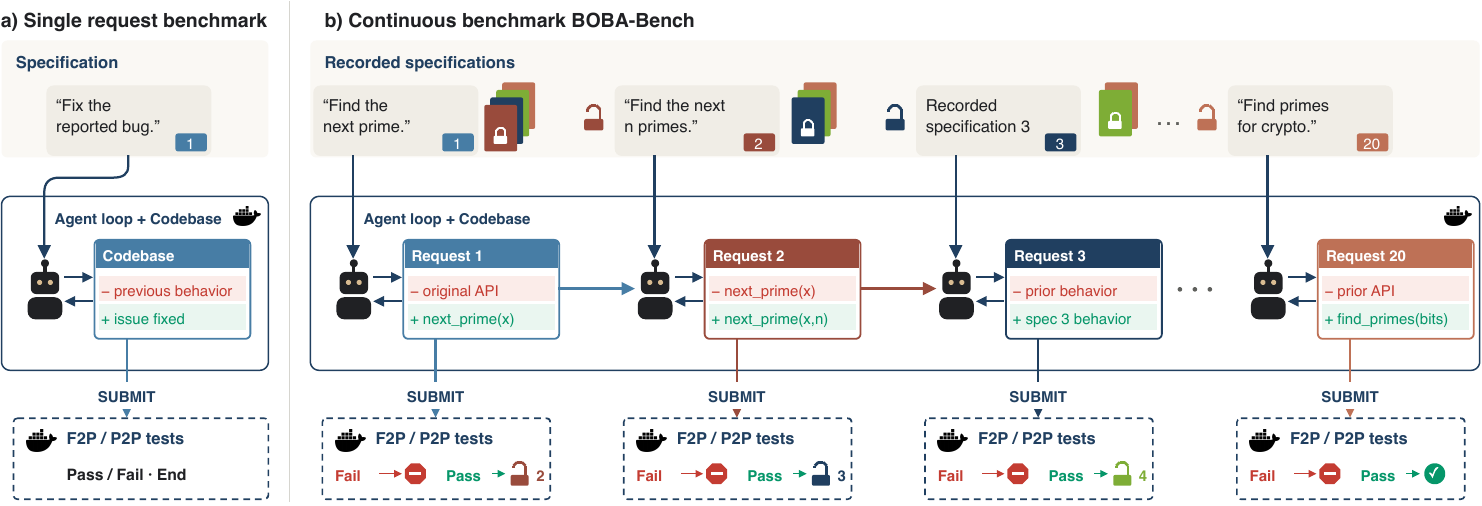}
\caption{\textbf{Autonomous evaluation grades the repository an agent leaves after 20 requests on its
own code, not one patch.} (a)~One specification, one submission, hidden F2P/P2P tests (SWE-bench
shown). (b)~\textsc{BOBA-Bench}: 20 specifications arrive one at a time on one persistent codebase,
each in a fresh conversation; after each submission a fresh verifier runs the hidden tests of
requests $1..k$ and records the verdict, and specification $k{+}1$ follows regardless.}
\label{fig:verification}
\end{figure}

To assess whether coding agents can be left alone, we develop a benchmark that repurposes the
commit histories of real repositories. First, we apply a \emph{chaining} process that turns a
stretch of a repository's history into a \emph{chain} of 20 requests, each specified from the tests the
developers wrote. Second, we implement an autonomous evaluation protocol that delivers the requests
one at a time on the agent's own code, grades each against hidden tests, and tells the agent
nothing about how it did. \textsc{BOBA-Bench} contains 280 chains from 140 Python repositories,
5,600 requests in all (\hyperref[tab:benchmarks]{Table~\ref*{tab:benchmarks}}).
\label{sec:stats}

\subsection{Chaining Process: From Commit Histories to Chains}
\label{sec:units}
\label{sec:construction}
\label{sec:qa}
\label{sec:mining}
\label{sec:specs}
\label{sec:chains}

\hyperref[fig:verification]{Figure~\ref*{fig:verification}} contrasts a single-request task, one specification on human-written
code graded once, with a chain. A chain is a sequence of
\emph{requests}, each derived from a run of consecutive commits in the repository's history that
change tests and source together, ending in one behavior change. Each request carries a
specification, written from the tests the developers added rather
than from their code, and a hidden \emph{contract}: the added tests must fail before the change
and pass after it, and the existing tests must keep passing. Taken jointly, the 20 requests
rebuild the developers' changes in their order on whatever code the agent has produced so far;
\emph{state} $k$ is the code after the first $k$.

\begin{figure}[b]%
\centering
\includegraphics[width=\columnwidth]{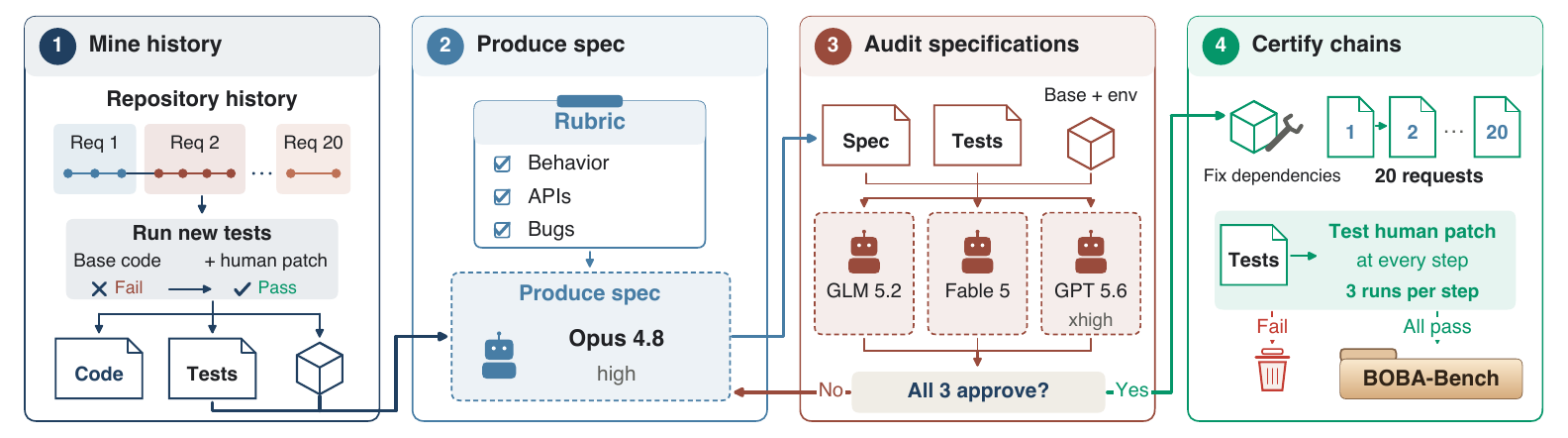}
\caption{\textbf{Requests are specified from the developers' tests, not their code, and certified
against their patches.} (1)~\emph{Mine history}: consecutive commits that change tests and source
together become a request if the new tests fail before and pass after the human patch.
(2)~\emph{Specify}: Claude Opus~4.8 writes each specification from commit messages, issues and the
test diff. (3)~\emph{Audit}: three model families must accept identical text. (4)~\emph{Certify}: 20
consecutive requests form a chain, sealed with frozen dependencies, in which the developers' patches
must pass every hidden test.}
\label{fig:benchmark-construction}
\end{figure}

\Needspace{5.5\baselineskip}%
We developed a semi-automatic chaining process to scale the construction of chains
(\hyperref[fig:benchmark-construction]{Figure~\ref*{fig:benchmark-construction}}). First, requests are mined from the histories of
Python repositories: wherever consecutive commits change tests and source together, the run is kept
as a request only if its new tests fail before the developers' change and pass after it. Second,
Claude Opus~4.8 writes each specification from the test diff, with commit messages and issue text as
aids.

Because specifications are written by a model from tests the agent never sees, a chain could fail
for reasons that are not the agent's. We therefore attempted every specification blind, from the
developers' code just before it: two models solve 78\% of the
released specifications between them, and the remainder were probed for behavior the tests demand
but the text omits, then rewritten or kept as hard but fair. Three model families then read every
specification against its hidden tests, rewriting it until all three accepted identical text.
Finally, each chain is sealed into one offline image with frozen dependencies, and
released only if the developers' own patches pass every hidden test at every
state in that image. We believe the process yields chains that are solvable from their text and on
which a miss can be attributed to the agent (details in \hyperref[app:qa]{Appendix~\ref*{app:qa}}).

\Needspace{12.5\baselineskip}%
\subsection{Autonomous Evaluation}
\label{sec:protocol}
\label{sec:grader}

\hyperref[fig:verification]{Figure~\ref*{fig:verification}}b depicts the loop. The \textbf{agent} works through
the 20 specifications in one offline container, each in a fresh conversation; the \textbf{grader}
runs in a separate container and sees only the git history the agent commits. On each request, the
agent receives the specification, the repository as the previous request left it and a submit
tool, edits and runs code and tests, and submits once it believes the specification satisfied. The grader then runs the hidden tests of requests $1..k$ on the submitted code
(\hyperref[app:qa-grader]{Appendix~\ref*{app:qa-grader}}): a submission whose own tests fail is a \emph{silent miss}, and a
test that passed before the submission and fails after it is a \emph{regression}. Whatever the
verdict, the next specification opens a fresh conversation on the same repository, and the chain
runs through all 20 requests.

The agent is told neither the verdict nor that it is on a chain: withholding feedback is what
unattended development looks like, so whether an agent can tell that its own work is correct is
itself measured.

A chain is complete if the repository after request 20 passes every hidden test of the chain and
its own;
estimated as the mean over eight independent attempts per chain, and the single-attempt value is
reported alongside. Because this judges only the final state, we also report, averaged likewise,
\textbf{strict pass@1}, every request's own tests passing at its own step with no regression
ever, which repairing everything at request 20 cannot score, and the \textbf{regression rate},
the share of chains with a regression at some request.

\begin{table}[t]%
\centering
\scriptsize
\setlength{\tabcolsep}{4.2pt}
\renewcommand{\arraystretch}{1.02}
\caption{Chain completion of ten coding agents on the 280 chains of \textsc{BOBA-Bench}: each request
in a fresh conversation on the repository the previous one left, all 20 requests run, eight
independent attempts per chain. pass@1: the probability that an attempt leaves the repository after
request 20 passing every hidden test of the chain, estimated from the eight attempts (single
attempt: the first attempt alone); strict pass@1: in addition, every request passed its own tests on
its own step and no passing test ever broke; regression rate: a passing test broke at some request,
repaired later or not; pass@8: at least one of the eight attempts passes; request completion: share
of all graded requests of the first attempt whose tests pass on the step they are issued.
Repository-clustered 95\% bootstrap intervals; an attempt without a final state counts as a failure.}
\label{tab:leaderboard}
\begin{tabular}{@{}lrlrrrrrr@{}}
\toprule
 & \multicolumn{6}{c}{Chains (\%)} & Requests & Calls per \\
\cmidrule(lr){2-7}
Model & pass@1 & [95\% CI] & single attempt & strict pass@1 & regression rate & pass@8 & completed (\%) & chain (median) \\
\midrule
Claude Opus~5.5 & 66.7 & [61.3, 71.9] & 66.8 & 55.8 & 22.9 & 83.9 & 84.0 & 216 \\
Claude Opus~5   & 61.6 & [56.2, 66.8] & 60.7 & 50.5 & 28.1 & 82.9 & 80.8 & 401 \\
GPT-6 Astra     & 53.2 & [46.7, 59.5] & 51.1 & 38.7 & 35.1 & 63.9 & 75.2 & 182 \\
Claude Fable~5  & 51.5 & [45.8, 57.1] & 51.8 & 38.2 & 40.5 & 72.1 & 73.2 & 292 \\
GLM-5.3         & 51.2 & [45.4, 57.1] & 46.4 & 39.8 & 30.1 & 76.1 & 73.0 & 560 \\
GLM-5.3-Flash   & 51.1 & [45.1, 57.0] & 51.1 & 40.9 & 30.1 & 77.5 & 73.9 & 739 \\
GPT-6.1-sol     & 50.3 & [43.9, 56.5] & 48.9 & 37.6 & 37.8 & 60.7 & 74.2 & 190 \\
GPT-5.6-sol     & 45.2 & [39.0, 51.3] & 45.7 & 33.3 & 42.2 & 60.0 & 71.8 & 299 \\
Kimi~K3         & 44.2 & [38.5, 50.0] & 45.4 & 33.7 & 42.8 & 68.2 & 68.8 & 388 \\
Claude Opus~4.8 & 39.2 & [33.3, 45.3] & 38.2 & 27.5 & 42.4 & 60.0 & 66.2 & 367 \\
\bottomrule
\end{tabular}
\end{table}

\subsection{Evaluation Settings}
\label{sec:settings}

We evaluate agents in five settings.

\paragraph{\textsc{\BobaMark{autonomous}Autonomous}} is the benchmark run described above: the agent is left alone and nothing stops
the chain before request 20.

\paragraph{\textsc{\BobaMark{agentcode}Agent code} and \textsc{\BobaMark{humancode}Human code}} isolate authorship. Request $k{+}1$ is solved on $k$
requests that all passed their contracts, implemented either by the agent itself, with retries until
each is certified, or by the developers, under identical specification, budget and grader; the
comparison asks whether agent code is a worse base than human code once both are verified.

\paragraph{\textsc{\BobaMark{continuation}Silent continuation}} isolates verification: request $k{+}1$ is issued on a silent miss as if it
had passed, and only failures of that request's own tests are counted, not those it inherits.

\paragraph{\textsc{\BobaMark{verified}Verified}} takes \textsc{\BobaMark{continuation}Silent continuation} a step further: after the miss, a message from
outside says that verification failed and the agent gets one repair before the next request.

\paragraph{\textsc{\BobaMark{review}Review}} asks whether an agent can supply that message itself: a model judges a finished
implementation without knowing its author, and accepts or rejects it.

\subsection{Scale and Parameters}
\label{sec:evalsetup}

We evaluate ten models on all 280 chains in the \textsc{\BobaMark{autonomous}Autonomous} setting,
eight attempts per chain: Claude Opus~5.5, Opus~5, Opus~4.8 and Fable~5
\citep{anthropic2026claudeopus55,anthropic2026claudeopus5,anthropic2026claudeopus48,anthropic2026claudefable5},
GPT-6 Astra, GPT-6.1-sol and GPT-5.6-sol
\citep{openai2026gpt6astra,openai2026gpt61sol,openai2026gpt56sol}, GLM-5.3 and GLM-5.3-Flash
\citep{zhipu2026glm53,zhipu2026glm53flash}, and Kimi~K3 \citep{moonshot2026kimik3}.
The harness is mini-swe-agent 2.4.6 \citep{yang2024sweagent}; a
chain may use unlimited model calls and 2\,h of tool execution over its 20 requests, and no
wall-clock limit (settings in \hyperref[app:qa-protocol]{Appendix~\ref*{app:qa-protocol}}). The four other settings use
Fable~5, GPT-6 Astra, GPT-5.6-sol, Kimi~K3 and GLM-5.2 \citep{zhipu2026glm52} on states of
the same chains (\hyperref[app:mf]{Appendices~\ref*{app:mf}}, \ref{app:cv}, \ref{app:eval} and \ref{app:trust}).

\section{Can Coding Agents Be Left Alone?}
\label{sec:agents}
\label{sec:evaluation}
\label{sec:eval-setup}
\label{sec:setup}
\label{sec:diagnostics}

\subsection{Completion Findings}
\label{sec:howfar}

\hyperref[fig:illusion]{Figure~\ref*{fig:illusion}} and \hyperref[tab:leaderboard]{Table~\ref*{tab:leaderboard}} summarize the results, with pass@1
estimated from eight attempts per chain. At a high level,
\textbf{every model sees its performance degrade from the requests it completes to the
chains it completes, with an average chain completion of 51\%}.
We name this phenomenon Built on
Broken Assumptions: models that complete 66-84\% of the individual requests they are
given\footnote{Counting only the requests up to and including a chain's first broken test gives
77-89\%.} complete only about half of the
20-request chains made of \emph{the exact same requests}, since
each request is built, in a fresh conversation, on whatever the earlier ones left behind.

At least one of a chain's eight attempts completes it 71\% of the time on average (pass@8), so the
loss is not all bad luck: the same chains fail every model (16\% are completed by none of the ten on the first
attempt and 16\% by all ten, against fewer than one expected under independence), and difficulty
lies in particular requests, of which a stronger model clears more, not different ones.

Strict pass@1 (\hyperref[tab:leaderboard]{Table~\ref*{tab:leaderboard}}) adds the condition that every request passed on its own
step and no passing test ever broke: in 23-43\% of attempts a later request breaks a test that
passed, usually one of the agent's own earlier requests (\hyperref[app:rootcauses]{Appendix~\ref*{app:rootcauses}}), and the
10-15 point gap between the two measures is chains repaired by request 20.
The per-request failure
rate does not fall as a chain progresses (3-6\%, \hyperref[app:leaderboard-additional]{Appendix~\ref*{app:leaderboard-additional}},
\hyperref[fig:chain-curves]{Figure~\ref*{fig:chain-curves}});
compounded over twenty requests it reproduces strict pass@1 but not pass@1, because what follows a
failure is seldom recovery: only 14-33\% of the tests that fail are ever repaired, and a request
issued on a broken repository fails about twice as often as one on a clean one, 9-19 points
more than the same request from the developers' state (\hyperref[app:resume-results]{Appendix~\ref*{app:resume-results}};
\hyperref[sec:mf]{\S\ref*{sec:mf}}, \hyperref[sec:verification]{\S\ref*{sec:verification}}).

A different harness does not help: under Codex CLI, the OpenHands SDK and pi, four models complete
no more chains than under mini-swe-agent, often fewer, and fail largely at the same requests
(\hyperref[tab:scaffold]{Table~\ref*{tab:scaffold}}); \textbf{the harness does not, on its own, allow models to verify their
own work}. Nor does memorized code explain the effect: three Claude models, including the leader,
memorize code of these repositories, but no more of benchmark code than of other code, and none
completes the repositories it memorized measurably more often (\hyperref[app:contamination]{Appendix~\ref*{app:contamination}}).
\label{sec:reliability}

\begin{figure}[t]
\centering
\includegraphics[width=\textwidth]{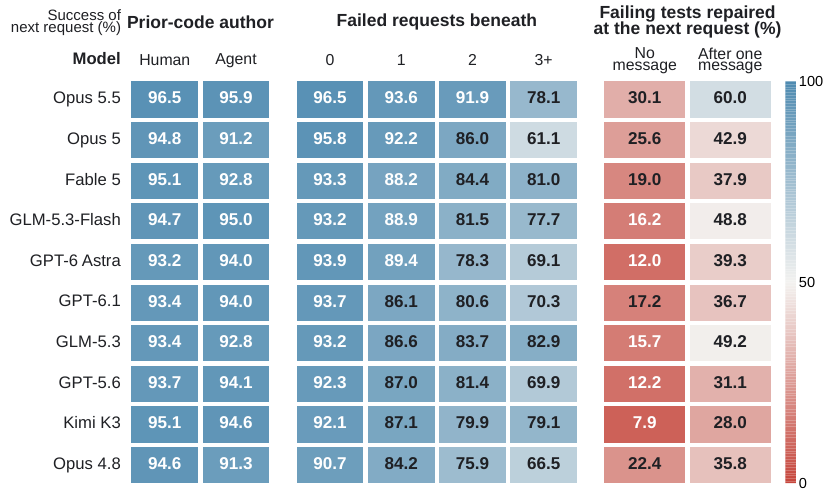}
\caption{\textbf{Verified foundations hold whoever built them, unverified failures beneath a request
erode its success, and one outside message raises their repair from 8--30\% to 28--60\%.} Success
rate (\%) of request $k{+}1$ by the state beneath it: who wrote the certified code (human or agent)
and how many earlier requests were still failing (0, 1, 2, 3+); last two columns: newly failing tests
repaired at the next request without and after one ``verification failed'' message (\hyperref[sec:verification]{\S\ref*{sec:verification}}).}
\label{fig:diagnostic-findings}
\end{figure}
\subsection{Authorship vs.\ Verification Analysis}
\label{sec:mf}
\label{sec:compound}

In \hyperref[tab:leaderboard]{Table~\ref*{tab:leaderboard}}, every request is built on the agent's own earlier work, whether or not
that work was correct; we now separate who wrote that code from whether it was verified.
\hyperref[fig:diagnostic-findings]{Figure~\ref*{fig:diagnostic-findings}} visually summarizes the results. First, looking at authorship,
we compare the \textsc{\BobaMark{agentcode}Agent code} and \textsc{\BobaMark{humancode}Human code} settings (\hyperref[app:mf]{Appendix~\ref*{app:mf}}). We see that models that build on their own verified code
do nearly as well as models that build on human code: across eleven models the gap runs from
$-3.6$ to $+0.8$ percent points, an average drop of 1 percent point, and excludes zero for three
(Claude Opus~5, Opus~4.8 and GLM-5.2, at 3 to 4 points). In summary, \textbf{when the code beneath a
request is verified, models leave code nearly as buildable as a human's.}

The verification analysis paints a different picture. In the \textsc{\BobaMark{continuation}Silent continuation} setting
(\hyperref[app:eval]{Appendix~\ref*{app:eval}}), failure of the next request's own tests rises sharply after a single unverified
mistake, with an average increase of 18 percent points (nearly tripling). Part of this is the request rather
than the state, since the requests that follow a miss are harder ones; attempted from the developers' own
state at the same point, the same requests still fail 9 to 19 points less often than from the unverified
state (\hyperref[app:resume-results]{Appendix~\ref*{app:resume-results}}). More interestingly, though better models
tend to fail less often on a verified state, all models tend to pay a similar price for an
unverified one. In other words, \textbf{on a state that nobody verified, all models we test exhibit very
high failure, with the odds of failing roughly doubling with every additional unverified mistake
beneath a request}. This refines our definition of the \emph{built on broken assumptions}
phenomenon:
when comparing verified and unverified states, we find that large performance degradations are
due in large part to mistakes going unverified, rather than to who wrote the code.

\Needspace{6.5\baselineskip}%
\hyperref[app:rootcauses]{Appendix~\ref*{app:rootcauses}} explores potential root causes for models building on broken
assumptions. We identify four causes: (1) agents break behavior the request did not touch, the
most common cause for all models but one (\hyperref[app:rc-regression]{Appendix~\ref*{app:rc-regression}}), (2) they
leave part of the specification unimplemented (\hyperref[app:rc-missed]{Appendix~\ref*{app:rc-missed}}), (3) they implement the wrong logic or leave the new code unreachable
(\hyperref[app:rc-logic]{Appendix~\ref*{app:rc-logic}}), and (4) they trust their own
tests, which barely lower the chance of a miss~\mbox{(\hyperref[app:rc-selfverif]{Appendix~\ref*{app:rc-selfverif}})}.%

\Needspace{6\baselineskip}%
\subsection{Cross-Model Review Experiment}
\label{sec:cv}

The \textsc{\BobaMark{autonomous}Autonomous} setting leaves every request unreviewed, which may not reflect how agents
are deployed in practice: a second agent could be asked to review each change, as prior work
proposes (\hyperref[sec:related]{\S\ref*{sec:related}}). We therefore ask whether an agent
can keep models from building on \mbox{broken assumptions}.%

\begin{figure}[t]
\centering
\includegraphics[width=\textwidth]{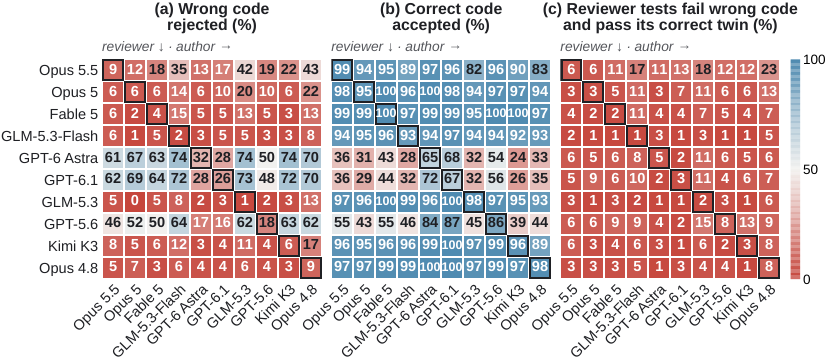}
\caption{\textbf{No model is a reliable reviewer: lenient reviewers accept most wrong code, strict ones
reject about half of the correct code.} (a) Wrong code rejected, (b) correct code accepted, (c) pairs
whose wrong twin fails and correct twin passes the reviewer's tests (\%). Framed cells: self-review.}
\label{fig:cross-model-review}
\end{figure}

In the cross-model review experiment, each of the ten models reviews 2{,}180 items in the
\textsc{\BobaMark{review}Review} setting: 109 wrong implementations per author from the \textsc{\BobaMark{autonomous}Autonomous} runs, each paired
with the same author's correct implementation of the same request from another attempt (\hyperref[app:cv]{Appendix~\ref*{app:cv}}).

\hyperref[fig:cross-model-review]{Figure~\ref*{fig:cross-model-review}} summarizes the results. In this setting none of the ten models
separates wrong from correct implementations well enough to serve as the check: lenient reviewers
(the four Claude models, GLM-5.3, GLM-5.3-Flash and Kimi~K3) accept 77--95\% of wrong
implementations, strict reviewers (GPT-6 Astra, GPT-6.1 and GPT-5.6) reject 41--58\% of correct
ones, and 4--19\% of matched pairs are discriminated (Youden $J\le0.15$; \hyperref[tab:cv-results]{Table~\ref*{tab:cv-results}}).
In other words, \textbf{a pipeline that relies on one of these agents to review another, with the
information given here, does not stop models building on broken assumptions}.

\Needspace{11\baselineskip}%
\section{Trust, but Verify: What an Outside Check Recovers}
\label{sec:verification}
\label{sec:trust}

\hyperref[sec:agents]{Section~\ref*{sec:agents}} leaves a practical question: agents build on broken assumptions and
cannot tell they have (\hyperref[sec:compound]{\S\ref*{sec:compound}}, \hyperref[sec:cv]{\S\ref*{sec:cv}}), yet nobody who deploys an agent
has our hidden grader, and few will write a test for every request. Anyone who runs the result,
however, can tell something is wrong without knowing what. \textbf{What is the least such a
watcher must say for the agent to recover}, so that verifying costs as little as
possible? We start from one bit, \emph{verification failed}, delivered once after a silent miss with
one repair round (the \textsc{\BobaMark{verified}Verified} setting of \hyperref[sec:settings]{\S\ref*{sec:settings}}, in one conversation per chain), and ask whether repeating
it, informing instead, or getting it wrong changes what it buys (\hyperref[app:trust]{Appendix~\ref*{app:trust}}).

\subsection{Implications for System Builders}
\label{sec:trust-harness}

The system around an agent has no test of its own, so it cannot judge a submission; only whoever
wrote the request can. What it can do is keep a line open in both directions, so that the verifier
sees the results and the model receives the verdict. Does one bit along that line suffice?

Told only that verification failed, \textbf{the models repair a third to a half of their misses in
one round} (\hyperref[tab:trust]{Table~\ref*{tab:trust}}), and \textbf{a repaired state is no longer a broken assumption}:
the next request fails on it no more often than on a certified state
(\hyperref[fig:diagnostic-findings]{Figure~\ref*{fig:diagnostic-findings}}), whereas an unrepaired one fails like any silent miss
(\hyperref[sec:compound]{\S\ref*{sec:compound}}). The signal, not the fresh start, does this: restarting on the same state
without the message does not help (\hyperref[app:trust]{Appendix~\ref*{app:trust}}). Applied after every failed submission
of a project, \textbf{completion rises
by 8-16 points for every model} (\hyperref[tab:trust]{Table~\ref*{tab:trust}}).
Under the benchmark's protocol the bit raises pass@1 by 1 to 22 points, seven of ten intervals
above zero (one attempt per arm, paired with the first attempt of the \textsc{\BobaMark{autonomous}Autonomous}
run; \hyperref[tab:trust-handoff]{Table~\ref*{tab:trust-handoff}}). The repair round, blind to the work it mends,
also raises the regression rate by 3 to 9 points and adds 47 to 294 calls per chain.

How much the verifier sees bounds how much it can say. For GPT-5.6 and Kimi~K3, repeating the bare
rejection after each failed repair helps with quickly diminishing returns, from 30 to 54-59\% over
eight rounds, with no round after the fourth adding more than 3 points and more than 40\% of the
misses remaining (\hyperref[app:trust]{Appendix~\ref*{app:trust}}). \textbf{Naming the failing tests once
achieves what insisting achieves in about three rounds}: insisting repeats the agent's belief about
what it built; naming the test replaces it with a fact.

In short, system builders should keep that line open in both directions and widen what the
verifier sees rather than multiply rounds: \textbf{more attempts
are an inefficient substitute for telling the agent what is wrong.} Multi-agent systems are especially exposed,
since models reviewing a finished implementation barely beat chance (\hyperref[sec:cv]{\S\ref*{sec:cv}}): the verifier's line must reach
the subagents directly, not stop at the agent that delegated.

\subsection{Implications for Model Builders}
\label{sec:trust-models}

Even a perfect supervisor recovers only part of the loss: more than half of the misses survive a
repair round, 22-59\% of projects still end short (\hyperref[tab:trust]{Table~\ref*{tab:trust}}) and 26-55\% of chains
under the benchmark's protocol (\hyperref[tab:trust-handoff]{Table~\ref*{tab:trust-handoff}}), because the agent
must find from one bit what it got wrong, cannot tell its wrong implementations from its correct
ones (\hyperref[sec:cv]{\S\ref*{sec:cv}}), and reports a wrong request complete a few times in~a~hundred (\hyperref[sec:howfar]{\S\ref*{sec:howfar}}).%

We therefore call on model builders to prioritize certifying a request as much as implementing it. \textbf{An agent that can be left alone should: (1) tell its wrong implementations from its correct
ones, where the best model we test barely beats chance; (2) repair a miss in one round when told only
that verification failed, where the models we test repair a third to a half; (3) achieve both without
a supervisor, so that a chain does not depend on who is watching.}

\subsection{Implications for Benchmark Builders}
\label{sec:trust-benchmarks}

\textbf{What lets an evaluation observe agents building on broken assumptions is not the harness}
(under three other harnesses no model completes more chains,
\hyperref[tab:scaffold]{Table~\ref*{tab:scaffold}}) \textbf{but three properties of the setting}: the agent's work persists across requests,
correctness is decided by a contract the agent does not see and cannot rewrite, and the requests
depend on one another. A single-request benchmark has none of the three, which is why the phenomenon
has gone unmeasured.

We hypothesize that agents in any such setting will build on broken assumptions, and
encourage benchmark builders to grade agents in chains, to report completion with and without a
supervisor, and to publish the hidden contracts alongside the specifications, as we do.

\subsection{Implications for Users}
\label{sec:trust-users}
\label{sec:false-alarm}

Users who delegate a sequence of requests should expect an agent to report a wrong request as
complete and build on it. We make two practical recommendations.

\paragraph{Check before you build on it.} If a request is the foundation of the next, run
it or try the feature before issuing the next request rather than after:
\textbf{a state built on a broken assumption fails the next request far more often
(\hyperref[sec:compound]{\S\ref*{sec:compound}}), whereas a repaired one is as reliable as one that was never wrong}
(\hyperref[fig:diagnostic-findings]{Figure~\ref*{fig:diagnostic-findings}}). Checking after the fact is harder: by then the next request
has already been built on the mistake.

\paragraph{Say what you saw.} A wrong ``wrong'' costs little on its
own: when we rejected correct submissions at random with probability 0.1 to 0.5 (GPT-5.6 and GPT-6
Astra, 100 chains; \hyperref[app:trust]{Appendix~\ref*{app:trust}}), about 98\% of the phantom repair rounds left the work
passing. But alarms add up over twenty requests (\hyperref[tab:trust]{Table~\ref*{tab:trust}}):
\textbf{rejecting one correct submission in ten costs at most 2 points of completion, and rejecting
every other one gives back most of the gain}, though never below the level without~a~\mbox{supervisor}.%

These are patches, not a principled approach; once agents can certify their own work, users will be
able to leave one alone with less risk of it building on broken assumptions, and recommendations
like these will no longer be needed. Until then, the watcher's check is the one that works.

\subsection{Release}
\label{sec:release}

Everything needed to reproduce and extend this work is available at
\url{https://github.com/amazon-agi/BOBA-bench}: the benchmark and its construction pipeline,
the data behind it (specifications, hidden contracts, gold patches and sealed images for every chain),
the trajectories of every evaluated run, and the grader and analysis scripts behind every table and
figure in this paper (Appendices \ref{app:qa-release} and \ref{app:repro}). Images are pinned by
digest and every artifact by hash, so results can be reproduced against the exact states we graded.
Source code stays under each repository's~\mbox{license (\hyperref[app:units]{Appendix~\ref*{app:units}})}.%

\section{Limitations}
\label{sec:limitations}

BOBA-Bench measures fixed-test correctness, not user intent, readability or production
readiness; it is Python-only, many repositories are small, and twenty requests do not imply a
twenty-link dependency chain. Headline pass@1 judges only the final repository, crediting chains
that missed or broke something on the way; strict pass@1 does not (\hyperref[tab:leaderboard]{Table~\ref*{tab:leaderboard}}).
Specifications are model-written and model-audited, not yet human-reviewed
(\hyperref[app:qa-audit]{Appendix~\ref*{app:qa-audit}}). The repositories come from public agent-task pools
\citep{badertdinov2025swerebench,fu2026davincienv} and nearly all changes predate the model
cutoffs; three Claude models, including the leader, reproduce 14--19\% of sampled change blocks
near-verbatim yet do not measurably complete more chains of the repositories they memorized
($\pm$12--18-point intervals bound but do not exclude an effect; \hyperref[sec:howfar]{\S\ref*{sec:howfar}},
\hyperref[app:contamination]{Appendix~\ref*{app:contamination}}).
\section{Conclusion}
\label{sec:conclusion}

In this work, we evaluate coding agents on twenty-request chains drawn from real repository
histories, and find that on a fixed set of requests, agents that complete most of them individually
complete only about half of the chains they form, and fewer than two in five without a single
request missed or earlier test broken along the way. Coding agents build on broken assumptions, which
materializes not as a loss of ability, since verified agent code is nearly as good a base as human code,
but as a failure of verification: a mistake the agent does not notice becomes the foundation of
everything after it, and agents asked to review barely tell right from wrong. Additional
experiments show that a single external ``that's wrong'' recovers part of the loss, whereas a fresh
start without it does not, and we call on model builders to make certifying a request as much a
goal as implementing it.

\newpage%
\subsection*{AI Use Statement}
Large language models are the subject of this study, and we also used them as tools; we say how,
following the categories of the ICLR 2027 AI Policy.
\emph{Synthetic data and dataset cleaning (used).} Language models wrote every specification from
the hidden test diff, commit messages, and linked issues and pull requests. Calls rotated over
Claude Opus~4.6, 4.7, and 4.8, with GLM-5.2 as a fallback during outages; the model behind each
specification was not recorded (\hyperref[app:qa-specgen]{Appendix~\ref*{app:qa-specgen}}). Claude Opus~4.8 attempted each
candidate specification blind, GPT-5.6 attempted those it did not solve, and Fable~5 and GPT-5.6
probed the rest for missing behavior (\hyperref[app:qa-specs]{Appendix~\ref*{app:qa-specs}}). GLM-5.2, Fable~5, and GPT-5.6
audited each specification against its hidden tests, and GPT-5.6 and Fable~5 rewrote rejected ones
(\hyperref[app:qa-audit]{Appendix~\ref*{app:qa-audit}}). GPT-5.6 repaired failing environment builds by editing only their
Dockerfiles, and sealing agents worked through a per-image checklist (\hyperref[app:qa-env]{Appendix~\ref*{app:qa-env}}).
GPT-5.6 assigned repository domain labels (\hyperref[app:units]{Appendix~\ref*{app:units}}). We treat these outputs as
generated data: the developers' hidden tests decide every grade, and every image must pass the 20
human patches at every state.
\emph{Qualitative and thematic analysis (used).} A judge model (GPT-5.6-sol) assigned one of 17
failure tags to each of the 4,787 failed requests of the first attempts, and a sandbox judge (Claude
Fable~5.1) re-executed the 1,733 it left unresolved in the sealed image (\hyperref[app:rootcauses]{Appendix~\ref*{app:rootcauses}});
one judge labels each stage, no site was relabeled by a second judge or by a human, and the agreement
we report between two model judges comes from an earlier census with the same prompt and schema. A
language model also labeled transcripts to check the self-verification rules of
\hyperref[app:trial-records]{Appendix~\ref*{app:trial-records}}. These labels are reported as indicative; every headline number comes
from the hidden tests, not from a model's judgment.
\emph{Implementing methods and writing code (used).} AI coding assistants helped write and edit code
under the authors' direction, including the agent scaffold, the grader, and the analysis scripts;
the authors reviewed this code, and the grader is validated by requiring every sealed image to pass
the 20 human patches at every state (\hyperref[app:qa]{Appendix~\ref*{app:qa}}).
\emph{Methodology, hypotheses, and interpretation (used as assistants).} AI assistants provided
feedback on experimental design and on the interpretation of results; the research questions,
the experimental protocol, and the conclusions drawn are the authors'.
\emph{Not used or not applicable.} We did not use generative AI to develop theoretical models,
formulate or prove mathematical claims, or translate text; the paper contains no proofs. Aside
from the labeling above, no model judged any result.
\emph{Writing (used).} AI assistants helped with literature search, feedback on the manuscript, and
drafting and editing of text, including figures and tables from the analysis outputs. The authors
reviewed all AI-assisted work, checked every number against its source, and take full
responsibility for the content of this paper, including text, claims, and artifacts produced with
the aid of generative AI.

\subsection*{Ethics Statement}
BOBA-Bench is built from the public histories of 140 open-source Python repositories, all under
open-source licenses: MIT (65), Apache-2.0 (36), BSD (19), GPL-3.0 (12), LGPL-3.0 (5), and one each under
GPL-2.0-or-later, LGPL-2.1, and AGPL-3.0 (\hyperref[app:units]{Appendix~\ref*{app:units}}).
Source code stays under each repository's license. No human subjects took part.
\hyperref[app:qa]{Appendix~\ref*{app:qa}} describes grader-defeat techniques found during construction so that
benchmark builders can defend against them; the grader rejects~them.%

\subsection*{Reproducibility Statement}
\hyperref[sec:benchmark]{Section~\ref*{sec:benchmark}} and \hyperref[app:qa]{Appendix~\ref*{app:qa}} describe construction end to end: mining and
contract validation, the verbatim specification prompt (\hyperref[app:qa-specgen]{Appendix~\ref*{app:qa-specgen}}), the
specification audit, environments and sealing, grader rules, and every filtering round, including
the post hoc 14-chain exclusion (\hyperref[app:qa-filtering]{Appendix~\ref*{app:qa-filtering}}). \hyperref[app:units]{Appendix~\ref*{app:units}}
reports task statistics. \hyperref[sec:evalsetup]{Section~\ref*{sec:evalsetup}} and \hyperref[app:eval]{Appendix~\ref*{app:eval}} give models,
settings, and budgets. \hyperref[app:mf]{Appendices~\ref*{app:mf}}, \ref{app:cv}, and~\ref{app:resume-results} give
the per-model estimates, and \hyperref[app:separation]{Appendix~\ref*{app:separation}} gives the paired tests with their seed.
\hyperref[app:repro]{Appendix~\ref*{app:repro}} lists the population of each study.
The 280 tasks with image digests, the grader, and the analysis scripts are released with the
benchmark (\hyperref[sec:release]{Section~\ref*{sec:release}}).

\newpage%
\bibliography{references,new_references,new_models}

\begin{thebibliography}{64}
\providecommand{\natexlab}[1]{#1}
\providecommand{\url}[1]{\texttt{#1}}
\expandafter\ifx\csname urlstyle\endcsname\relax
  \providecommand{\doi}[1]{doi: #1}\else
  \providecommand{\doi}{doi: \begingroup \urlstyle{rm}\Url}\fi

\bibitem[Advani(2026)]{advani2026falsesuccess}
Laksh Advani.
\newblock From confident closing to silent failure: Characterizing false
  success in {LLM} agents.
\newblock \emph{arXiv preprint arXiv:2606.09863}, June 2026.
\newblock URL \url{https://arxiv.org/abs/2606.09863}.
\newblock FAGEN Workshop at ICML 2026.

\bibitem[Aleithan et~al.(2024)Aleithan, Xue, Mohajer, Nnorom, Uddin, and
  Wang]{aleithan2024swebenchplus}
Reem Aleithan, Haoran Xue, Mohammad~Mahdi Mohajer, Elijah Nnorom, Gias Uddin,
  and Song Wang.
\newblock {SWE-Bench+:} enhanced coding benchmark for {LLMs}.
\newblock \emph{arXiv preprint arXiv:2410.06992}, October 2024.
\newblock URL \url{https://arxiv.org/abs/2410.06992}.

\bibitem[{Anthropic}(2025)]{anthropic2025claudecode}
{Anthropic}.
\newblock {Claude 3.7 Sonnet and Claude Code}.
\newblock \url{https://www.anthropic.com/news/claude-3-7-sonnet}, February
  2025.
\newblock Announcement, February 24, 2025; documentation at
  \url{https://code.claude.com/docs}.

\bibitem[{Anthropic}(2026{\natexlab{a}})]{anthropic2026claudefable5}
{Anthropic}.
\newblock {Claude Fable 5 and Claude Mythos 5}.
\newblock \url{https://www.anthropic.com/news/claude-fable-5-mythos-5}, June
  2026{\natexlab{a}}.
\newblock Announcement, June 9, 2026; system card at
  \url{https://anthropic.com/claude-fable-5-mythos-5-system-card}.

\bibitem[{Anthropic}(2026{\natexlab{b}})]{anthropic2026claudeopus48}
{Anthropic}.
\newblock {Introducing Claude Opus 4.8}.
\newblock \url{https://www.anthropic.com/news/claude-opus-4-8}, May
  2026{\natexlab{b}}.
\newblock Announcement, May 28, 2026; system card at
  \url{https://www.anthropic.com/claude-opus-4-8-system-card}.

\bibitem[{Anthropic}(2026{\natexlab{c}})]{anthropic2026claudeopus5}
{Anthropic}.
\newblock {Introducing Claude Opus 5}.
\newblock \url{https://www.anthropic.com/news/claude-opus-5}, July
  2026{\natexlab{c}}.
\newblock Announcement, July 24, 2026; system card at
  \url{https://www.anthropic.com/claude-opus-5-system-card}.

\bibitem[{Anthropic}(2026{\natexlab{d}})]{anthropic2026claudeopus55}
{Anthropic}.
\newblock {Introducing Claude Opus 5.5}.
\newblock \url{https://www.anthropic.com/news/claude-opus-5-5}, September
  2026{\natexlab{d}}.
\newblock Announcement, September 22, 2026; system card at
  \url{https://anthropic.com/claude-opus-5-5-system-card}.

\bibitem[Badertdinov et~al.(2025)Badertdinov, Golubev, Nekrashevich, Shevtsov,
  Karasik, Andriushchenko, Trofimova, Litvintseva, and
  Yangel]{badertdinov2025swerebench}
Ibragim Badertdinov, Alexander Golubev, Maksim Nekrashevich, Anton Shevtsov,
  Simon Karasik, Andrei Andriushchenko, Maria Trofimova, Daria Litvintseva, and
  Boris Yangel.
\newblock {SWE-rebench:} an automated pipeline for task collection and
  decontaminated evaluation of software engineering agents.
\newblock In \emph{Advances in Neural Information Processing Systems (NeurIPS),
  Datasets and Benchmarks Track}, 2025.
\newblock URL \url{https://arxiv.org/abs/2505.20411}.
\newblock arXiv:2505.20411, May 2025.

\bibitem[Badertdinov et~al.(2026)Badertdinov, Nekrashevich, Shevtsov, and
  Golubev]{badertdinov2026swerebenchv2}
Ibragim Badertdinov, Maksim Nekrashevich, Anton Shevtsov, and Alexander
  Golubev.
\newblock {SWE-rebench V2:} {Language-Agnostic} {SWE} task collection at scale.
\newblock In \emph{International Conference on Machine Learning (ICML)}, 2026.
\newblock URL \url{https://arxiv.org/abs/2602.23866}.
\newblock arXiv:2602.23866, Feb 2026.

\bibitem[Chen et~al.(2026)Chen, Xu, Wei, Chen, and Zhao]{sweci2026}
Jialong Chen, Xander Xu, Hu~Wei, Chuan Chen, and Bing Zhao.
\newblock {SWE-CI:} evaluating agent capabilities in maintaining codebases via
  continuous integration.
\newblock \emph{arXiv preprint arXiv:2603.03823}, March 2026.
\newblock URL \url{https://arxiv.org/abs/2603.03823}.

\bibitem[Chen et~al.(2021)Chen, Tworek, Jun, Yuan, de~Oliveira~Pinto, Kaplan,
  Edwards, Burda, Joseph, Brockman, Ray, Puri, Krueger, Petrov, Khlaaf, Sastry,
  Mishkin, Chan, Gray, Ryder, Pavlov, Power, Kaiser, Bavarian, Winter, Tillet,
  Such, Cummings, Plappert, Chantzis, Barnes, Herbert-Voss, Guss, Nichol,
  Paino, Tezak, Tang, Babuschkin, Balaji, Jain, Saunders, Hesse, Carr, Leike,
  Achiam, Misra, Morikawa, Radford, Knight, Brundage, Murati, Mayer, Welinder,
  McGrew, Amodei, McCandlish, Sutskever, and Zaremba]{chen2021codex}
Mark Chen, Jerry Tworek, Heewoo Jun, Qiming Yuan, Henrique~Ponde
  de~Oliveira~Pinto, Jared Kaplan, Harri Edwards, Yuri Burda, Nicholas Joseph,
  Greg Brockman, Alex Ray, Raul Puri, Gretchen Krueger, Michael Petrov, Heidy
  Khlaaf, Girish Sastry, Pamela Mishkin, Brooke Chan, Scott Gray, Nick Ryder,
  Mikhail Pavlov, Alethea Power, Lukasz Kaiser, Mohammad Bavarian, Clemens
  Winter, Philippe Tillet, Felipe~Petroski Such, Dave Cummings, Matthias
  Plappert, Fotios Chantzis, Elizabeth Barnes, Ariel Herbert-Voss,
  William~Hebgen Guss, Alex Nichol, Alex Paino, Nikolas Tezak, Jie Tang, Igor
  Babuschkin, Suchir Balaji, Shantanu Jain, William Saunders, Christopher
  Hesse, Andrew~N. Carr, Jan Leike, Josh Achiam, Vedant Misra, Evan Morikawa,
  Alec Radford, Matthew Knight, Miles Brundage, Mira Murati, Katie Mayer, Peter
  Welinder, Bob McGrew, Dario Amodei, Sam McCandlish, Ilya Sutskever, and
  Wojciech Zaremba.
\newblock Evaluating large language models trained on code.
\newblock \emph{arXiv preprint arXiv:2107.03374}, July 2021.
\newblock URL \url{https://arxiv.org/abs/2107.03374}.

\bibitem[Chen et~al.(2024)Chen, Lin, Schärli, and Zhou]{chen2023selfdebug}
Xinyun Chen, Maxwell Lin, Nathanael Schärli, and Denny Zhou.
\newblock Teaching large language models to {Self-Debug}.
\newblock In \emph{International Conference on Learning Representations
  (ICLR)}, 2024.
\newblock URL \url{https://arxiv.org/abs/2304.05128}.
\newblock arXiv:2304.05128, Apr 2023.

\bibitem[Cuadron et~al.(2025)Cuadron, Yu, Liu, and Gupta]{cuadron2025saber}
Alejandro Cuadron, Pengfei Yu, Yang Liu, and Arpit Gupta.
\newblock {SABER:} small actions, big errors -- safeguarding mutating steps in
  {LLM} agents.
\newblock \emph{arXiv preprint arXiv:2512.07850}, November 2025.
\newblock URL \url{https://arxiv.org/abs/2512.07850}.

\bibitem[Deng et~al.(2026)Deng, Chen, Yu, Fan, Liu, Yang, Parikh, Kannan, Cong,
  Wang, Zhang, Prasanna, Tang, and Wang]{swemilestone2026}
Gangda Deng, Zhaoling Chen, Zhongming Yu, Haoyang Fan, Yuhong Liu, Yuxin Yang,
  Dhruv Parikh, Rajgopal Kannan, Le~Cong, Mengdi Wang, Qian Zhang, Viktor
  Prasanna, Xiangru Tang, and Xingyao Wang.
\newblock {SWE-Milestone:} evaluating {AI} agents on continuous software
  evolution.
\newblock In \emph{International Conference on Machine Learning (ICML)}, 2026.
\newblock URL \url{https://arxiv.org/abs/2603.13428}.
\newblock arXiv:2603.13428, Mar 2026.

\bibitem[Deng et~al.(2025)Deng, Da, Pan, He, Ide, Garg, Lauffer, Park, Pasari,
  Rane, Sampath, Krishnan, Kundurthy, Hendryx, Wang, Bharadwaj, Holm, Aluri,
  Zhang, Jacobson, Liu, and Kenstler]{deng2025swebenchpro}
Xiang Deng, Jeff Da, Edwin Pan, Yannis~Yiming He, Charles Ide, Kanak Garg,
  Niklas Lauffer, Andrew Park, Nitin Pasari, Chetan Rane, Karmini Sampath, Maya
  Krishnan, Srivatsa Kundurthy, Sean Hendryx, Zifan Wang, Vijay Bharadwaj, Jeff
  Holm, Raja Aluri, Chen Bo~Calvin Zhang, Noah Jacobson, Bing Liu, and Brad
  Kenstler.
\newblock {SWE-Bench} pro: Can {AI} agents solve {Long-Horizon} software
  engineering tasks?
\newblock \emph{arXiv preprint arXiv:2509.16941}, September 2025.
\newblock URL \url{https://arxiv.org/abs/2509.16941}.

\bibitem[Duma et~al.(2026)Duma, Wróblewski, Bobińska, Winiarska, and
  Przymus]{duma2026reviewsaiprs}
Kacper Duma, Patryk Wróblewski, Jagoda Bobińska, Julia Winiarska, and Piotr
  Przymus.
\newblock These aren't the reviews you're looking for: How humans review
  {AI-Generated} pull requests.
\newblock \emph{arXiv preprint arXiv:2605.02273}, May 2026.
\newblock URL \url{https://arxiv.org/abs/2605.02273}.
\newblock EASE 2026.

\bibitem[Fu et~al.(2026)Fu, Wu, Wu, Peng, Huang, Sun, Zeng, Jiang, Zhang, Li,
  Hu, Liu, Hou, and Liu]{fu2026davincienv}
Dayuan Fu, Shenyu Wu, Yunze Wu, Zerui Peng, Yaxing Huang, Jie Sun, Ji~Zeng,
  Mohan Jiang, Lin Zhang, Yukun Li, Jiarui Hu, Liming Liu, Jinlong Hou, and
  Pengfei Liu.
\newblock {daVinci-Env:} open {SWE} environment synthesis at scale.
\newblock \emph{arXiv preprint arXiv:2603.13023}, March 2026.
\newblock URL \url{https://arxiv.org/abs/2603.13023}.

\bibitem[Garousi(2026)]{garousi2026oversightoverload}
Vahid Garousi.
\newblock Human oversight and overload: Two hidden and costly burdens of
  {AI-Assisted} software engineering.
\newblock \emph{arXiv preprint arXiv:2606.05770}, June 2026.
\newblock URL \url{https://arxiv.org/abs/2606.05770}.

\bibitem[Huang et~al.(2026)Huang, Lee, Tng, and Ge]{huang2026deepswe}
Wenqi Huang, Charley Lee, Leonard Tng, and Serena Ge.
\newblock {DeepSWE:} measuring frontier coding agents on original,
  {Long-Horizon} engineering tasks.
\newblock \emph{arXiv preprint arXiv:2607.07946}, July 2026.
\newblock URL \url{https://arxiv.org/abs/2607.07946}.

\bibitem[Jimenez et~al.(2024)Jimenez, Yang, Wettig, Yao, Pei, Press, and
  Narasimhan]{jimenez2024swebench}
Carlos~E. Jimenez, John Yang, Alexander Wettig, Shunyu Yao, Kexin Pei, Ofir
  Press, and Karthik Narasimhan.
\newblock {SWE-bench:} can language models resolve {Real-World} {GitHub}
  issues?
\newblock In \emph{International Conference on Learning Representations
  (ICLR)}, 2024.
\newblock URL \url{https://arxiv.org/abs/2310.06770}.
\newblock arXiv:2310.06770, Oct 2023.

\bibitem[Jin et~al.(2026)Jin, Tung, Li, Shi, She, Jia, Zhao, Xia, Du, Pei,
  Zhang, Qi, Duan, Lee, and Jin]{jin2026chainswe}
Qirui Jin, Lingching Tung, Kenan Li, Qiyang Shi, Yushi She, Huanzhong Jia,
  Harrison Zhao, Kejing Xia, Zhenbang Du, Jiaxin Pei, Zhenyu Zhang, Zhen Qi,
  Yuyan Duan, Wenke Lee, and Zijian Jin.
\newblock {ChainSWE:} benchmarking coding agents on {Multi-Bug} software
  maintenance.
\newblock \emph{arXiv preprint arXiv:2607.02606}, July 2026.
\newblock URL \url{https://arxiv.org/abs/2607.02606}.

\bibitem[Krentsel et~al.(2026)Krentsel, Agarwal, Cemri, Liu, Sankhe, Mao,
  Zaharia, and Stoica]{krentsel2026reality}
Alexander Krentsel, Shubham Agarwal, Mert Cemri, Shu Liu, Sidharth Sankhe,
  Ziming Mao, Matei Zaharia, and Ion Stoica.
\newblock Reality is the final verifier: On two key gaps in agentic software
  engineering.
\newblock \emph{arXiv preprint arXiv:2609.12039}, September 2026.
\newblock URL \url{https://arxiv.org/abs/2609.12039}.

\bibitem[Laban et~al.(2025)Laban, Hayashi, Zhou, and
  Neville]{laban2025lostmultiturn}
Philippe Laban, Hiroaki Hayashi, Yingbo Zhou, and Jennifer Neville.
\newblock {LLMs} get lost in {Multi-Turn} conversation.
\newblock \emph{arXiv preprint arXiv:2505.06120}, May 2025.
\newblock URL \url{https://arxiv.org/abs/2505.06120}.

\bibitem[Lam et~al.(2026)Lam, Wang, Liu, Xiao, sing Li, tse Huang, Zhuo, and
  Lyu]{swechain2026}
Man~Ho Lam, Chaozheng Wang, Hange Liu, Jingyu Xiao, Haau sing Li, Jen tse
  Huang, Terry~Yue Zhuo, and Michael~R. Lyu.
\newblock {SWE-Chain:} benchmarking coding agents on chained {Release-Level}
  package upgrades.
\newblock \emph{arXiv preprint arXiv:2605.14415}, May 2026.
\newblock URL \url{https://arxiv.org/abs/2605.14415}.

\bibitem[Le et~al.(2025)Le, Thai, Manh, Nhat, and Bui]{le2025sweevo}
Tue Le, Minh V.~T. Thai, Dung~Nguyen Manh, Huy~Phan Nhat, and Nghi D.~Q. Bui.
\newblock {SWE-EVO:} benchmarking coding agents in {Long-Horizon} software
  evolution scenarios.
\newblock \emph{arXiv preprint arXiv:2512.18470}, December 2025.
\newblock URL \url{https://arxiv.org/abs/2512.18470}.

\bibitem[Liu et~al.(2026{\natexlab{a}})Liu, Qiu, Íñigo Goiri, Fonseca,
  Bianchini, and Choukse]{liu2026copilottraces}
Banruo Liu, Haoran Qiu, Íñigo Goiri, Rodrigo Fonseca, Ricardo Bianchini, and
  Esha Choukse.
\newblock Agentic coding in the wild: Characterizing {GitHub} copilot traces at
  production scale.
\newblock \emph{arXiv preprint arXiv:2608.00101}, July 2026{\natexlab{a}}.
\newblock URL \url{https://arxiv.org/abs/2608.00101}.

\bibitem[Liu et~al.(2026{\natexlab{b}})Liu, Liu, Sun, Luo, and
  Guo]{liu2026converged}
Fengshuo Liu, Ying Liu, Ruize Sun, Lie Luo, and Siyuan Guo.
\newblock Coding agents have converged: Why the {SWE-bench} leaderboard can no
  longer order its top entries, and what to measure instead.
\newblock In \emph{International Conference on Advanced Data Mining and
  Applications (ADMA)}, 2026{\natexlab{b}}.
\newblock URL \url{https://arxiv.org/abs/2609.17394}.
\newblock arXiv:2609.17394, Sep 2026.

\bibitem[Mehta(2026)]{confidentwrong2026}
Aman Mehta.
\newblock Confident and wrong: Silent semantic failures in coding agents.
\newblock \emph{arXiv preprint arXiv:2603.25764}, March 2026.
\newblock URL \url{https://arxiv.org/abs/2603.25764}.

\bibitem[{Moonshot AI}(2026)]{moonshot2026kimik3}
{Moonshot AI}.
\newblock {Kimi K3: Open Frontier Intelligence}.
\newblock \url{https://github.com/MoonshotAI/Kimi-K3}, July 2026.
\newblock Model card and open weights, released July 16, 2026.

\bibitem[Mündler et~al.(2024)Mündler, Müller, He, and
  Vechev]{mundler2024swtbench}
Niels Mündler, Mark~Niklas Müller, Jingxuan He, and Martin Vechev.
\newblock {SWT-Bench:} testing and validating {Real-World} {Bug-Fixes} with
  code agents.
\newblock In \emph{Advances in Neural Information Processing Systems
  (NeurIPS)}, 2024.
\newblock URL \url{https://arxiv.org/abs/2406.12952}.
\newblock arXiv:2406.12952, Jun 2024.

\bibitem[Olausson et~al.(2024)Olausson, Inala, Wang, Gao, and
  Solar-Lezama]{olausson2024selfrepair}
Theo~X. Olausson, Jeevana~Priya Inala, Chenglong Wang, Jianfeng Gao, and
  Armando Solar-Lezama.
\newblock Is {Self-Repair} a silver bullet for code generation?
\newblock In \emph{International Conference on Learning Representations
  (ICLR)}, 2024.
\newblock URL \url{https://arxiv.org/abs/2306.09896}.
\newblock arXiv:2306.09896, Jun 2023.

\bibitem[{OpenAI}(2024)]{openai2024swebenchverified}
{OpenAI}.
\newblock Introducing {SWE-bench} {Verified}.
\newblock \url{https://openai.com/index/introducing-swe-bench-verified/};
  dataset
  \url{https://huggingface.co/datasets/princeton-nlp/SWE-bench_Verified},
  August 2024.

\bibitem[{OpenAI}(2025)]{openai2025codex}
{OpenAI}.
\newblock {Introducing Codex}.
\newblock \url{https://openai.com/index/introducing-codex/}, May 2025.
\newblock Announcement, May 16, 2025; cloud software-engineering agent, with
  the open-source Codex CLI at \url{https://github.com/openai/codex}.

\bibitem[{OpenAI}(2026{\natexlab{a}})]{openai2026gpt56sol}
{OpenAI}.
\newblock {GPT-5.6: Frontier intelligence that scales with your ambition}.
\newblock \url{https://openai.com/index/gpt-5-6/}, July 2026{\natexlab{a}}.
\newblock Released July 9, 2026; GPT-5.6 Sol is the flagship model of the
  family; system card at \url{https://deploymentsafety.openai.com/gpt-5-6}.

\bibitem[{OpenAI}(2026{\natexlab{b}})]{openai2026gpt61sol}
{OpenAI}.
\newblock {Introducing GPT-6.1 Sol}.
\newblock \url{https://openai.com/index/introducing-gpt-6-1-sol/}, September
  2026{\natexlab{b}}.
\newblock Released September 29, 2026; system card addendum at
  \url{https://deploymentsafety.openai.com/gpt-6-1-sol}.

\bibitem[{OpenAI}(2026{\natexlab{c}})]{openai2026gpt6astra}
{OpenAI}.
\newblock {GPT-6 Astra: A new generation of intelligence}.
\newblock \url{https://openai.com/index/gpt-6-astra/}, September
  2026{\natexlab{c}}.
\newblock Released September 3, 2026; system card at
  \url{https://deploymentsafety.openai.com/gpt-6-astra}.

\bibitem[{OpenAI}(2026{\natexlab{d}})]{openai2026signalnoise}
{OpenAI}.
\newblock Separating signal from noise in coding evaluations.
\newblock
  \url{https://openai.com/index/separating-signal-from-noise-coding-evaluations/},
  July 2026{\natexlab{d}}.
\newblock Audit of SWE-Bench Pro: ${\approx}30\%$ of tasks judged broken,
  including underspecified and misleading prompts.

\bibitem[{OpenAI}(2026{\natexlab{e}})]{openai2026swebenchverified}
{OpenAI}.
\newblock Why {SWE-bench} {Verified} no longer measures frontier coding
  capabilities.
\newblock
  \url{https://openai.com/index/why-we-no-longer-evaluate-swe-bench-verified/},
  February 2026{\natexlab{e}}.

\bibitem[Orlanski et~al.(2026)Orlanski, Roy, Yun, Shin, Gu, Ge, Adila, Roberts,
  Sala, and Albarghouthi]{slopcodebench2026}
Gabriel Orlanski, Devjeet Roy, Alexander Yun, Changho Shin, Alex Gu, Albert Ge,
  Dyah Adila, Nicholas Roberts, Frederic Sala, and Aws Albarghouthi.
\newblock {SlopCodeBench:} benchmarking how coding agents degrade over
  {Long-Horizon} iterative tasks.
\newblock \emph{arXiv preprint arXiv:2603.24755}, March 2026.
\newblock URL \url{https://arxiv.org/abs/2603.24755}.

\bibitem[Patel et~al.(2026)Patel, Hou, Purohit, Xu, Pan, He, and
  Chen]{patel2026codethread}
Shaswat Patel, Betty~Li Hou, Arun Purohit, Kai Xu, Jane Pan, He~He, and Valerie
  Chen.
\newblock Is agent code less maintainable than human code?
\newblock \emph{arXiv preprint arXiv:2606.21804}, June 2026.
\newblock URL \url{https://arxiv.org/abs/2606.21804}.

\bibitem[Paul et~al.(2025)Paul, Zhu, and Bayley]{paul2025smells}
Debalina~Ghosh Paul, Hong Zhu, and Ian Bayley.
\newblock Investigating the smells of {LLM} generated code.
\newblock \emph{arXiv preprint arXiv:2510.03029}, October 2025.
\newblock URL \url{https://arxiv.org/abs/2510.03029}.

\bibitem[Peng et~al.(2026)Peng, Lyu, Dong, Dong, and Lin]{peng2026residual}
Chao Peng, Zhiheng Lyu, Peijie Dong, Hande Dong, and Qiang Lin.
\newblock Benchmarking the residual: What {Long-Horizon} evaluations add beyond
  matched {Short-Task} performance.
\newblock \emph{arXiv preprint arXiv:2607.27283}, July 2026.
\newblock URL \url{https://arxiv.org/abs/2607.27283}.

\bibitem[Rahman et~al.(2026)Rahman, Kim, Parmar, Heydari, Liu, Lee, Zhang,
  Hosseini, Metwally, Yang, Mirzasoleiman, Liu, Izmailov, Gabriel, Malhotra,
  Patel, McDuff, and Palangi]{traverse2026}
Salman Rahman, Yubin Kim, Mihir Parmar, A.~Ali Heydari, Genglin Liu, Simon~A.
  Lee, Weizhi Zhang, Arian Hosseini, Ahmed~A. Metwally, Yuzhe Yang, Baharan
  Mirzasoleiman, Xin Liu, Pavel Izmailov, Saadia Gabriel, Mark Malhotra,
  Shwetak Patel, Daniel McDuff, and Hamid Palangi.
\newblock Locating hidden failures makes {Long-Horizon} agents more reliable.
\newblock \emph{arXiv preprint arXiv:2609.17930}, September 2026.
\newblock URL \url{https://arxiv.org/abs/2609.17930}.

\bibitem[Shastry et~al.(2026)Shastry, Senrayan, Satapara, Panda, and
  Devaguptapu]{shastry2026swesteps}
KN~Ajay Shastry, Ganesh Senrayan, Shrey Satapara, Pranoy Panda, and Chaitanya
  Devaguptapu.
\newblock Beyond isolated tasks: A framework for evaluating coding agents on
  sequential software evolution.
\newblock \emph{arXiv preprint arXiv:2604.03035}, April 2026.
\newblock URL \url{https://arxiv.org/abs/2604.03035}.

\bibitem[Shen et~al.(2026)Shen, Chen, Xu, Ma, Chen, and
  Li]{shen2026evocodebench}
Haiyang Shen, Xuanzhong Chen, Wendong Xu, Yun Ma, Liang Chen, and Kuan Li.
\newblock {EvoCode-Bench:} evaluating coding agents in {Multi-Turn} iterative
  interactions.
\newblock \emph{arXiv preprint arXiv:2605.24110}, May 2026.
\newblock URL \url{https://arxiv.org/abs/2605.24110}.

\bibitem[Smyth et~al.(2026)Smyth, Mantilla-Ramos, Notsawo, Helbling, Tosato,
  Merzouk, Dziri, Gidel, and Tosato]{smyth2026overclaim}
Nolan Smyth, Yorguin-Jose Mantilla-Ramos, Pascal Jr~Tikeng Notsawo, Saskia
  Helbling, Alberto Tosato, Mohamed~Amine Merzouk, Nouha Dziri, Gauthier Gidel,
  and Tommaso Tosato.
\newblock Quantifying overclaiming propensity in frontier {LLM} agents.
\newblock \emph{arXiv preprint arXiv:2609.20812}, September 2026.
\newblock URL \url{https://arxiv.org/abs/2609.20812}.

\bibitem[Sobal et~al.(2026)Sobal, Yang, Zhang, Xia, and
  Soatto]{sobal2026staminabench}
Vlad Sobal, Shuo Yang, Yuting Zhang, Wei Xia, and Stefano Soatto.
\newblock {StaminaBench:} {Stress-Testing} coding agents over 100 interaction
  turns.
\newblock \emph{arXiv preprint arXiv:2606.19613}, June 2026.
\newblock URL \url{https://arxiv.org/abs/2606.19613}.

\bibitem[Sun et~al.(2026)Sun, St{\aa}hl, Sandahl, and
  Kessler]{sun2026nonfunctional}
Xin Sun, Daniel St{\aa}hl, Kristian Sandahl, and Christoph Kessler.
\newblock Beyond resolved rate: A {Non-Functional} quality study.
\newblock \emph{arXiv preprint arXiv:2607.18462}, July 2026.
\newblock URL \url{https://arxiv.org/abs/2607.18462}.

\bibitem[Tao et~al.(2026)Tao, Peng, Wang, Wang, Cheng, Yao, Wu, Ge, Li, and
  Gao]{execcritic2026}
Leitian Tao, Baolin Peng, Haorui Wang, Hang Wang, Hao Cheng, Wenlin Yao,
  Qianhui Wu, Tao Ge, Sharon Li, and Jianfeng Gao.
\newblock {ExecCritic:} learn to test, test to improve for coding agents.
\newblock \emph{arXiv preprint arXiv:2609.09133}, September 2026.
\newblock URL \url{https://arxiv.org/abs/2609.09133}.

\bibitem[Wang et~al.(2026)Wang, Chen, Wang, Tao, Yang, Jiang, Yap, Shang, Li,
  and Bai]{swereview2026}
Ruoyu Wang, Jierun Chen, Shaowei Wang, Chaofan Tao, Sidi Yang, Yuxin Jiang,
  Kim-Hui Yap, Lifeng Shang, Xiaohui Li, and Haoli Bai.
\newblock {SWE-Review:} closing the loop on issue resolution with agentic code
  review.
\newblock \emph{arXiv preprint arXiv:2607.06065}, July 2026.
\newblock URL \url{https://arxiv.org/abs/2607.06065}.

\bibitem[Wang et~al.(2025)Wang, Li, Song, Xu, Tang, Zhuge, Pan, Song, Li,
  Singh, Tran, Li, Ma, Zheng, Qian, Shao, Muennighoff, Zhang, Hui, Lin,
  Brennan, Peng, Ji, and Neubig]{wang2024openhands}
Xingyao Wang, Boxuan Li, Yufan Song, Frank~F. Xu, Xiangru Tang, Mingchen Zhuge,
  Jiayi Pan, Yueqi Song, Bowen Li, Jaskirat Singh, Hoang~H. Tran, Fuqiang Li,
  Ren Ma, Mingzhang Zheng, Bill Qian, Yanjun Shao, Niklas Muennighoff, Yizhe
  Zhang, Binyuan Hui, Junyang Lin, Robert Brennan, Hao Peng, Heng Ji, and
  Graham Neubig.
\newblock {OpenHands}: An open platform for {AI} software developers as
  generalist agents.
\newblock In \emph{International Conference on Learning Representations
  (ICLR)}, 2025.
\newblock URL \url{https://arxiv.org/abs/2407.16741}.
\newblock arXiv:2407.16741, Jul 2024.

\bibitem[Winninger(2026)]{winninger2026steerability}
Thomas Winninger.
\newblock Steerability via constraints: a substrate for scalable oversight of
  coding agents.
\newblock \emph{arXiv preprint arXiv:2607.02389}, July 2026.
\newblock URL \url{https://arxiv.org/abs/2607.02389}.
\newblock Deep Learning for Code Workshop at ICML 2026.

\bibitem[Yang et~al.(2024)Yang, Jimenez, Wettig, Lieret, Yao, Narasimhan, and
  Press]{yang2024sweagent}
John Yang, Carlos~E. Jimenez, Alexander Wettig, Kilian Lieret, Shunyu Yao,
  Karthik Narasimhan, and Ofir Press.
\newblock {SWE-agent:} {Agent-Computer} interfaces enable automated software
  engineering.
\newblock In \emph{Advances in Neural Information Processing Systems
  (NeurIPS)}, 2024.
\newblock URL \url{https://arxiv.org/abs/2405.15793}.
\newblock arXiv:2405.15793, May 2024.

\bibitem[Yang et~al.(2026)Yang, Lieret, Ma, Thakkar, Pedchenko, Sootla,
  McMilin, Yin, Hou, Synnaeve, Yang, and Press]{yang2026programbench}
John Yang, Kilian Lieret, Jeffrey Ma, Parth Thakkar, Dmitrii Pedchenko, Sten
  Sootla, Emily McMilin, Pengcheng Yin, Rui Hou, Gabriel Synnaeve, Diyi Yang,
  and Ofir Press.
\newblock {ProgramBench:} can language models rebuild programs from scratch?
\newblock \emph{arXiv preprint arXiv:2605.03546}, May 2026.
\newblock URL \url{https://arxiv.org/abs/2605.03546}.

\bibitem[Ye et~al.(2026)Ye, Zou, Yu, and Shi]{ye2026sabotage}
Jingheng Ye, Huiqi Zou, Simon Yu, and Weiyan Shi.
\newblock Coding with ``enemy'': Can human developers detect {AI} agent
  sabotage?
\newblock \emph{arXiv preprint arXiv:2606.05647}, June 2026.
\newblock URL \url{https://arxiv.org/abs/2606.05647}.

\bibitem[Yu et~al.(2025)Yu, Zhu, He, and Kang]{yu2025utboost}
Boxi Yu, Yuxuan Zhu, Pinjia He, and Daniel Kang.
\newblock {UTBoost:} rigorous evaluation of coding agents on {SWE-Bench}.
\newblock In \emph{Annual Meeting of the Association for Computational
  Linguistics (ACL)}, 2025.
\newblock URL \url{https://arxiv.org/abs/2506.09289}.
\newblock arXiv:2506.09289, Jun 2025.

\bibitem[Yu et~al.(2026)Yu, Liu, Jiang, Jia, Wang, Qian, and
  Chen]{yu2026habituation}
Haoran Yu, Lifei Liu, Xiaochong Jiang, Yuwen Jia, Su~Wang, Pin Qian, and Yihang
  Chen.
\newblock Habituation at the gate: Rising approval and declining scrutiny in
  human review of {AI} agent code.
\newblock \emph{arXiv preprint arXiv:2606.22721}, June 2026.
\newblock URL \url{https://arxiv.org/abs/2606.22721}.
\newblock KDD 2026 Workshop on Agentic Software Engineering (SE 3.0).

\bibitem[Zhao et~al.(2025)Zhao, Jiang, Lee, Chiu, Cardie, Gall{\'e}, and
  Rush]{zhao2024commit0}
Wenting Zhao, Nan Jiang, Celine Lee, Justin~T Chiu, Claire Cardie, Matthias
  Gall{\'e}, and Alexander~M Rush.
\newblock {Commit0:} library generation from scratch.
\newblock In \emph{International Conference on Learning Representations
  (ICLR)}, 2025.
\newblock URL \url{https://arxiv.org/abs/2412.01769}.
\newblock arXiv:2412.01769, Dec 2024.

\bibitem[{Zhipu AI}(2026{\natexlab{a}})]{zhipu2026glm52}
{Zhipu AI}.
\newblock {GLM-5.2: Built for Long-Horizon Tasks}.
\newblock \url{https://z.ai/blog/glm-5.2}, June 2026{\natexlab{a}}.
\newblock Released June 16, 2026.

\bibitem[{Zhipu AI}(2026{\natexlab{b}})]{zhipu2026glm53}
{Zhipu AI}.
\newblock {GLM-5.3: Frontier Coding with Emergent Cyber Capabilities}.
\newblock \url{https://z.ai/blog/glm-5.3}, August 2026{\natexlab{b}}.
\newblock Released August 14, 2026; model card at
  \url{https://huggingface.co/zai-org/GLM-5.3}.

\bibitem[{Zhipu AI}(2026{\natexlab{c}})]{zhipu2026glm53flash}
{Zhipu AI}.
\newblock {GLM-5.3-Flash: Frontier Intelligence, Flash Cost}.
\newblock \url{https://z.ai/blog/glm-5.3-flash}, August 2026{\natexlab{c}}.
\newblock Released August 26, 2026; open weights and model card at
  \url{https://huggingface.co/zai-org/GLM-5.3-Flash} (the BF16 weights we
  served: \url{https://huggingface.co/zai-org/GLM-5.3-Flash-BF16}).

\bibitem[Zhong et~al.(2025)Zhong, Raghunathan, and
  Carlini]{zhong2025impossiblebench}
Ziqian Zhong, Aditi Raghunathan, and Nicholas Carlini.
\newblock {ImpossibleBench:} measuring {LLMs'} propensity of exploiting test
  cases.
\newblock \emph{arXiv preprint arXiv:2510.20270}, October 2025.
\newblock URL \url{https://arxiv.org/abs/2510.20270}.

\bibitem[Zhou et~al.(2026)Zhou, Zhang, Wang, Hao, Wang, Han, Yang, Wu, Pan,
  Fan, Tu, and Zhang]{zhou2026featurebench}
Qixing Zhou, Jiacheng Zhang, Haiyang Wang, Rui Hao, Jiahe Wang, Minghao Han,
  Yuxue Yang, Shuzhe Wu, Feiyang Pan, Lue Fan, Dandan Tu, and Zhaoxiang Zhang.
\newblock {FeatureBench:} benchmarking agentic coding for complex feature
  development.
\newblock In \emph{International Conference on Learning Representations
  (ICLR)}, 2026.
\newblock URL \url{https://arxiv.org/abs/2602.10975}.
\newblock arXiv:2602.10975, Feb 2026.

\bibitem[Zhu et~al.(2025)Zhu, Liu, Li, Tian, Yang, Zhang, Han, Xie, Cui, Zhang,
  Ma, Yu, Ramesh, Wu, Liu, Lu, Zou, and You]{zhu2025agenterrortaxonomy}
Kunlun Zhu, Zijia Liu, Bingxuan Li, Muxin Tian, Yingxuan Yang, Jiaxun Zhang,
  Pengrui Han, Qipeng Xie, Fuyang Cui, Weijia Zhang, Xiaoteng Ma, Xiaodong Yu,
  Gowtham Ramesh, Jialian Wu, Zicheng Liu, Pan Lu, James Zou, and Jiaxuan You.
\newblock Where {LLM} agents fail and how they can learn from failures.
\newblock \emph{arXiv preprint arXiv:2509.25370}, September 2025.
\newblock URL \url{https://arxiv.org/abs/2509.25370}.

\end{thebibliography}
\bibliographystyle{iclr2027_conference}

\newpage
\appendix

\section{Task Statistics}
\label{app:units}

All distributions in this appendix are over the 280 chains of \hyperref[sec:stats]{Section~\ref*{sec:stats}} (5,600 graded states, 4,404
unique segments, 140 repositories); the unique-segment medians are within 3\% of the occurrence medians for every size metric.
Intervals are repository-clustered percentile bootstraps (2,000 replicates) for means and
shares and Wilson 95\% intervals where stated, because chains from one repository are not
independent.

\begin{table}[p]\centering\small
\caption{Repositories (n=140). Size is measured at the base commit of the repository's first chain; stars queried 2026-09-09 (packaging 2026-08-27..30). Domain labels: GPT-5.6 from public metadata (fixed 13-way list; a keyword heuristic agrees on 53/140). Test style: pytest is the runner for every chain; style is inferred from the chains' own F2P test ids and test patches.}
\label{tab:units-repos}
\resizebox{\linewidth}{!}{\begin{tabular}{@{}lrrrrrr@{}}\toprule
 & p10 & p25 & median & p75 & p90 & max \\\midrule
GitHub stars & 6 & 14 & 46 & 195 & 603 & 5,835 \\
code LOC at first chain base & 79 & 215 & 424 & 2,146 & 4,914 & 33,731 \\
Python LOC & 76 & 212 & 420 & 1,907 & 4,801 & 33,684 \\
files & 8 & 11 & 18 & 37 & 66 & 325 \\
test LOC & 9 & 29 & 97 & 538 & 1,317 & 11,574 \\
chains per repository & 1 & 2 & 2 & 2 & 3 & 5 \\
\bottomrule\end{tabular}}
\par\smallskip\begin{minipage}{\linewidth}\footnotesize
\textbf{Chains per repository:} 34 repos with 1, 81 repos with 2, 18 repos with 3, 5 repos with 4, 2 repos with 5. 
\textbf{Primary language (GitHub):} Python 139, LookML 1; median Python share of code LOC at chain base 100\% (12 repos below 90\%). 
\textbf{Secondary languages present (repos):} Shell 15, JavaScript 10, CSS 7, HTML 5, C 2, TypeScript 2, C++ 1, Lua 1, SQL 1. 
\textbf{Domain:} api-client-sdk 21, language-parsing-analysis 17, scientific-numerical 15, general-utility 14, devtools-testing-packaging 13, data-engineering-formats 13, infra-cloud-ops 12, iot-hardware-home 10, web-backend-apps 9, media-games-graphics 6, text-docs-nlp 4, security-identity 4, finance-business 2. 
\textbf{Test style:} unittest.TestCase classes 63, pytest functions 60, pytest test classes 16, mixed 1; 51\% of all F2P test ids are class-based. 
\textbf{Stars:} $<$10: 20, 10--49: 51, 50--99: 20, 100--499: 31, 500--999: 8, 1000--4999: 9, $\geq$5000: 1; 6 repositories archived; licences (all 140 open source): MIT 65, Apache-2.0 36, BSD 19 (3-clause 13, 2-clause 6), GPL-3.0 12, LGPL-3.0 5, and one each GPL-2.0-or-later, LGPL-2.1 and~\mbox{AGPL-3.0}.%
\end{minipage}
\end{table}

\begin{table}[p]\centering\small
\caption{Per-segment statistics over the 5,600 segment occurrences (280 chains $\times$ 20). Mean is followed by its repository-clustered bootstrap 95\% CI. Source commits: commits folded into the segment by the generation walk (see text). Contract nodes at $k$ = F2P$_k$ $\cup$ P2P$_k$ $\cup$ still-valid \mbox{earlier F2P}.}%
\label{tab:units-segments}
\resizebox{\linewidth}{!}{\begin{tabular}{@{}lrrrrrrl@{}}\toprule
 & p10 & p25 & median & p75 & p90 & max & mean [95\% CI] \\\midrule
\multicolumn{8}{@{}l}{\emph{Reference (gold) patch}} \\
files changed & 1 & 2 & 3 & 7 & 13 & 224 & 6.0 [5.3, 6.7] \\
hunks & 2 & 4 & 7 & 15 & 30 & 1,654 & 15.5 [12.4, 19.3] \\
lines added & 6 & 16 & 50 & 146 & 384 & 81,022 & 276.8 [179.8, 404.1] \\
lines removed & 1 & 3 & 14 & 51 & 163 & 44,855 & 134.2 [83.8, 201.1] \\
lines changed (+/$-$) & 9 & 24 & 73 & 215 & 596 & 94,155 & 411.1 [267.3, 579.1] \\
new files & 0 & 0 & 0 & 1 & 2 & 135 & 1.0 [0.8, 1.2] \\
new functions (net new \texttt{def} names) & 0 & 0 & 1 & 3 & 7 & 280 & 3.2 [2.7, 3.7] \\
new classes & 0 & 0 & 0 & 1 & 2 & 469 & 1.3 [0.7, 2.5] \\
\multicolumn{8}{@{}l}{\emph{Hidden test patch and contract}} \\
test files changed & 1 & 1 & 1 & 2 & 4 & 73 & 2.5 [2.2, 2.8] \\
test lines added & 4 & 12 & 31 & 79 & 178 & 24,704 & 109.0 [77.4, 165.4] \\
new test functions & 0 & 0 & 1 & 3 & 8 & 160 & 3.3 [2.9, 3.7] \\
F2P tests & 1 & 1 & 2 & 7 & 18 & 707 & 8.7 [6.8, 11.5] \\
P2P tests & 7 & 22 & 45 & 91 & 208 & 1,303 & 93.3 [63.2, 136.3] \\
earlier F2P still demanded (accumulated) & 2 & 9 & 23 & 45 & 87 & 835 & 41.9 [32.6, 54.6] \\
contract tests at $k$ & 14 & 29 & 51 & 100 & 216 & 1,312 & 102.0 [70.5, 145.2] \\
superseded earlier F2P (not demanded) & 0 & 0 & 1 & 7 & 19 & 2,050 & 12.2 [5.8, 23.5] \\
\multicolumn{8}{@{}l}{\emph{Specification}} \\
specification words & 190 & 309 & 493 & 780 & 1,197 & 3,693 & 606.6 [577.1, 641.5] \\
spec words per test line added & 4.0 & 7.2 & 14.6 & 32.2 & 80.9 & 1,419.0 & 36.2 [32.4, 40.3] \\
spec words per F2P test & 33.9 & 80.4 & 178.0 & 352.6 & 598.0 & 3,068.0 & 267.8 [247.4, 289.4] \\
\multicolumn{8}{@{}l}{\emph{Source history}} \\
source commits folded into the segment & 1 & 1 & 3 & 6 & 12 & 163 & 5.2 [4.7, 5.8] \\
merge commits among them & 0 & 0 & 0 & 1 & 2 & 35 & 0.8 [0.6, 0.9] \\
distinct authors (e-mail) & 1 & 1 & 1 & 2 & 3 & 35 & 1.6 [1.5, 1.7] \\
segment duration, days (prev.\ target $\to$ target) & 0.00 & 0.04 & 0.95 & 8.69 & 49.29 & 2,895.97 & 21.25 [16.63, 26.19] \\
\bottomrule\end{tabular}}
\par\smallskip\begin{minipage}{\linewidth}\footnotesize
\textbf{Shares (Wilson 95\%):} P2P non-empty 96.0\% [95.5, 96.5]; more than one source commit 68.9\% [67.7, 70.1]; contains a merge commit 33.3\%; more than one author 33.1\%; creates a new file 28.7\%; adds a function or class 62.4\%; touches a non-Python file 56.0\% (pure non-Python 0.20\%); adds a new test file 23.9\%; some earlier F2P superseded at $k$ 64.6\%; specification rewritten by a model during the spec audit 24.5\%.
\end{minipage}
\end{table}

\begin{table}[p]\centering\small
\caption{Per-chain statistics (n=280 chains of 20 segments). Dependency depth = longest path in the DAG whose edges $j \to k$ ($j<k$) mark a segment whose gold or test patch references a symbol introduced, or imports a module created, by an earlier segment's gold patch (symbol/import definition), or whose gold patch touches a file the earlier gold patch touched (file definition).}
\label{tab:units-chains}
\resizebox{\linewidth}{!}{\begin{tabular}{@{}lrrrrrrl@{}}\toprule
 & p10 & p25 & median & p75 & p90 & max & mean [95\% CI] \\\midrule
gold lines added (20 segments) & 781 & 1,246 & 2,177 & 4,274 & 7,013 & 121,416 & 5,536.6 [3,682.0, 8,007.4] \\
gold lines removed & 278 & 442 & 775 & 1,836 & 3,396 & 56,611 & 2,684.6 [1,643.7, 4,033.9] \\
net gold lines & 242 & 499 & 987 & 2,311 & 4,674 & 89,154 & 2,852.0 [1,552.4, 4,631.0] \\
distinct files touched by gold & 11 & 16 & 25 & 45 & 74 & 297 & 38.0 [32.3, 44.7] \\
distinct test files touched & 4 & 6 & 10 & 18 & 30 & 309 & 16.8 [13.2, 21.5] \\
new files created & 2 & 4 & 10 & 24 & 46 & 201 & 19.1 [15.6, 23.4] \\
new functions & 15 & 24 & 42 & 77 & 155 & 568 & 63.3 [53.2, 74.7] \\
test lines added & 541 & 803 & 1,197 & 1,904 & 3,896 & 82,930 & 2,179.9 [1,546.1, 3,126.1] \\
F2P tests (sum over segments) & 47 & 67 & 106 & 187 & 320 & 3,632 & 174.2 [137.7, 226.7] \\
contract tests at $k{=}20$ & 30 & 44 & 72 & 128 & 254 & 1,300 & 126.4 [93.2, 167.9] \\
specification words (sum) & 7,794 & 9,583 & 11,502 & 13,696 & 16,518 & 34,430 & 12,132.9 [11,500.6, 12,804.8] \\
spec words per test line added & 3.6 & 6.5 & 9.8 & 14.1 & 18.6 & 38.5 & 10.9 [9.8, 12.0] \\
spec words per F2P test & 38.4 & 60.5 & 109.2 & 170.3 & 243.7 & 678.8 & 128.9 [114.7, 144.1] \\
segments with non-empty P2P (of 20) & 18 & 19 & 20 & 20 & 20 & 20 & 19.2 [18.8, 19.5] \\
dependency depth, symbol/import (max 19) & 2 & 4 & 6 & 8 & 11 & 17 & 6.2 \\
dependency depth, file overlap (max 19) & 14 & 16 & 18 & 19 & 19 & 19 & 17.0 \\
code LOC at chain base & 163 & 331 & 1,232 & 3,155 & 5,867 & 33,889 & 2,833.4 [1,980.3, 3,993.8] \\
files at chain base & 9 & 14 & 28 & 50 & 85 & 336 & 42.4 [34.4, 53.5] \\
net gold lines / code LOC at base & 0.07 & 0.24 & 0.90 & 4.61 & 10.31 & 125.97 & 4.62 [3.23, 6.21] \\
source commits folded into the chain & 43 & 57 & 85 & 129 & 196 & 422 & 104.8 [94.5, 115.4] \\
commits before the chain base & 1 & 4 & 40 & 116 & 200 & 529 & 80.6 [66.0, 96.7] \\
commit-date span, days & 18 & 44 & 200 & 520 & 1,143 & 3,426 & 425.0 [336.0, 522.5] \\
median segment duration, days & 0.05 & 0.19 & 0.99 & 4.54 & 14.97 & 77.48 & 5.09 [3.46, 6.96] \\
distinct human authors & 1 & 2 & 4 & 9 & 15 & 39 & 6.5 [5.5, 7.5] \\
\bottomrule\end{tabular}}
\par\smallskip\begin{minipage}{\linewidth}\footnotesize
\textbf{Shares:} all 20 segments carry P2P 66.8\% [61.1, 72.0]; single human author 14.6\% [11.0, 19.3]. \textbf{Distinct human authors per chain:} 1: 41, 2: 47, 3--4: 62, 5--9: 72, $\geq$10: 58 chains. \textbf{Commit-date span:} $<$1m: 54, 1--3m: 46, 3--6m: 30, 6--12m: 46, 1--2y: 57, 2--4y: 29, $\geq$4y: 18 chains.
\end{minipage}
\end{table}

\begin{table}[p]\centering\small
\caption{Segment dependency structure over the 5,320 segment occurrences at positions 2--20. ``Any earlier'': some $j<k$ in the same chain satisfies the definition; ``$k{-}1$'': the immediately preceding segment does. CIs are repository-clustered bootstrap 95\%.}
\label{tab:units-dependency}
\footnotesize\setlength{\tabcolsep}{3.5pt}
\resizebox{\linewidth}{!}{\begin{tabular}{@{}lrrcc@{}}\toprule
definition & any earlier $j<k$ & $k{-}1$ & nearest lag (med.) & chain depth med.\ (IQR) \\\midrule
gold patches share a file & 98.1\% [97.2, 98.8] & 89.2\% [86.8, 91.5] & 1 & 18 (16--19) \\
references a function/class touched by $j$ & 79.3\% [75.5, 82.8] & 40.1\% [36.7, 43.5] & 1 & 8 (5--11) \\
references a function/class introduced by $j$ & 72.3\% [68.3, 76.2] & 28.2\% [25.6, 31.0] & 2 & 6 (4--8) \\
\quad strict (names $\geq 4$ chars, non-generic) & 68.2\% [64.5, 71.9] & 25.5\% [23.1, 28.0] & 2 & 5 (3--7) \\
\quad gold-patch references only & 64.6\% [60.7, 68.6] & 24.4\% [21.8, 27.0] & 2 & 5 (3--8) \\
\quad test-patch references only & 54.1\% [49.8, 58.2] & 15.0\% [13.2, 16.9] & 3 & 3 (2--5) \\
imports a module created by $j$ & 19.0\% [15.8, 22.2] & 4.0\% [3.1, 4.9] & 3 & 1 (0--2) \\
symbol introduced OR module import (headline) & 73.3\% [69.6, 77.0] & 28.8\% [26.1, 31.6] & 2 & 6 (4--8) \\
file OR symbol OR import & 98.9\% [98.4, 99.3] & 90.5\% [88.2, 92.4] & 1 & 18 (17--19) \\
\bottomrule\end{tabular}}
\par\smallskip\begin{minipage}{\linewidth}\footnotesize
\textbf{Share of segments depending on the segment exactly $\ell$ positions earlier} (symbol/import): $\ell{=}1$: 29\%, $\ell{=}2$: 27\%, $\ell{=}3$: 24\%, $\ell{=}5$: 23\%, $\ell{=}10$: 21\%, $\ell{=}15$: 22\%, $\ell{=}19$: 27\%. File overlap: $\ell{=}1$: 89\%, $\ell{=}2$: 85\%, $\ell{=}5$: 78\%, $\ell{=}10$: 72\%, $\ell{=}19$: 66\%.
\end{minipage}
\end{table}

\begin{table}[p]\centering\small
\caption{A \textsc{BOBA-Bench} segment next to a single-request task. Median (IQR) per instance; identical parser for all rows. SWE-bench Verified from \texttt{princeton-nlp/SWE-bench\_Verified} (500); SWE-bench Pro public set from \texttt{ScaleAI/SWE-bench\_Pro} (731; Go 280, Python 266, JS 165, TS 20). Specification = problem statement (Verified) / problem statement + requirements + interface (Pro) / frozen segment specification (\textsc{BOBA-Bench}).}
\label{tab:units-comparison}
\resizebox{\linewidth}{!}{\begin{tabular}{@{}lrrrrr@{}}\toprule
 & \textsc{BOBA-Bench} segment & \textsc{BOBA-Bench} chain (20 seg.) & SWE-bench Verified & SWE-bench Pro & SWE-bench Pro (Py) \\
 & n=5,600 & n=280 & n=500 & n=731 & n=266 \\\midrule
gold files changed & 3 (2--7) & 25 (16--45) & 1 (1--1) & 4 (2--7) & 3 (2--4) \\
gold hunks & 7 (4--15) & -- & 1 (1--2) & 9 (5--16) & 7 (4--12) \\
gold lines added & 50 (16--146) & 2,177 (1,246--4,274) & 4 (2--10) & 63 (33--135) & 46 (30--108) \\
gold lines removed & 14 (3--51) & 775 (442--1,836) & 2 (1--4) & 20 (7--52) & 15 (5--41) \\
gold lines changed & 73 (24--215) & -- & 7 (3--13) & 94 (50--207) & 74 (45--142) \\
new functions & 1 (0--3) & -- & 0 (0--0) & 0 (0--0) & 1 (0--3) \\
test files changed & 1 (1--2) & -- & 1 (1--1) & 1 (1--2) & 1 (1--2) \\
test lines added & 31 (12--79) & 1,197 (803--1,904) & 15 (7--27) & 46 (17--116) & 40 (14--99) \\
F2P tests & 2 (1--7) & 106 (67--187) & 1 (1--2) & 3 (1--7) & 4 (2--8) \\
P2P tests & 45 (22--91) & -- & 50 (19--111) & 0 (0--30) & 12 (1--51) \\
specification words & 493 (309--780) & 11,502 (9,583--13,696) & 143 (85--234) & 418 (320--542) & 425 (318--573) \\
P2P non-empty & 96.0\% & -- & 97.8\% & 49.1\% & 77.8\% \\
\bottomrule\end{tabular}}
\end{table}

\begin{table}[p]\centering\small
\caption{Stratification bins (segment occurrences per bin; assignments in \texttt{data/segment\_bins.jsonl}). Size terciles are computed over the 5,600 occurrences; position blocks are balanced by construction.}
\label{tab:units-bins}
\resizebox{\linewidth}{!}{\begin{tabular}{@{}llr@{}}\toprule
stratum & bins & n \\\midrule
gold lines changed (terciles) & small $<36$ / medium $36$--$144$ / large $\geq 145$ & 1,832 / 1,896 / 1,872 \\
source commits per segment & 1 / 2--4 / $\geq 5$ & 1,741 / 2,081 / 1,778 \\
P2P protection & empty / non-empty & 223 / 5,377 \\
symbol/import dependency ($k\geq 2$) & none / on $k{-}1$ / only on $j<k{-}1$ & 1,421 / 1,533 / 2,366 \\
position block & 1--5 / 6--10 / 11--15 / 16--20 & 1,400 / 1,400 / 1,400 / 1,400 \\
\bottomrule\end{tabular}}
\end{table}

\section{Matched Foundations: Protocol and Controls}
\label{app:mf}

\subsection{Per-model results}
\label{app:mf-results}

\hyperref[tab:mf-results]{Tables~\ref*{tab:mf-results}} and~\ref{tab:mf-results-new} give the per-model estimates behind
\hyperref[sec:mf]{Section~\ref*{sec:mf}} on the 140 repositories, for the five models of the earlier study and for the
six added with the leaderboard: next-request success from the agent's certified prefix and from the human prefix over eight
attempts per boundary, the agent-minus-human gap with 95\% repository-bootstrap intervals
(boundaries averaged within repository, then across repositories), the co-primary pass@8 gap,
the gap by depth, the sensitivity that excludes boundaries whose next request already passed at
$k$, and two descriptive counts: the share of boundaries where both prefixes pass all eight
attempts, and the number of boundaries where the agent prefix passes at least four fewer
attempts than the human prefix. The six added models were run with the same windows, depths, retry
cap, attempt counts and grader, under the settings of their leaderboard runs.

\begin{table}[!htbp]
\centering
\scriptsize
\setlength{\tabcolsep}{1.5pt}
\newcommand{\lab}[1]{\parbox[t]{1.35in}{\raggedright\strut #1\strut}}
\caption{\textbf{Matched Foundations per model} (140 repositories). Hidden-tests protocol; 95\% repository-bootstrap CIs in brackets.}
\label{tab:mf-results}
\begin{tabular}{@{}lccccc@{}}
\toprule
 & Fable~5 & GPT-6 Astra & Kimi~K3 & GPT-5.6 & GLM-5.2 \\
\midrule
\lab{Boundaries (repositories)} & 311 (124) & 298 (124) & 332 (130) & 283 (121) & 286 (120) \\
\lab{Repositories never certified (of 140)} & 46 & 55 & 40 & 64 & 63 \\
\addlinespace
\lab{Success, agent prefix (\%)} & 92.8 [89.0,96.2] & 94.0 [90.5,97.0] & 94.6 [92.1,96.9] & 94.1 [91.0,96.8] & 89.6 [85.6,93.2] \\
\lab{Success, human prefix (\%)} & 95.1 [92.3,97.4] & 93.2 [89.7,96.2] & 95.1 [93.2,96.7] & 93.7 [91.1,96.0] & 92.2 [88.8,95.1] \\
\lab{Gap, agent $-$ human (pp)} & $-$2.3 [$-$5.2,$+$0.3] & $+$0.8 [$-$1.9,$+$3.3] & $-$0.5 [$-$2.7,$+$1.6] & $+$0.4 [$-$2.2,$+$2.8] & $-$2.6 [$-$5.2,$-$0.2] \\
\lab{Sign-flip $p$} & 0.11 & 0.59 & 0.67 & 0.76 & 0.04 \\
\lab{pass@8 gap (pp)} & $-$3.6 [$-$7.1,$-$0.5] & $+$0.0 [$-$2.7,$+$2.7] & $+$1.3 [$+$0.3,$+$2.6] & $-$1.0 [$-$4.1,$+$1.8] & $-$1.1 [$-$3.5,$+$0.7] \\
\addlinespace
\lab{Gap after 4 requests (pp)} & $-$1.4 [$-$4.1,$+$1.0] & $+$1.7 [$-$1.0,$+$4.6] & $-$1.3 [$-$3.6,$+$0.6] & $-$0.4 [$-$2.8,$+$1.8] & $-$2.4 [$-$4.9,$-$0.3] \\
\lab{Gap after 9 requests (pp)} & $+$1.4 [$-$2.4,$+$5.2] & $+$2.0 [$+$0.0,$+$4.9] & $+$1.0 [$-$1.5,$+$3.9] & $+$3.8 [$+$0.1,$+$7.9] & $-$1.9 [$-$6.5,$+$2.3] \\
\lab{Gap after 14 requests (pp)} & $-$3.3 [$-$7.9,$+$1.1] & $-$1.4 [$-$7.5,$+$4.2] & $+$2.8 [$-$0.7,$+$7.0] & $+$1.7 [$-$4.8,$+$8.7] & $+$0.8 [$-$2.5,$+$4.4] \\
\addlinespace
\lab{Gap excl.\ pre-solved (pp)} & $-$1.4 [$-$4.0,$+$1.0] & $+$0.2 [$-$2.4,$+$2.5] & $-$1.1 [$-$3.3,$+$0.8] & $-$0.0 [$-$1.7,$+$1.7] & $-$2.4 [$-$4.9,$-$0.2] \\
\lab{Pre-solved boundaries excluded} & 50 & 38 & 52 & 34 & 40 \\
\lab{Boundaries 8/8 from both prefixes (\%)} & 85 & 91 & 79 & 86 & 72 \\
\lab{Boundaries with agent $\le$ human $-$ 4/8} & 8 & 5 & 4 & 6 & 7 \\
\bottomrule
\end{tabular}
\end{table}

\begin{table}[!htbp]
\centering
\scriptsize
\setlength{\tabcolsep}{1.5pt}
\newcommand{\labb}[1]{\parbox[t]{1.2in}{\raggedright\strut #1\strut}}
\caption{\textbf{Matched Foundations per model, the six models added with the leaderboard} (140 repositories). Hidden-tests protocol; 95\% repository-bootstrap CIs in brackets.}
\label{tab:mf-results-new}
\resizebox{\textwidth}{!}{%
\begin{tabular}{@{}lcccccc@{}}
\toprule
 & Opus~5.5 & Opus~5 & Opus~4.8 & GPT-6.1 & GLM-5.3 & GLM-5.3-Flash \\
\midrule
\labb{Boundaries (repositories)} & 321 (123) & 294 (118) & 293 (121) & 291 (122) & 328 (131) & 329 (127) \\
\labb{Repositories never certified (of 140)} & 41 & 52 & 57 & 61 & 43 & 40 \\
\addlinespace
\labb{Success, agent prefix (\%)} & 95.9 [93.0,98.3] & 91.2 [87.0,94.8] & 91.3 [87.6,94.6] & 94.0 [90.3,97.2] & 92.8 [89.3,95.8] & 95.0 [93.0,96.7] \\
\labb{Success, human prefix (\%)} & 96.5 [94.3,98.2] & 94.8 [91.7,97.4] & 94.6 [91.5,97.3] & 93.4 [90.0,96.4] & 93.4 [90.7,95.6] & 94.7 [92.9,96.3] \\
\labb{Gap, agent $-$ human (pp)} & $-$0.6 [$-$3.2,$+$1.7] & $-$3.6 [$-$6.6,$-$1.1] & $-$3.3 [$-$6.6,$-$0.1] & $+$0.6 [$-$2.4,$+$3.5] & $-$0.6 [$-$3.5,$+$1.9] & $+$0.3 [$-$1.5,$+$2.2] \\
\labb{Sign-flip $p$} & 0.69 & 0.01 & 0.05 & 0.70 & 0.67 & 0.77 \\
\labb{pass@8 gap (pp)} & $-$1.1 [$-$3.8,$+$0.8] & $-$2.0 [$-$4.8,$+$0.3] & $+$0.3 [$-$1.8,$+$2.6] & $-$1.1 [$-$4.1,$+$1.5] & $-$1.3 [$-$3.8,$+$0.8] & $-$0.1 [$-$1.2,$+$0.8] \\
\addlinespace
\labb{Gap after 4 requests (pp)} & $-$2.8 [$-$5.8,$-$0.3] & $-$2.4 [$-$5.5,$+$0.2] & $-$2.2 [$-$5.4,$+$0.9] & $+$1.7 [$-$1.5,$+$5.3] & $-$2.2 [$-$5.0,$+$0.2] & $-$0.4 [$-$2.5,$+$1.8] \\
\labb{Gap after 9 requests (pp)} & $+$1.4 [$-$2.5,$+$5.6] & $-$2.0 [$-$5.0,$+$0.8] & $-$2.8 [$-$7.8,$+$1.9] & $+$1.5 [$-$1.1,$+$4.9] & $+$3.1 [$+$0.0,$+$6.8] & $+$1.4 [$-$2.0,$+$5.0] \\
\labb{Gap after 14 requests (pp)} & $+$3.8 [$+$1.2,$+$7.1] & $-$3.2 [$-$8.0,$+$1.4] & $-$2.2 [$-$6.3,$+$1.6] & $-$1.0 [$-$7.0,$+$5.0] & $+$1.6 [$-$1.8,$+$5.6] & $+$2.0 [$-$1.6,$+$6.0] \\
\addlinespace
\labb{Gap excl.\ pre-solved (pp)} & $-$1.2 [$-$3.9,$+$1.1] & $-$3.6 [$-$6.8,$-$1.0] & $-$3.4 [$-$6.7,$-$0.1] & $+$0.2 [$-$2.4,$+$2.8] & $-$1.2 [$-$4.3,$+$1.3] & $+$0.1 [$-$2.2,$+$2.2] \\
\labb{Pre-solved boundaries excluded} & 53 & 59 & 35 & 38 & 61 & 53 \\
\labb{Boundaries 8/8 from both prefixes (\%)} & 88 & 84 & 81 & 90 & 80 & 74 \\
\labb{Boundaries with agent $\le$ human $-$ 4/8} & 8 & 10 & 12 & 5 & 7 & 7 \\
\bottomrule
\end{tabular}}
\end{table}

\subsection{Ablation: earlier requests' tests visible in the repository versus hidden}
\label{app:mf-visible-tests}

Matched Foundations was run twice with one difference: whether the human tests of the
already-accepted requests were present in the repositories the model worked in. In the first
run, a constructed request was graded by applying its hidden test patch inside the agent's own
container and the resulting tree was committed, so the certified agent prefix at boundary $k$
carried the human tests of requests $1,\dots,k$; the same patches were applied to the human
prefix, so both arms saw them. For request $k{+}1$ these are its P2P tests (behavior that
must keep working), never its F2P tests, which stayed hidden in both runs. In the second
run, the one reported in the main text, the agent's tree is snapshotted before any hidden
material enters the container and a throwaway copy is graded, so no hidden test appears in
either prefix, matching the benchmark. Windows, depths, retry cap, scaffold and model settings
are identical, so the boundaries complete in both runs form a paired ablation of P2P-test
visibility (\hyperref[tab:mf-visible-tests]{Table~\ref*{tab:mf-visible-tests}}). Visible P2P tests
lift both arms by 4--5\,pp for the two models with clean runs in both conditions (GPT-5.6,
Fable~5), but the agent-minus-human gap does not move: the differences are $-0.4$ to
$+1.5$\,pp with every interval spanning zero. The visible tests supplied regression protection
and interface leakage (names, signatures and conventions the next contract assumes), and
supplied them equally from either foundation. GLM-5.2's visible-tests construction ran through
a proxy incident and Kimi~K3's on a prefix-cached fleet that its hidden-tests run did not have,
so their arm-level differences are not interpretable as a visibility effect; their gap
differences are. GPT-6 Astra was run only with tests hidden.

\begin{table}[!htbp]
\centering
\scriptsize
\setlength{\tabcolsep}{2.2pt}
\newcommand{\lab}[1]{\parbox[t]{1.5in}{\raggedright\strut #1\strut}}
\caption{\textbf{Ablation: human tests of the accepted requests visible in both prefixes versus hidden} Paired on the boundaries complete under both conditions; pp with 95\% repository-bootstrap CIs; gap = agent-minus-human next-request pass rate.}
\label{tab:mf-visible-tests}
\begin{tabular}{@{}lcccc@{}}
\toprule
 & Fable~5 & GPT-5.6 & GLM-5.2 & Kimi~K3 \\
\midrule
\lab{Paired boundaries (repositories)} & 296 (121) & 280 (120) & 230 (104) & 263 (108) \\
\addlinespace
\lab{Agent arm, visible $-$ hidden} & $+$5.4 [$+$2.1,$+$9.3] & $+$4.0 [$+$1.7,$+$6.9] & $-$4.1 [$-$7.8,$-$0.2] & $-$0.4 [$-$3.1,$+$2.4] \\
\lab{Human arm, visible $-$ hidden} & $+$3.9 [$+$1.5,$+$6.8] & $+$4.4 [$+$2.6,$+$6.5] & $-$5.1 [$-$8.1,$-$2.1] & $-$0.9 [$-$3.0,$+$1.1] \\
\addlinespace
\lab{Gap, tests visible} & $-$0.4 [$-$1.5,$+$0.6] & $+$0.2 [$-$0.8,$+$1.2] & $-$1.3 [$-$3.9,$+$1.0] & $+$0.4 [$-$1.5,$+$2.6] \\
\lab{Gap, tests hidden} & $-$1.9 [$-$4.8,$+$0.7] & $+$0.5 [$-$2.2,$+$2.9] & $-$2.3 [$-$5.2,$+$0.3] & $-$0.0 [$-$2.0,$+$1.8] \\
\lab{Gap difference, visible $-$ hidden} & $+$1.5 [$-$1.1,$+$4.5] & $-$0.4 [$-$2.7,$+$2.2] & $+$0.9 [$-$2.1,$+$4.2] & $+$0.5 [$-$2.1,$+$3.2] \\
\bottomrule
\end{tabular}
\end{table}

\section{Cross-Verification: Protocol and Census}
\label{app:cv}

\subsection{Per-model results}
\label{app:cv-results}

\hyperref[tab:cv-results]{Table~\ref*{tab:cv-results}} gives the per-model estimates behind \hyperref[sec:cv]{Section~\ref*{sec:cv}}. Each of the
ten models contributes 109 matched pairs as an author: a wrong implementation from its first attempt
at some request, whose own tests failed or which broke an earlier test, and the same request passed
from a correct state by the same author in another attempt; both states are regraded in the sealed
environment and the git history is rewritten to carry no identifying metadata. Every review runs in
a fresh container with the reviewer's leaderboard settings and a budget of 60 model calls, 45
minutes of tool execution and 90 minutes of wall time; the reviewer may write and run tests and
returns accept, reject or abstain. The review rows are equal-weight aggregates over the ten
authors, with repository-clustered bootstrap 95\% intervals; a review that ends without a decision
counts as neither an acceptance nor a rejection. The reviewers' own test suites run and exercise the
code in 94--98\% of reviews, and conditioning on such a suite moves no rate by more than two points.
The number of pairs per author is set by the three authors with the fewest correct twins; 101 of
Opus~5.5's, 89 of GPT-6 Astra's and 85 of GPT-6.1's wrong implementations stand on a repository that
carried no earlier failure, as do all 109 of every other author's, and restricting to those moves no
rate by more than one point.

The three GPT reviewers are strict with every author but themselves: each accepts 68--82\% of its
own wrong implementations against 37--52\% of the other authors' and rejects 14--34\% of its own
correct ones against 45--61\% of the others' (differences of $+30$ to $+36$ and $-27$ to $-31$
points, repository-clustered intervals excluding zero). The lenient reviewers leave less room for
the pattern: Opus~5.5 accepts 91\% of its own wrong implementations against 75\% of the others'
($+15$ points, interval [$+9,+21$]), GLM-5.3 98\% against 94\% ($+4$ [$+2,+7$]), Opus~5 94\% against
88\% ($+5$ [$+0,+10$]), and the other four are within $\pm5$ points with intervals covering zero.

\begin{table}[!htbp]
\centering
\scriptsize
\setlength{\tabcolsep}{3pt}
\caption{\textbf{Cross-model review per reviewer} (ten models; 109 matched pairs per author, 21{,}800 reviews). Equal weight over the ten authors; repository-clustered bootstrap 95\% intervals in brackets. Youden $J$ = wrong rejected $+$ correct accepted $-$ 1. Pairs discriminated: the reviewer rejects the wrong implementation and accepts its correct twin. Suites: the reviewer's own tests fail the wrong implementation and pass the twin, over the valid suites with a known counterpart ($n$).}
\label{tab:cv-results}
\begin{tabular}{@{}lccccrr@{}}
\toprule
Reviewer & Wrong accepted (\%) & Correct rejected (\%) & Youden $J$ & Pairs discr.\ (\%) & Suites (\%) & $n$ \\
\midrule
Opus~5.5 & 77.0 [73.7,80.0] & 7.8 [5.7,10.1] & $+$0.15 [$+$0.12,$+$0.19] & 18.8 & 12.8 & 1090 \\
Opus~5 & 88.9 [86.5,91.0] & 2.8 [1.8,4.1] & $+$0.08 [$+$0.06,$+$0.10] & 9.7 & 6.7 & 1090 \\
Fable~5 & 92.9 [90.9,94.7] & 1.2 [0.5,2.0] & $+$0.05 [$+$0.04,$+$0.07] & 6.5 & 4.9 & 1086 \\
GLM-5.3-Flash & 90.5 [87.6,92.9] & 1.2 [0.5,2.0] & $-$0.02 [$-$0.05,$+$0.01] & 3.7 & 1.8 & 1085 \\
GPT-6 Astra & 40.4 [36.2,44.3] & 58.4 [54.0,62.7] & $+$0.01 [$-$0.01,$+$0.04] & 8.6 & 5.8 & 1090 \\
GPT-6.1 & 41.6 [37.6,45.5] & 57.1 [52.6,61.5] & $+$0.01 [$-$0.01,$+$0.03] & 8.3 & 6.3 & 1090 \\
GLM-5.3 & 94.4 [92.5,95.9] & 1.2 [0.5,1.9] & $+$0.01 [$-$0.01,$+$0.03] & 3.9 & 2.1 & 1090 \\
GPT-5.6 & 54.9 [51.1,58.4] & 41.5 [37.9,45.2] & $+$0.04 [$+$0.00,$+$0.07] & 12.9 & 8.2 & 1090 \\
Kimi~K3 & 92.4 [90.2,94.3] & 3.5 [2.2,5.0] & $+$0.04 [$+$0.02,$+$0.06] & 6.7 & 4.0 & 1090 \\
Opus~4.8 & 94.8 [92.8,96.4] & 1.6 [0.8,2.5] & $+$0.03 [$+$0.02,$+$0.05] & 4.8 & 3.3 & 1088 \\
\bottomrule
\end{tabular}
\end{table}

\subsection{Self-authored test suites}
\label{app:cv-selfsuites}

\hyperref[tab:cv-selfsuites]{Table~\ref*{tab:cv-selfsuites}} gives the self-authored suites of four models of the earlier study (one
target per repository, eight suites each): a suite the model writes from the specification alone,
executed against the correct human implementation and against wrong implementations of the same
request. Sensitivity is estimable only on repositories where the model itself produced a wrong
implementation. Such a suite fails 13--84\% of the wrong sibling implementations but passes the
correct human implementation in only 30--56\% of cases.

\begin{table}[!htbp]
\centering
\scriptsize
\setlength{\tabcolsep}{1.5pt}
\newcommand{\lab}[1]{\parbox[t]{1.5in}{\raggedright\strut #1\strut}}
\caption{\textbf{Self-authored suites} (four models of the earlier study; 95\% repository-clustered \mbox{bootstrap intervals}).}%
\label{tab:cv-selfsuites}
\begin{tabular}{@{}lcccc@{}}
\toprule
 & Fable~5 & Kimi~K3 & GPT-5.6 & GLM-5.2 \\
\midrule
\lab{Self-authored suites executable (\%)} & 89.4 & 85.3 & 75.8 & 87.3 \\
\lab{Suites passing the correct human implementation (\%)} & 55.9 [47.9,63.7] & 35.5 [28.7,42.5] & 30.1 [23.4,37.0] & 42.2 [35.0,49.6] \\
\lab{Suites failing wrong sibling implementations (\%)} & 13.0 [1.0,28.9] & 68.5 [42.5,92.7] & 50.8 [23.3,78.3] & 83.5 [67.0,97.5] \\
\lab{Repositories with a wrong sibling (sensitivity denominator)} & 7 & 8 & 10 & 20 \\
\bottomrule
\end{tabular}
\end{table}

\section{Benchmark Construction and Quality Assurance}
\label{app:qa}

This appendix expands the construction of \hyperref[sec:construction]{Section~\ref*{sec:construction}} in the same order
(mining, contracts, windows, specifications, audit, environments), then
the protocol and grader of \hyperref[sec:protocol]{Section~\ref*{sec:protocol}}, and
records the attrition from 25,519 verified segments to the 4,404 in the benchmark.

\subsection{Segment mining}
\label{app:qa-mining}

Candidate repositories are Python projects for which SWE-rebench
\citep{badertdinov2025swerebench,badertdinov2026swerebenchv2} or OpenSWE
\citep{fu2026davincienv} publish a working test image: 5,124 images over a pool of 10,319
repositories. For each repository the generator iterates the commits reachable from the
default branch in oldest-first commit-date order (\texttt{git rev-list --reverse}), not a
first-parent path, and groups consecutive commits into a \emph{candidate segment} as soon
as the accumulated change touches both test files and source files; files are classified by
path, and no file belongs to both patches. The changed tests become the hidden
\emph{test patch}, everything else the \emph{gold patch}. A candidate is verified by
execution (\hyperref[app:qa-contracts]{\S\ref*{app:qa-contracts}}); a candidate that fails verification is not discarded
but absorbed into the next candidate (fold-forward), so segments are contiguous and the
target commit of segment $i$ is the base of segment $i{+}1$. No size limit is applied. The
walk produced 25,519 verified segments across 1,491 repositories.

Because the walk orders commits by date rather than by integration, the base of a segment
is an ancestor of its target for 92.3\% of boundary pairs (Wilson 95\% CI [91.6, 93.0]);
the remaining 7.7\% hop between branches, and the gold patch is in every case the tree diff
between the two boundary states. Only 71.7\% of boundaries are first-parent-reachable. The
4,404 unique segments of the benchmark fold 23,407 source commits in walk order (median 3, mean
5.3, maximum 163; 25,505 by \texttt{rev-list base..target}, 19,145 first-parent).
Chains start at the first tested commit of a repository, so a median of about 40 commits precede
the first chain base and are covered by no~\mbox{segment}.%

\subsection{Executable contracts}
\label{app:qa-contracts}

Verification runs in fresh containers: the full pytest suite at the base with the test
patch applied, then again with the gold patch added, three runs in each state so that a
flaky test cannot enter the contract. \emph{Fail-to-pass} (F2P) nodes are
tests in the changed test files that do not pass before and pass after; \emph{pass-to-pass}
(P2P) nodes pass in both states. Runner errors are separated and retried rather than
counted, so a broken collection can never manufacture an F2P, and at least one F2P node is
required. The two node lists are the segment's \emph{contract}. Throughout the benchmark a
state passes iff every listed node reports the literal outcome \texttt{passed}; skipped,
xfail/xpass, errored and absent nodes all fail. Because the human patch has been shown to
satisfy the contract in the same environment, a later failure is attributable to the
submission, not to the tests or the environment.

\subsection{From segments to chains}
\label{app:qa-windows}

\paragraph{Attrition before windowing.} The 25,519 verified segments were processed as two
repository-disjoint pools; 14,921 belong to the primary pool. A chain requires
20 consecutive verified segments in one repository, so only the 6,316 primary-pool segments (141
repositories) lying in continuity runs of at least 20 were candidates. Strict
retained-behavior eligibility removed 1,178 of these, and 240 more fell with 22
repositories under the 20-run floor, leaving 4,898 segments in 119 repositories. Interior
removals were made chain-safe: the nearest retained predecessor was extended to the next
retained base (112 interior collapses, 65 boundary runs), and for each changed range the
target metadata, source-only gold patch, hidden tests, exact F2P/P2P lists, Dockerfile
metadata and specification were regenerated; every retained segment kept at least two
passing F2P tests. All 4,898 pass the strict offline Oracle. To these were added 1,842
Oracle-verified segments from continuous runs of the second pool (drawn from its 8,368
Oracle-passing segments).

\paragraph{Windows and task materialization.} Chains are cut only inside repository-local
runs in which every consecutive segment passed the Oracle: non-overlapping windows of
exactly 20, plus one final window aligned to the run's end when its length is not a multiple
of 20; shorter fragments are discarded, never bridged. This gave 439 windows over 193
repositories (119 from the primary pool, 74 from the second; 6,740 unique segments, 8,780 occurrences) before
filtering; specification unanimity (\hyperref[app:qa-audit]{\S\ref*{app:qa-audit}}) left 397 windows over 176
repositories and rounds 4--15 (\hyperref[app:qa-filtering]{\S\ref*{app:qa-filtering}}) left 280 over 140. A \emph{task}
is an immutable base commit; 20 ordered specifications, a later one taking precedence where
it intentionally supersedes earlier behavior; 20 frozen hidden-test patches with exact
F2P/P2P lists per state; 20 frozen source-only gold patches used only by the Oracle; and an
immutable, digest-addressed environment. All 22,400 contract files in the benchmark are
byte-identical to the freeze. The overlapping final windows make 1,196 of the 5,600 segment
occurrences repeat a segment of the previous window.

\begin{table}[htbp]
\centering\small
\caption{\textbf{Attrition from verified changes to released requests.} Counts are unique
requests unless stated. The two pools are repository-disjoint.}
\label{tab:attrition}
\resizebox{\linewidth}{!}{%
\begin{tabular}{@{}lrr@{}}
\toprule
Step & Requests & Repositories\\
\midrule
Verified changes from the history walk & 25,519 & 1,491\\
\quad in the primary pool & 14,921 & --\\
\quad in continuity runs of at least 20 & 6,316 & 141\\
\quad removed by retained-behavior eligibility & 1,178 & --\\
\quad removed with runs below 20 & 240 & 22\\
\quad remaining primary-pool requests & 4,898 & 119\\
Added from the second pool (of 8,368 that pass the reference check) & 1,842 & 74\\
Candidate windows: 439 chains (8,780 request occurrences) & 6,740 & 193\\
After the specification audit: 397 chains & 6,066 & 176\\
After the filtering rounds (\hyperref[tab:filtering]{Table~\ref*{tab:filtering}}): 280 chains & 4,404 & 140\\
\bottomrule
\end{tabular}}
\end{table}

\subsection{Specification authoring and solvability}
\label{app:qa-specs}

One call to Claude Opus~4.8 (high reasoning) receives the segment's commit messages, linked
issue and PR bodies and the full test diff, never the gold patch. It must treat the tests as
the sole source of truth, name the public API they import, state behavior and edge cases
rather than implementation, add nothing the tests do not cover, and never mention tests or
assertions. The verbatim prompt and generating function are in \hyperref[app:qa-specgen]{\S\ref*{app:qa-specgen}}.

Every released segment carries an offline \emph{Oracle} proof: the gold target, checked out
in a history-scrubbed, network-free container, passes its exact F2P/P2P lists.

\paragraph{Blind solvability campaign (Stage~A, 7--17 August 2026).} The 14,921 candidate
specifications of the primary pool were attempted blind from the human prefix in a sequential design: Claude
Opus~4.8 attempted every candidate eight times; GPT-5.6-sol (xhigh) then attempted eight
times only those Opus never solved (\hyperref[tab:solvability-campaign]{Table~\ref*{tab:solvability-campaign}}). Specifications
that both models failed were probed by Claude Fable~5 and GPT-5.6-sol for behavior the
hidden tests demand but the text omits. Inside the material that could still become
benchmark chains, every flagged specification (294) was rewritten and re-attempted blind
four times by GPT-5.6-sol (75 solved; 219 validations were cut short by harness failures and
remain \texttt{repaired\_pending}); flagged specifications outside that material were not
rewritten, since their segments left with their runs. Never-solved specifications with no
gap found were kept as \texttt{hard\_but\_fair} on the strength of the strict offline Oracle
and the two spec-gap verdicts. Non-solve was recorded as evidence and never treated as
invalidity.

\begin{table}[h]
\centering\small
\caption{Blind solvability campaign on the 14,921 candidate specifications (August 2026).}
\label{tab:solvability-campaign}
\begin{tabular}{lr}
\toprule
Outcome & Candidates \\
\midrule
Solved $\geq 1/8$ by Claude Opus~4.8 & 8,695 (58.3\%) \\
Opus 0/8, then solved $\geq 1/8$ by GPT-5.6-sol & 816 (5.5\%) \\
0/8 by both & 5,145 (34.5\%) \\
Incomplete (51 GPT runs, 214 harness E2BIG) & 265 (1.8\%) \\
\bottomrule
\end{tabular}
\end{table}

\paragraph{Composition of the benchmark.} Selection favored solved specifications (37\% of
blind-solved candidates reached the benchmark against 17\% of dual-failed ones), but one
benchmark specification in five was never solved blind by either model in 16 attempts
(\hyperref[tab:solvability-release]{Table~\ref*{tab:solvability-release}}). These are spread evenly over chain positions (47--67
per position) and touch 204 of the 280 chains (76 chains have none, the median chain has one
or two, 9 chains consist entirely of never-solved specifications). They were retained on the
strength of the strict offline Oracle and model-based adequacy adjudication: in the
primary pool they carry the v1.2 statuses \texttt{hard\_but\_fair} (575),
\texttt{repaired\_validated} (45: specification rewritten, then solved in blind pass@4),
\texttt{repaired\_pending\_validation} (66) and \texttt{repaired\_pending\_pass4} (2); the
1,305 second-pool segments have no status field but the same campaign classes apply (966
Opus-solved, 68 GPT-only, 205 dual-failed, 66 incomplete). When reached in a chain, these
never-solved specifications are passed slightly less often than blind-solved ones, by 0.9 to 3.6
points per model under the handoff protocol and 1.2 to 3.1 points in the continuous-conversation
runs (\hyperref[app:trust]{Appendix~\ref*{app:trust}}), so the request-to-chain gap of \hyperref[sec:howfar]{Section~\ref*{sec:howfar}} does not
rest on them.
``Never solved'' refers to the
specification as it stood in August 2026; the review of \hyperref[app:qa-audit]{\S\ref*{app:qa-audit}} later rewrote
22.7\% of specifications.

\begin{table}[h]
\centering\small
\caption{Blind-solvability class of the benchmark's 4,404 specifications (280 chains).}
\label{tab:solvability-release}
\begin{tabular}{lr}
\toprule
Class & Specifications \\
\midrule
Solved by Claude Opus~4.8 & 3,183 (72.3\%) \\
Solved only by GPT-5.6-sol & 263 (6.0\%) \\
Never blind-solved by either & 886 (20.1\%) \\
Incomplete campaign runs & 72 (1.6\%) \\
\bottomrule
\end{tabular}
\end{table}

\begin{table}[htbp]
\centering\small
\caption{Adequacy status of the never-solved and rewritten primary-pool specifications of the~\mbox{benchmark}.}%
\label{tab:solvability-status}
\begin{tabular}{lr}
\toprule
Status & Specifications \\
\midrule
Hard but fair: never solved blind, no specification gap found & 575 \\
Rewritten, then solved in blind pass@4 & 45 \\
Rewritten, blind validation not finished & 68 \\
\bottomrule
\end{tabular}
\end{table}

\subsection{Adequacy audit (Stage~B, 21--30 August 2026)}
\label{app:qa-audit}

\paragraph{Criterion.} Does the specification give a competent engineer all
\emph{publicly relevant} behavior needed to satisfy the hidden tests (APIs, inputs and
outputs, ordering, exceptions, edge cases, compatibility) without dictating internals?

\paragraph{Reviewers and sessions.} Three reviewers from different model families
(GLM-5.2, Claude Fable~5, GPT-5.6 at xhigh reasoning) judge each specification with the base
code, hidden tests and F2P/P2P lists available and the gold patch withheld, with no token,
message, tool or time limit (model-generated compaction only); provider errors are
infrastructure, never votes. Because segment $k$ presumes all before
it, reviews run as sealed sequential sessions: the 20 base states are encrypted with
per-segment keys, only segment $k$ is decrypted, and a structured verdict (decision,
missing contracts with evidence, contradictions, confidence) must be submitted before the
key is destroyed and $k{+}1$ is revealed. A specification is accepted only when all three
return \textsc{specific\_enough} for the same SHA-256 of its text.

\paragraph{Waves and freeze.} The review ran after windowing, on the specifications inside
the 439 candidate chains: 6,740 materialized, 6,716 after one abnormal segment was removed,
6,688 after two \texttt{google/gcp\_scanner} chains were lost to a provider content filter
(provider failures are never votes). In wave~1, 4,989 (74.6\%) were unanimous at once;
reviewer strictness differed (INSUFFICIENT rates 10.6\% GLM-5.2, 12.0\% Fable~5, 21.5\%
GPT-5.6). The 8,539 missing-contract findings on the remaining 1,699 fed rewrites by GPT-5.6
(1,459 targets after source filtering, 1,389 accepted, 95.2\%) and then by Claude Fable~5
(70 of 101), each re-entering fresh sealed review by all three reviewers, followed by a
hygiene pass for leaked test language. The 62 still contested were discarded; because a
discard splits its run and fragments shorter than 20 are dropped, 560 collateral segments
and 15 repositories left with them, and the runs were re-windowed: 397 chains over 176
repositories with 6,066 specifications, of which 4,607 (75.9\%) keep their original text,
1,389 (22.9\%) are GPT-5.6 rewrites and 70 (1.2\%) Fable~5 rewrites. Every specification in
the benchmark is unanimously accepted, and a lint finds none of its 4,404
specifications mentioning hidden tests, graders, gold or node IDs. A specification can be
\emph{hard but fair} from \hyperref[app:qa-specs]{\S\ref*{app:qa-specs}} (never solved blind, adjudicated adequate)
and unanimously accepted here: 575 benchmark specifications are exactly that
(\hyperref[tab:solvability-status]{Table~\ref*{tab:solvability-status}}).

\paragraph{Disclosed limits.} Authors and reviewers saw the hidden tests, so specification
and tests are not independent. The rewritten specifications are passed at the same rate as the
unchanged ones by every evaluated model ($-0.5$ to $+1.9$ points), and the families of the three
reviewers gain no more from them than the other models (difference in differences $-1.0$ [$-1.8$, $-0.2$]
points in the continuous-conversation runs, $+0.8$ [$-0.3$, $+1.9$] under the handoff protocol).
No human-labeled adequacy sample exists yet: a pre-registered review of 80 request slots from 66
chains (30 never solved blind, 20 first failures of chains that no model completed, 18 that the
failure census flags as contradicting an earlier request's tests, 12 certified controls; two
reviewers per item and an adjudicator, blind to the developers' patch and to every outcome) is
prepared, with its strata, seed and analysis fixed in advance, and will be reported.

\begin{table}[htbp]
\centering\small
\caption{\textbf{Specification audit, from candidate windows to the re-windowed chains.}}
\label{tab:audit-waves}
\begin{tabular}{@{}lr@{}}
\toprule
Step & Specifications\\
\midrule
In the 439 candidate chains & 6,740\\
After removing one abnormal request & 6,716\\
After a provider content filter blocked two chains & 6,688\\
Accepted unanimously in the first wave & 4,989 (74.6\%)\\
Contested after the first wave (8,539 missing-contract findings) & 1,699\\
Rewrites by GPT-5.6: accepted / targets & 1,389 / 1,459 (95.2\%)\\
Rewrites by Fable~5: accepted / targets & 70 / 101\\
Discarded as still contested & 62\\
Other requests lost when runs split (15 repositories) & 560\\
Re-windowed: 397 chains over 176 repositories & 6,066\\
\bottomrule
\end{tabular}
\end{table}

\subsection{Specification-generation prompt and code}
\label{app:qa-specgen}

The specification of every segment is produced by one call of the function below
(\texttt{\_build\_spec\_incremental} in \texttt{pipeline/dind\_worker.py} of the SWE-Build
generator), sent as a single user message to a chat-completions endpoint at temperature
0.3 with \texttt{max\_tokens} equal to the remaining context window, no tools and no
repository access. Its inputs are the segment's commit messages, its linked GitHub issues
and PRs, and the segment's test diff; the gold source patch is never an input. The proxy
was configured for Claude Opus~4.8; the campaign record describes round-robin over Opus
4.6/4.7/4.8 with GLM-5.2 as a manual fallback during outages, and the per-segment author
model was not recorded. Over-long lines of the function were re-wrapped for the page
(adjacent string literals and parenthesized continuations; the parse tree is identical to
the archived file). If the call overflows the context window it is repeated with the
narrative blocks dropped and only the test diff kept. Both files are archived under
\texttt{benchmark\_construction/spec\_generation/} in the release; the template's
\mbox{SHA-256}~is~{\small\texttt{e667281d068ad30f2a698f9b4b7ac494e3a98d4656ac0e2fb491715f8345beea}}.%

\paragraph{Prompt template.} Line breaks added for width and the em dash shown as
\texttt{--} (the typewriter font has no em dash); \texttt{\{context\}} is filled as shown in
the second listing.
{\scriptsize
\begin{verbatim}
You are writing a task specification for a software engineer who must
implement a change to a codebase. The TESTS ARE THE SOURCE OF TRUTH: the
spec must describe exactly the behavior the test changes below require --
nothing more, nothing less. Anything not exercised by the tests does not
exist for this task; do NOT invent requirements from the commit messages
or issues that the tests don't actually verify (treat those as weak
hints only, and ignore them where they disagree with the tests).

CRITICAL CONSTRAINT: the engineer will NOT be able to see the tests.
Your specification is the ONLY thing they get. Therefore, derived PURELY
from what the tests exercise:
- Describe the required behavior -- expected inputs, outputs, edge cases,
  error conditions -- for exactly what the tests check.
- NAME the public API surface the tests import and call: module paths,
  class names, function/method names and signatures (the engineer must
  create these so the hidden tests can import and call them).
- Do NOT reference the tests themselves: no test file/function names, no
  assertion snippets, no 'the test expects...'. Describe the BEHAVIOR
  the tests imply, not the tests.
- Do NOT describe the implementation -- state WHAT is required, not HOW.
- Do NOT add behavior the tests don't cover, even if commit
  messages/issues mention it.

{context}
Write the self-contained, test-grounded problem statement now:
\end{verbatim}}

\paragraph{Context blocks and overflow fallback.}
{\scriptsize
\begin{verbatim}
--- {context} is built from these blocks, each included only when
non-empty ---
=== COMMIT MESSAGES (in order) ===
- <commit message 1>
- <commit message 2>
...

=== LINKED GITHUB ISSUES / PRs ===
Issue #<n>: <title>
<body>
PR #<m>: <title>
<body>
...

=== TEST CHANGES (hidden from the solver -- describe their behavior,
never reference them) ===
<full test diff of the segment>

--- On context overflow the call is repeated with the narrative dropped
and only this block ---
=== TEST CHANGES (source of truth -- describe the behavior they require,
never reference them) ===
<full test diff of the segment>
\end{verbatim}}

\paragraph{Generating function.}
{\scriptsize
\begin{verbatim}
def _build_spec_incremental(per_commit_dir, github_issues_by_number, vllm_url,
                            overall_test_patch=""):
    """Build ONE self-contained problem statement per TESTABLE UNIT (segment).

    SWE-Build change: the unit of spec generation is the testable segment, not
    individual commits. We aggregate ALL of the segment's signal into a single
    LLM call:
      - every commit message in the segment
      - every linked GitHub issue / PR body
      - the segment's full test diff (the behavioral ground truth)

    The solver agent NEVER sees the tests, so the spec must translate test
    behavior into self-contained prose (name the public API surface the tests
    exercise, describe expected behavior/edge cases) WITHOUT referencing test
    files, test function names, or assertions.

    Signature preserved for the caller. Returns (list_of_specs, issue_generated).
    """
    # ---- aggregate commit messages ----
    messages = []
    if per_commit_dir.exists():
        for cdir in sorted(per_commit_dir.iterdir()):
            if not cdir.is_dir():
                continue
            mf = cdir / "message.txt"
            if mf.exists():
                m = mf.read_text().strip()
                if m:
                    messages.append(m)
    # ---- aggregate linked issues/PRs (dedup by number) ----
    issue_text = ""
    issue_used = False
    seen = set()
    if per_commit_dir.exists():
        for cdir in sorted(per_commit_dir.iterdir()):
            rf = cdir / "issue_refs.txt"
            if not (cdir.is_dir() and rf.exists()):
                continue
            for ref in rf.read_text().strip().split("\n"):
                ref = ref.strip()
                if not ref or ref in seen:
                    continue
                try:
                    num = int(ref)
                except ValueError:
                    continue
                seen.add(ref)
                iss = github_issues_by_number.get(num)
                if iss:
                    label = "Issue" if iss.get("is_issue") else "PR"
                    issue_text += (f"\n{label} #{num}: {iss.get('title','')}\n"
                                   f"{iss.get('body','')}\n")
                    issue_used = True

    if not (messages or issue_text
            or (overall_test_patch and overall_test_patch.strip())):
        return [], False

    context = ""
    if messages:
        context += ("=== COMMIT MESSAGES (in order) ===\n"
                    + "\n".join(f"- {m}" for m in messages) + "\n\n")
    if issue_text:
        context += "=== LINKED GITHUB ISSUES / PRs ===" + issue_text + "\n"
    if overall_test_patch and overall_test_patch.strip():
        context += ("=== TEST CHANGES (hidden from the solver -- describe their behavior, "
                    "never reference them) ===\n" + overall_test_patch + "\n")

    prompt = (
        "You are writing a task specification for a software engineer who must implement "
        "a "
        "change to a codebase. The TESTS ARE THE SOURCE OF TRUTH: the spec must describe "
        "exactly "
        "the behavior the test changes below require -- nothing more, nothing less. "
        "Anything not "
        "exercised by the tests does not exist for this task; do NOT invent requirements "
        "from the "
        "commit messages or issues that the tests don't actually verify (treat those as "
        "weak hints "
        "only, and ignore them where they disagree with the tests).\n\n"
        "CRITICAL CONSTRAINT: the engineer will NOT be able to see the tests. Your "
        "specification "
        "is the ONLY thing they get. Therefore, derived PURELY from what the tests "
        "exercise:\n"
        "- Describe the required behavior -- expected inputs, outputs, edge cases, error "
        "conditions -- "
        "for exactly what the tests check.\n"
        "- NAME the public API surface the tests import and call: module paths, class "
        "names, "
        "function/method names and signatures (the engineer must create these so the "
        "hidden tests "
        "can import and call them).\n"
        "- Do NOT reference the tests themselves: no test file/function names, no "
        "assertion snippets, "
        "no 'the test expects...'. Describe the BEHAVIOR the tests imply, not the "
        "tests.\n"
        "- Do NOT describe the implementation -- state WHAT is required, not HOW.\n"
        "- Do NOT add behavior the tests don't cover, even if commit messages/issues "
        "mention it.\n\n"
        f"{context}\n"
        "Write the self-contained, test-grounded problem statement now:"
    )
    result = _worker_llm_call(vllm_url, prompt, max_tokens=None)
    if result == CONTEXT_OVERFLOW or not result:
        # Tests are the source of truth -- on overflow, DROP the narrative (commit
        # messages + issues), KEEP the test diff, and retry. (Old behavior dropped
        # the tests first, which is backwards for a test-driven spec.)
        if overall_test_patch and overall_test_patch.strip():
            test_only_context = ("=== TEST CHANGES (source of truth -- describe the "
                                 "behavior they require, never reference them) ===\n"
                                 + overall_test_patch + "\n")
            result = _worker_llm_call(
                vllm_url, prompt.replace(context, test_only_context), max_tokens=None)
        if result == CONTEXT_OVERFLOW or not result:
            return [], issue_used
    return [result.strip()], issue_used
\end{verbatim}}

\subsection{Environments, the repair loop and the two-container verifier}
\label{app:qa-env}

Each repository receives one image built from source, with the dependencies of every
commit in the chain installed offline from a frozen closure of ${\approx}3{,}000$ wheels;
history is scrubbed to the base and gold is never present. A task is pinned to an image
digest, and it is the digest, not a rebuild, that is certified. Acceptance is a model-free
\emph{gold walk}: replaying the 20 gold patches with the hidden tests must pass at every
state. 225 of 397 environments passed at once; GPT-5.6-sol (xhigh, at most 80 turns),
allowed to edit only the Dockerfile and accepted only by the model-free gold walk, repaired
167 more, and the 5 unrepairable windows were dropped.

Auditing the repaired images exposed a failure unrelated to the agents under test: pytest
hooks that rewrote a passing base outcome to \texttt{failed} to fake the F2P precondition,
hidden node IDs baked into image layers, dependencies branching on hidden-test names, and
\texttt{git apply} wrappers that special-cased the gold path. The acceptance instrument had
run inside the artifact it certified. All 32 flagged environments were dropped (the release
has 200 baseline and 94 repaired environments, pre-repair Dockerfiles included) and the
verifier was rebuilt so this cannot recur: agent and verifier run in separate containers
and only \texttt{/testbed/.git} crosses between them; the verifier starts from a fresh copy
of the digest with hidden tests in a root-only layer, runs candidate tests as an
unprivileged user in disposable worktrees, and captures outcomes through a registered
in-memory plugin rather than a report file.

\paragraph{Sealing.} The source images carry the repository's full history, so the gold for
every later segment is reachable from \texttt{HEAD}, and installed package metadata names
the future version. Each image is sealed in an offline rebuild of its repository layer: the
object store is re-created from scratch with only the history reachable from the base (not
pruned, since unreachable objects linger after \texttt{git gc}), the checkout is the base
with a single branch ref and no other refs, packed-refs, reflogs, commit-graph files or
hooks; installed Python metadata (dist-info, SCM version files, entry points) is reset to
the base version; build products, caches and stray copies of the project from the future
head, and any credentials, are removed. Content that base tests need (tracked files,
submodules, ignored runtime assets) is kept and declared in a root-owned fingerprint that
the grader trusts, so nested \texttt{.git} markers planted by an agent gain nothing.
Verification runs at four layers: a scripted check in a fresh container (\texttt{HEAD} equals
the base, no refs beyond the branch, \texttt{git fsck} finds nothing unreachable, the
worktree is byte-identical to the base tree, imports resolve into \texttt{/testbed},
metadata at the base version, no credential files); a per-image checklist worked through by
the sealing agents for what a script can miss (the former head unresolvable, \texttt{git
status} shows only allowed entries, \texttt{/tmp} empty, layer size plausible, image
configuration unchanged, fingerprint consistent with the base tree); the gold walk on the sealed image at every state; and the
verifier itself, which at grading time rebuilds the test-time \texttt{.git} from the chain
start, flags unreachable commits, refuses undeclared nested repositories, and withstands the
adversarial probes of \hyperref[app:qa-certification]{Appendix~\ref*{app:qa-certification}}, including attempts to recover the
gold from \texttt{.git}.

\subsection{Protocol details}
\label{app:qa-protocol}

The agent receives the base repository in a persistent workspace, works through shell commands
inside a network-disabled container started from the pinned digest, and at each request is told
to work only on that request in the existing repository, not to modify tests, and that its work
will be checkpointed and verified after submission. On submission only \texttt{/testbed/.git}
is copied to a fresh, digest-identical verifier container, which restores grader-owned
paths, applies hidden test patches $1..k$ and grades state $k$ (\hyperref[app:qa-grader]{\S\ref*{app:qa-grader}}). Whatever
the verdict, the conversation is discarded: request $k{+}1$ opens a fresh one, holding only the
system prompt and specification $k{+}1$, on the repository as the agent left it, with no earlier
message, summary or verdict carried over. Every chain runs all 20 requests; reward is~1 iff the
repository after request 20 passes the whole contract of state 20; neither the horizon nor the
budget is disclosed. Budgets are cumulative over the chain: unlimited model calls, 7,200~s of
tool-execution time, 64,000 output tokens per response and no wall-clock limit; a chain that
exhausts the tool time or the context window is graded for the request in which this happens,
ends there and scores~0. The harness is mini-swe-agent 2.4.6 for every model; the Claude models
run at effort high, the GPT models at reasoning effort xhigh, GLM-5.3 and Kimi~K3 at high, and
GLM-5.3-Flash, whose open weights we serve ourselves, at temperature 1.0.
stages: the verifier initially stopped a chain at its first regression, and every such chain was
later resumed from the archived repository state of that request, with the budget counters
restored, and run to request 20 under the same protocol. The resume diagnostics of \hyperref[sec:agents]{Sections~\ref*{sec:agents}}--\ref{sec:verification} run on the same substrate
from the silent misses of first attempts made in one continuous conversation per chain, continued
after the miss (\hyperref[app:trust]{Appendix~\ref*{app:trust}}), and differ in what follows the miss: they continue
silently past it, return a failure verdict with one repair opportunity, or restart with fresh
context on the same state.

\subsection{Grader rules}
\label{app:qa-grader}

The sealed chain grader, which certifies every task (\hyperref[app:qa-certification]{\S\ref*{app:qa-certification}}), anchors at the
base SHA and requires exactly 20 direct-parent commits, one per segment with the prescribed
subject; merges, extra or duplicate commits, replace refs, grafts and changed git controls are
rejected, and security-sensitive changes are accepted only when byte-identical to gold. The
per-request verifier runs through the same sandboxed gates. At request $k$ it overlays frozen test
patches $1..k$ on two trees, the one the agent received (its state after request $k{-}1$; the base
for $k{=}1$) and the one it submitted, and runs the same nodes on both: F2P$_k$, P2P$_k$ and every
earlier F2P node that the human gold still satisfies at $k$. Behavior the human history
itself later superseded (24.5\% of earlier F2P nodes by state~20; 255 of 280 tasks have
some) is recorded and not demanded.
Nodes that pass on the received tree, other than F2P$_k$, are \emph{protected}; F2P$_k$ and every
earlier request's node still failing on the received tree are \emph{fixable}, so a missed
requirement is carried into the following requests rather than written off. The request is
completed iff every fixable node passes on the submitted tree, and a protected node that no longer
passes is a regression. Every node is \texttt{passed}, \texttt{failed}, \texttt{uncollectable} or
\texttt{missing}, and only the literal \texttt{passed} counts, so a skipped, erroring, deleted or
renamed test fails closed; test paths are restored from the frozen copies before every run,
\texttt{conftest}, \texttt{sitecustomize}, \texttt{.pth} and pytest configuration are protected,
candidate code runs unprivileged in a disposable copy of the tree, and a run that alters any of
them fails every node. At request 20 the pass-to-pass list already holds the repository's
original tests and the hidden tests of requests 1--19, so the nodes run are the whole accumulated
contract and their outcome on the submitted tree is the attempt's final-state verdict, from
which pass@1 is estimated.

\subsection{Filtering rounds}
\label{app:qa-filtering}

Once frozen, nothing is edited: a task that fails any check is dropped and archived, and a
pass on rerun proves nondeterminism, never acceptability. Fifteen rounds took 439 windows
to 280: 42 for contested specifications, provider-filtered or malformed chains, and
re-windowing (R1--R3); 32 for the environment manipulation above (R8); 27 for specification
hygiene, gold patches touching test-classified paths, or non-reproducible repaired
environments (R5, R6); 17 for gold-walk, packaging, apply and window-composition failures
(R4, R7); 20 for failing the first two-container gate (R12); 4 for nondeterminism on an
identical digest (R9--R11); 2 for verifier time above 1,800~s (R13); 1 for a frozen node
failing under the final grader (R14); and, last, 14 chains in six repositories identified post
hoc from the evaluation runs (R15): in each, a model failed the same hidden test when it started from its
own implementations of the earlier segments but passed it when it started from the human
ones, so the test presupposed the human implementation of an earlier segment rather than its
specification. The benchmark is 280 chains, 140 repositories and 4,404 distinct segments.

\begin{table}[htbp]
\centering\small
\caption{\textbf{Chains dropped by the filtering rounds}, from 439 candidate windows to 280 chains.}
\label{tab:filtering}
\resizebox{\linewidth}{!}{%
\begin{tabular}{@{}lr@{}}
\toprule
Reason & Chains\\
\midrule
Contested specifications, provider-filtered or malformed chains, and re-windowing & 42\\
Gold-walk, packaging, apply and window-composition failures & 17\\
Specification hygiene, gold patches on test paths, non-reproducible repaired environments & 27\\
Environment manipulation (\hyperref[app:qa-env]{\S\ref*{app:qa-env}}) & 32\\
Nondeterminism on an identical digest & 4\\
Failure at the first two-container gate & 20\\
Verifier time above 1,800~s & 2\\
A frozen node failing under the final grader & 1\\
Hidden tests presupposing the human implementation of an earlier segment (post hoc, six repositories) & 14\\
\midrule
Total & 159\\
\bottomrule
\end{tabular}}
\end{table}

\subsection{Certification and threat model}
\label{app:qa-certification}

The 280 tasks pass a digest-bound gold walk at all 20 states, the final grader gate and an
independent no-retry repeat; the Oracle scores~1 through the same two-container path an
agent uses. Twenty adversarial canaries score~0: payloads committed before the chain, extra,
duplicate or merge commits, replace refs, grafts, hooks, test renames or edits, symlinks,
gitlinks, grader-aware source and grader tokens, among others. Five legitimate variants
score~1: gold, an untracked conftest, uncommitted source edits, tampering with an external
pytest, and an already-satisfied empty checkpoint. The residual limit is stated: no
executable-test benchmark can stop an implementation from branching on functional inputs.
After the fact, the 26,355 agent trajectories of the continuous-conversation runs
(\hyperref[app:qa-protocol]{\S\ref*{app:qa-protocol}}), covering 447,787 graded steps, were scanned with
14 detector families and a judged sample; no grader defeat was confirmed.

\subsection{Release}
\label{app:qa-release}

\textsc{BOBA-Bench} contains 280 tasks from 140 repositories, 4,404 unique segments and 5,600
graded states; task images are mirrored to an immutable archive registry identified by
SHA-256. Temporal exposure and memorization are measured, not assumed absent
(\hyperref[app:contamination]{Appendix~\ref*{app:contamination}}).

\section{Prior Exposure and Memorization}
\label{app:contamination}

\paragraph{Direct access and prior exposure are different.}
The evaluation image contains the base checkout and no reachable future repository
history. The seal removes future git objects, refs, reflogs, hooks, caches, and credentials,
and resets source-control version metadata to the base.
Scripted checks and the reference-patch walk test the sealed image
(\hyperref[app:qa-certification]{Appendix~\ref*{app:qa-certification}}).
These controls prevent direct retrieval through the evaluation environment; they do not
erase code learned before evaluation.

\paragraph{Source dates.}
Among the dated model cutoffs we compared, only one of the benchmark's 4,404 unique requests
postdates the earliest, and none postdates the others.
Source dates therefore cannot show that the benchmark is unseen.

\paragraph{Patch-continuation probe.}
The model receives the first 40\% of a reference-patch hunk and attempts its remainder.
The near-verbatim endpoint uses a 0.9 similarity threshold.
Fable~5 reproduces 15.1\% of benchmark hunks near-verbatim (interval 8.9--22.2\%) and
14.5\% of same-repository hunks from before its cutoff (\hyperref[tab:contamination]{Table~\ref*{tab:contamination}}).
Its matched contrast is $+0.6$ points (interval [$-8.5,+9.7$]).
Nearly all benchmark requests also predate the model cutoffs, so the probe cannot rule out
exposure to either set.
\hyperref[tab:contamination]{Table~\ref*{tab:contamination}} gives all ten leaderboard models and GLM-5.2. Three Claude models,
including the leader, reproduce benchmark hunks near-verbatim: Claude Opus~5.5 19.2\% (interval
15.0--23.4\%; 27 hunks exactly), Fable~5 15.1\% and Claude Opus~5 14.1\% (interval 9.8--18.4\%, on
the two thirds of requests it does not refuse), with Claude Opus~4.8 at 4.9\%; the GPT, GLM and Kimi
models reproduce 0.7--1.7\%, the level of same-repository code written after their cutoffs
(0.0--1.5\%). None shows benchmark-specific recall: the matched contrast against same-repository
pre-cutoff hunks is $+4.0$ points (interval [$-1.7,+10.2$]) for Opus~5.5, $+2.6$ [$-3.9,+9.4$] for
Opus~5, $+1.6$ [$-0.8,+4.1$] for Opus~4.8 and within $\pm1.2$ points for the rest.

\begin{figure}[!htbp]
\centering
\includegraphics[width=\textwidth]{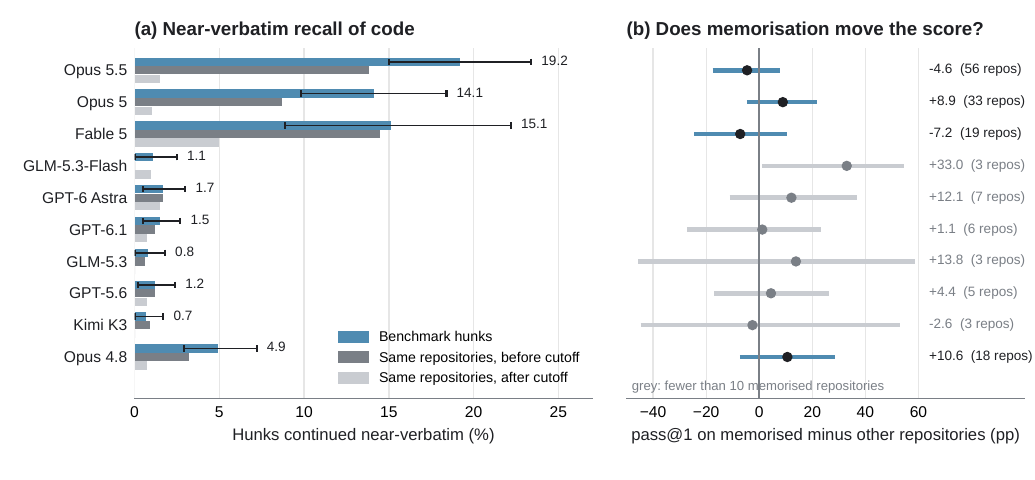}
\caption{\textbf{Three Claude models memorize these repositories, none recalls benchmark hunks more than other pre-cutoff code, and no model with ten
or more memorized repositories passes them measurably more often.} (a) Share of hunks a model continues near-verbatim from their first 40\% (similarity
$\geq$0.9): benchmark hunks, with 95\% repository-bootstrap intervals, and same-repository hunks from before and after the model's cutoff. (b) Handoff
pass@1 on the repositories a model reproduces near-verbatim minus pass@1 on the rest, with repository-clustered 95\% intervals; grey where fewer than ten
repositories are memorized.}
\label{fig:memorisation}
\end{figure}

\paragraph{Memorized text and the score.}
On the handoff trajectories themselves the memorized text does not appear in the solutions more than
in other models': the step patches of all ten models have a mean normalized edit similarity of
0.22--0.30 to the reference patches, 1.8--5.2\% of steps are near-verbatim (similarity $\geq$0.8),
and the request pass rate on the 10\% of steps most similar to the reference minus the 10\% least
similar is within $-1.4$ to $+2.7$ points for every model, every interval covering zero; of the 27
hunks Opus~5.5 reproduces exactly in the probe, its own solution at that request contains on average
20\% of their lines. Within each model, the repositories it reproduces near-verbatim are not passed
more often than the rest: pass@1 differs by $-4.6$ points (interval [$-17.6,+7.9$]) for Opus~5.5
(56 repositories), $+8.9$ [$-4.6,+21.9$] for Opus~5 (33), $+10.6$ [$-7.3,+28.4$] for Opus~4.8 (18)
and $-7.2$ [$-24.4,+10.5$] for Fable~5 (19). \hyperref[fig:memorisation]{Figure~\ref*{fig:memorisation}} shows both. Memorization of the repositories is real; evidence that
it moves the ranking is absent within these bounds, and nonsignificance is not~\mbox{equivalence}.%

\begin{table}[!htbp]
\centering\small
\caption{\textbf{Benchmark-hunk reproduction} (280-chain set). These are probe samples, not
whole-chain rates: benchmark hunks continued near-verbatim from their first 40\%, the same
rate for same-repository hunks from before the model's cutoff, and their difference paired by
repository with a 95\% repository-bootstrap interval. Sampled = requests the model answered
(Opus~5 refuses to reproduce source text verbatim in a third of them; GLM-5.3 and GLM-5.3-Flash ran
to the token cap in the rest of theirs).}
\label{tab:contamination}
\begin{tabular}{@{}lrrrl@{}}
\toprule
Model & Near-verbatim / sampled & Rate (\%) & Pre-cutoff (\%) & Contrast (pp)\\
\midrule
Opus~5.5 & 78 / 407 & 19.2 & 13.8 & $+4.0$ [$-1.7$, $+10.2$]\\
Opus~5 & 38 / 270 & 14.1 & 8.7 & $+2.6$ [$-3.9$, $+9.4$]\\
Fable~5 & 19 / 126 & 15.1 & 14.5 & $+0.6$ [$-8.5$, $+9.7$]\\
GLM-5.3-Flash & 3 / 274 & 1.1 & 0.0 & $+1.2$ [$+0.0$, $+3.2$]\\
GPT-6 Astra & 7 / 407 & 1.7 & 1.7 & $-0.1$ [$-2.1$, $+1.7$]\\
GPT-6.1 & 6 / 407 & 1.5 & 1.2 & $+0.3$ [$-1.3$, $+1.8$]\\
GLM-5.3 & 3 / 395 & 0.8 & 0.6 & $+0.0$ [$-1.0$, $+1.0$]\\
GPT-5.6 & 5 / 406 & 1.2 & 1.2 & $+0.0$ [$-1.6$, $+1.6$]\\
Kimi~K3 & 3 / 407 & 0.7 & 0.9 & $-0.4$ [$-1.7$, $+0.8$]\\
Opus~4.8 & 20 / 406 & 4.9 & 3.2 & $+1.6$ [$-0.8$, $+4.1$]\\
GLM-5.2 & 0 / 97 & 0.0 & 0.0 & $+0.0$ [$+0.0$, $+0.0$]\\
\bottomrule
\end{tabular}
\end{table}

We report the probe as limited evidence about reproduction, not proof of no prior
exposure.

\section{Evaluation Details}
\label{app:eval}

\subsection{Primary run and harness comparison}
\label{app:run-config}

The primary harness is mini-swe-agent \citep{yang2024sweagent}, version 2.4.6 for every model. Each
request starts a fresh conversation on the repository the previous request left; a chain permits
unlimited model calls, two hours of cumulative tool execution over its 20 requests and no
wall-clock limit, with 64,000 output tokens per response; the budget is not disclosed to the
agent, and a chain that exhausts it or the context window ends there and counts as a failure.
Claude Opus~5.5, Opus~5, Opus~4.8 and Fable~5 ran at effort \texttt{high}, GPT-6 Astra, GPT-6.1
and GPT-5.6 at \texttt{xhigh} reasoning, GLM-5.3 and Kimi~K3 at \texttt{high} reasoning, all
through Amazon Bedrock, and GLM-5.3-Flash (BF16 weights) was served with SGLang on our own GPUs at
temperature 1.0 with its default thinking.
These settings do not equalize compute across models.

A chain without a request-20 state counts as a failure: a verifier error on a request, a
tool-time or context-window exit before request 20, or a run whose model requests stopped
returning. \hyperref[app:leaderboard-additional]{Appendix~\ref*{app:leaderboard-additional}} separates these from final states that fail a
test.

\hyperref[tab:scaffold]{Table~\ref*{tab:scaffold}} compares harnesses on a stratified subset of 58 chains from 52
repositories: a 60-chain draw from the release minus the two chains of an excluded repository.
Claude Opus~5.5, GPT-5.6, Kimi~K3 and GLM-5.3-Flash ran these chains under Codex CLI, the
OpenHands SDK and pi, each request a fresh invocation of the harness on the persisting repository,
paired with their mini-swe-agent chains. No harness completes more chains than mini-swe-agent: the
paired difference in pass@1 runs from 0 (GPT-5.6 under the OpenHands SDK) to $-$29 points
(GLM-5.3-Flash under Codex CLI); the interval covers zero for seven of the twelve arms and lies
below it for five, among them all three GLM-5.3-Flash arms, whose regression rate rises from 26\%
to 36-62\% under the richer tool sets. Strict pass@1 stays within 5 points of mini-swe-agent for
the other three models under every harness but pi, 67-90\% of chains get the same verdict under a
harness as under mini-swe-agent, and the first failure falls at the same request in 139 of the 264
chains that fail under both. An earlier comparison in one conversation per chain, with GPT-5.6 and
Kimi~K3 and a chain counted complete only when every request passed on its own step, found no
paired difference excluding zero (intervals ruling out effects above about 21 points) and the same
verdict under all three harnesses for 44 and 48 of the 58 chains; GLM-5.2's OpenHands SDK run
ended 28 of its 58 attempts at the wall limit and is not comparable.

\subsection{Chain completion per model}
\label{app:leaderboard}
\hyperref[tab:leaderboard]{Table~\ref*{tab:leaderboard}} gives the per-model figures behind \hyperref[fig:illusion]{Figure~\ref*{fig:illusion}} and
\hyperref[sec:howfar]{Section~\ref*{sec:howfar}}. pass@1 is the probability that an attempt leaves the repository after
request 20 passing every hidden test of the chain, the original tests and those of all 20 requests,
estimated as the mean over the eight independent attempts of each chain; the single-attempt column
is the first attempt alone. Strict pass@1 requires in addition that every request passed its own
tests on the request's own step and that no passing test ever broke, and the regression rate is the
share of attempts in which a passing test broke at some request, whether or not it was repaired
later; both are means over the eight attempts, with the first attempt's values in
\hyperref[app:leaderboard-additional]{Appendix~\ref*{app:leaderboard-additional}}. The request completion rate is the share of all graded
requests of the first attempt whose tests all pass on the step they are issued, pooled over a
model's chains. An attempt without a request-20 state counts as a failure. Intervals are
repository-clustered bootstrap 95\%; the pairwise tests, on the first attempt, are in
\hyperref[app:separation]{Appendix~\ref*{app:separation}}.

\subsection{Resume arms: per-model results}
\label{app:resume-results}

\hyperref[tab:resume-results]{Table~\ref*{tab:resume-results}} gives the per-model estimates behind \hyperref[sec:compound]{Section~\ref*{sec:compound}}.
Starting points are every request a model submitted as complete that failed its hidden contract
($k\le19$); the control is the benchmark's own continuation after a certified pass of the same
request, standardized to the arm's distribution of $k$. Own-work failure at $k{+}1$ counts
failures of $k{+}1$'s own tests and excludes failures inherited from the open defect. Intervals are
Wilson 95\% for rates and repository-clustered bootstrap 95\% for contrasts (resampling arm and
control together); the repaired and unrepaired subsets of arm E are outcome-selected. GPT-6 Astra and GLM-5.2
have no arm F. The starting states come from the five models' first attempts under one
conversation per chain, as in our earlier release; arm D continues that conversation, whereas
arm F opens a fresh one on the same state, which is what the handoff protocol does at every
request.

The control of \hyperref[tab:resume-results]{Table~\ref*{tab:resume-results}} is a continuation after a certified pass,
standardized to the arm's positions but not to its requests, so part of the gap between arm D and
the control could be the difficulty of the requests that follow a miss. A second control removes
this: for every starting point of Fable~5, GPT-6 Astra and Kimi~K3, the same model attempts
request $k{+}1$ from the developers' own state at $k$ (their patches $1..k$, no hidden tests in the
tree), in a fresh mini-swe-agent conversation, two attempts per state, graded with the frozen
contract of $k{+}1$; the developers' state passes the contract of $k$ and fails the tests of
$k{+}1$ before every attempt (120, 130 and 256 states). On the same requests the developers' state
fails 13.1, 13.5 and 5.6\% of the time against 21.8, 28.5 and 25.0\% after an untreated miss, a
paired difference of 8.7 [$-$2.2, 20.4], 15.0 [5.7, 24.8] and 19.4 [12.4, 27.2] points: about
9 of Fable~5's 18 and of GPT-6 Astra's 24 points against the standardized control are request
difficulty or selection and the rest is the state, and for Kimi~K3 the whole gap is the state.
Where request $k{+}1$ does not touch the code of the missed request the difference is 25.8
[5.7, 51.5], 21.4 [0.0, 42.9] and 14.0 [4.9, 26.9] points. The gold-state attempts are fresh
conversations (arm F bounds what that changes), Fable~5 and Kimi~K3 were served through a different
route and reasoning setting than in arm D, and two attempts per state give coarse per-request rates;
intervals are repository-clustered~bootstrap~95\%.%

\begin{table}[h]
\centering
\scriptsize
\setlength{\tabcolsep}{1.5pt}
\newcommand{\lab}[1]{\parbox[t]{1.35in}{\raggedright\strut #1\strut}}
\caption{\textbf{Resume arms per model.} D: silent continuation; E: one external ``verification failed'' message and one repair before request $k{+}1$; F: fresh conversation on the same broken state; gold state: the same request $k{+}1$ attempted from the developers' state at $k$.}
\label{tab:resume-results}
\begin{tabular}{@{}lccccc@{}}
\toprule
 & Fable~5 & GPT-6 Astra & Kimi~K3 & GPT-5.6 & GLM-5.2 \\
\midrule
\lab{Starting points / control units} & 470 / 174 & 130 / 161 & 836 / 595 & 939 / 572 & 1451 / 1594 \\
\addlinespace
\lab{Control (verified state): $k{+}1$ own-work failure (\%)} & 3.8 & 4.3 & 9.6 & 16.9 & 11.9 \\
\lab{Arm D: after one untreated miss (\%)} & 21.8 [18.3,25.8] & 28.5 [21.4,36.7] & 25.0 [22.2,28.1] & 30.4 [27.5,33.4] & 29.3 [27.0,31.7] \\
\lab{Multiplier, D / control} & 5.8 [2.2,38.0] & 6.6 [2.2,24.1] & 2.6 [1.6,4.8] & 1.8 [1.2,2.9] & 2.5 [1.8,3.4] \\
\lab{Difference, D $-$ control (pp)} & 18.0 [8.3,28.8] & 24.2 [13.1,32.1] & 15.4 [7.3,23.5] & 13.4 [5.4,22.2] & 17.3 [11.5,23.4] \\
\lab{Gold state: $k{+}1$ own-work failure, same requests (\%)} & 13.1 & 13.5 & 5.6 & --- & --- \\
\lab{Difference, D $-$ gold state, paired (pp)} & 8.7 [$-$2.2,20.4] & 15.0 [5.7,24.8] & 19.4 [12.4,27.2] & --- & --- \\
\addlinespace
\lab{Arm F: fresh conversation, same state, F $-$ D paired (pp)} & 4.1 [0.7,8.2] & --- & 0.4 [$-$1.9,2.7] & $-$1.2 [$-$5.1,2.5] & --- \\
\addlinespace
\lab{Arm E: repair@1 (\%)} & 44.6 [40.1,49.1] & 45.4 [37.1,54.0] & 29.4 [26.4,32.6] & 30.0 [27.2,33.0] & 37.4 [34.9,39.9] \\
\lab{Arm E: $k{+}1$ failure, repaired states (\%)} & 5.7 & 5.1 & 3.4 & 12.4 & 8.5 \\
\lab{Arm E: repaired states ($n$)} & 209 & 59 & 232 & 282 & 542 \\
\lab{Arm E: $k{+}1$ failure, unrepaired states (\%)} & 22.7 & 28.2 & 25.0 & 30.0 & 31.1 \\
\lab{Arm E: unrepaired states ($n$)} & 260 & 71 & 559 & 657 & 909 \\
\lab{Arm E: $k{+}1$ failure, all states (\%)} & 15.1 [12.2,18.7] & 17.7 [12.1,25.2] & 18.7 [16.1,21.6] & 24.7 [22.1,27.6] & 22.7 [20.6,24.9] \\
\bottomrule
\end{tabular}
\end{table}

\subsection{Own-work failure by number of open defects}
\label{app:dose-response}

\hyperref[tab:dose-response]{Table~\ref*{tab:dose-response}} pools every graded request from $k{+}1$ to 20 of the silent continuations
of \hyperref[sec:compound]{Section~\ref*{sec:compound}} and groups them by the number of defects still open beneath the request
when it was issued (the original miss plus any later own-work failures). The odds ratio is the slope of
a logistic fit of own-work failure on the count (capped at five), with a 95\% repository-bootstrap
interval. Requests with many open defects sit late in the chain and come from continuations that
were already failing, so the progression is descriptive, not a controlled dose.

\begin{table}[h]
\centering
\scriptsize
\setlength{\tabcolsep}{1.5pt}
\newcommand{\lab}[1]{\parbox[t]{1.5in}{\raggedright\strut #1\strut}}
\caption{\textbf{Own-work failure at requests $k{+}1$ to 20 by number of open defects beneath the request} (silent continuations, 140 repositories).}
\label{tab:dose-response}
\begin{tabular}{@{}lccccc@{}}
\toprule
 & Fable~5 & GPT-6 Astra & Kimi~K3 & GPT-5.6 & GLM-5.2 \\
\midrule
\lab{No open defect beneath the request (\%)} & 8.0 & 8.0 & 12.1 & 12.0 & 10.5 \\
\lab{One (\%)} & 12.7 & 16.2 & 18.1 & 21.5 & 21.7 \\
\lab{Two (\%)} & 36.1 & 29.5 & 38.8 & 28.3 & 32.2 \\
\lab{Three (\%)} & 33.7 & 40.0 & 48.6 & 43.8 & 49.0 \\
\lab{Four (\%)} & 66.1 & 74.1 & 80.1 & 57.3 & 55.0 \\
\lab{Five or more (\%)} & 92.5 & 84.3 & 90.0 & 70.2 & 67.1 \\
\addlinespace
\lab{Units, none} & 893 & 212 & 994 & 1394 & 2268 \\
\lab{Units, one} & 2419 & 672 & 4388 & 4968 & 7902 \\
\lab{Units, two} & 781 & 244 & 1340 & 2270 & 3426 \\
\lab{Units, three} & 409 & 75 & 451 & 726 & 1413 \\
\lab{Units, four} & 127 & 27 & 176 & 391 & 845 \\
\lab{Units, five or more} & 106 & 51 & 260 & 420 & 1276 \\
\addlinespace
\lab{Odds ratio per additional open defect} & 2.28 [1.69,3.86] & 2.20 [1.69,2.79] & 2.32 [1.97,3.00] & 1.72 [1.49,2.05] & 1.72 [1.56,1.92] \\
\bottomrule
\end{tabular}
\end{table}

\paragraph{Inference cost.}
\label{app:cost}
\hyperref[tab:inference-cost]{Table~\ref*{tab:inference-cost}} reports standardized public-API cost estimates for the first
attempts of our earlier single-conversation runs on the 280 chains (140 repositories), the runs
whose token usage we have extracted, including ordinary failures; for the handoff first attempts
of \hyperref[tab:leaderboard]{Table~\ref*{tab:leaderboard}}, that table gives model calls per chain.
For every model, we assume that 98\% of input tokens are cache reads and price the remaining
2\% at short-TTL cache-write rates where available, otherwise at uncached-input rates; these
are estimates under a common caching scenario, not measured cache-hit rates or hosting bills.
Means and medians use attempts with recorded usage; missing usage is not assigned zero cost.

\begin{table}[H]
\centering
\small
\setlength{\tabcolsep}{9pt}
\caption{\textbf{Standardized first-attempt API cost (USD), single-conversation runs.} Coverage gives the number of first
attempts with recorded token usage out of the 280 chains. All models use the same caching
assumption described above.}
\label{tab:inference-cost}
\begin{tabular}{@{}lrrr@{}}
\toprule
Model & Usage / 280 & Mean USD & Median USD \\
\midrule
Fable~5     & 273 & 16.68 & 14.60 \\
GPT-6 Astra & 280 & 22.03 & 17.01 \\
Kimi~K3     & 280 & 9.88 & 8.12 \\
GPT-5.6     & 279 & 12.34 & 10.19 \\
GLM-5.2    & 266 & 9.03 & 6.43 \\
\bottomrule
\end{tabular}
\end{table}

The estimate uses the recorded total input $I$ and output $O$ as
$[I(0.98p_C+0.02p_R)+Op_O]/10^6$, with prices per million tokens for cache reads ($p_C$),
remaining input ($p_R$), and output ($p_O$). Completion tokens already include reasoning.
We use the public rate cards accessed on 6 October 2026, retaining the GPT-5.6 long-context
split above 272,000 prompt tokens and using Astra's flat standard schedule.
These references exclude runner/verifier compute, infrastructure retries and benchmark
construction. Multi-attempt search costs are reported separately in the released cost tables,
accumulating attempts through the first success or exhaustion.

\paragraph{Verifier compute.}
A separate compute model charges a hidden check at four virtual CPUs and \$0.05 per vCPU-hour,
multiplying that rate by mean verifier wall time. This gives approximately \$0.001 per check;
the median recorded check takes 16--17 seconds. Fable uses Harbor-timed continuation checks
because its continuous-worker timestamps also include setup overhead. The estimate excludes
runner provisioning and idle time, storage, network, and test construction and maintenance.
It is not a measured system bill or an estimate of human-review cost. The overall cost of
verification with repair additionally depends on repair rechecks and the downstream work that
successful repairs enable; it is not established by the per-check estimate.

\subsection{Noisy supervisor: intervals}
\label{app:false-alarm-ci}

\begin{figure}[h]
\centering
\includegraphics[width=0.55\textwidth]{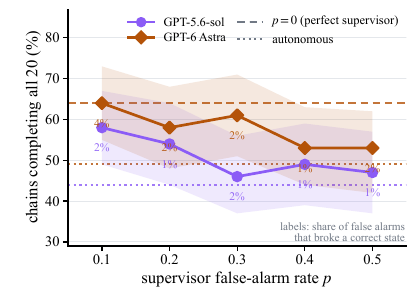}
\caption{\textbf{Occasional false alarms cost little; one in two gives back most of the supervisor's gain, yet
completion never drops below the unsupervised level.} Completion at 20 against the supervisor's
false-alarm rate $p$ (\hyperref[tab:trust]{Table~\ref*{tab:trust}}; GPT-5.6-sol and GPT-6 Astra, 100 chains per level), with
repository-clustered 95\% bootstrap bands. Dashed: perfect supervisor ($p{=}0$); dotted: autonomous (the
same chains' first attempt, no supervisor). Labels: false alarms that broke a correct state (\%).}
\label{fig:false-alarm-ci}
\end{figure}

\hyperref[fig:false-alarm-ci]{Figure~\ref*{fig:false-alarm-ci}} plots the noisy-supervisor results in \hyperref[tab:trust]{Table~\ref*{tab:trust}} with repository-clustered
95\% \mbox{bootstrap bands}.%

\subsection{Root causes of failures}
\label{app:rootcauses}

A request fails when one of its own hidden tests does not pass after it (a silent miss) or when a
protected test of the original repository or of an earlier request that passed before it fails
after it (a regression); a test that was already failing when the request was issued is a carried
failure and is not charged to the request (\hyperref[app:failure-anatomy]{Appendix~\ref*{app:failure-anatomy}}). To ask why requests
fail, we labeled every failed request of the ten models' first attempts on the 140 repositories,
4,787 sites in all (235 to 706 per model, \hyperref[tab:rootcauses]{Table~\ref*{tab:rootcauses}}): 1,645 are the first failure
of their chain, the one failure that happens on a state in which every earlier request passed, and
3,142 are later failures, 1,908 of them on a repository already carrying an earlier failure. Of the
4,787, 3,384 are misses, 1,153 regressions with every own test passing, and 250 both. A judge model
(GPT-5.6-sol at its highest reasoning setting) reads, for each site, the specification as delivered,
the agent's diff and final message, the failing hidden tests with their output, the hidden test patch,
the gold patch and the compacted trajectory, is told that the conversation started fresh at this
request and which tests were already failing, and assigns one of 17 failure tags adapted from the
DeepSWE taxonomy. \hyperref[tab:rootcauses]{Table~\ref*{tab:rootcauses}} gives the distribution of its tags. A sandbox judge
(Claude Fable~5.1) then re-executed, in the sealed image on the agent's own repository, the 1,733 sites
the primary judge had left as benchmark-side suspects or untagged: it confirmed 795 as test-side (a
hidden test stricter than, or in conflict with, the specification), reassigned 743 to an agent-side
cause (184 integration errors, 171 missed requirements, 157 regressions, 153 wrong logic, 78 other)
and withheld a tag on 195; 1,679 of these 1,733 requests were passed from a correct state by another
model or attempt, as were 4,614 of the 4,787 sites. One judge labels each stage, and no handoff site
was relabeled by a second judge or by a human. The agreement between two model judges was measured on
an earlier census of the same benchmark with the same prompt and schema: on 275 sites labeled by both
GPT-5.6-sol and Claude Fable~5, the two chose the same tag on 55\%, $\kappa$ 0.47 [0.40, 0.54], and
the same coarse class on 62\%; where they differed they described the same mechanism and disagreed on
the tag. The labels are therefore indicative, not exact. The four paragraphs below describe the
agent-side causes.

\begin{table}[h]
\centering
\scriptsize
\setlength{\tabcolsep}{2.2pt}
\newcommand{\rclab}[1]{\parbox[t]{1.7in}{\raggedright\strut #1\strut}}
\caption{\textbf{Labels of the 4,787 failed requests of the first attempts on the 140 repositories}
(primary judge; \% of the column's sites). First: the first failure of a chain; Later: every failure
after it; All and the model columns: every failure. Benchmark-side tags are the primary judge's
suspicion from static evidence; their re-execution by the sandbox judge is reported in the text.}
\label{tab:rootcauses}
\resizebox{\linewidth}{!}{%
\begin{tabular}{@{}l*{13}{r}@{}}
\toprule
Label & First & Later & All & Opus~5.5 & Opus~5 & GLM-5.3-Flash & Fable~5 & GPT-6 Astra & GPT-6.1 & GLM-5.3 & GPT-5.6 & Kimi~K3 & Opus~4.8 \\
\midrule
\rclab{Regression: pre-existing behavior broken} & 44.0 & 19.3 & 27.8 & 31 & 31 & 23 & 39 & 25 & 28 & 22 & 30 & 32 & 22 \\
\rclab{Missed requirement: part of the specification unimplemented} & 10.6 & 15.9 & 14.1 & 9 & 12 & 18 & 17 & 9 & 8 & 13 & 11 & 17 & 21 \\
\rclab{Wrong logic} & 10.6 & 10.6 & 10.6 & 9 & 13 & 12 & 6 & 11 & 10 & 18 & 10 & 10 & 9 \\
\rclab{Integration error: new code unreachable from the tested path} & 5.4 & 7.7 & 6.9 & 4 & 5 & 5 & 5 & 6 & 6 & 10 & 8 & 5 & 11 \\
\rclab{Other agent-side: wrong file, syntax, unverified assumption, knowledge gap} & 3.6 & 4.9 & 4.4 & 4 & 2 & 3 & 3 & 6 & 4 & 3 & 2 & 4 & 10 \\
\rclab{Benchmark-side, suspected: test stricter than the specification, broken or ambiguous} & 24.3 & 38.0 & 33.3 & 40 & 35 & 36 & 25 & 41 & 41 & 33 & 36 & 30 & 24 \\
\rclab{No tag} & 1.6 & 3.6 & 2.9 & 2 & 3 & 3 & 5 & 3 & 4 & 1 & 3 & 1 & 4 \\
\midrule
Sites & 1645 & 3142 & 4787 & 235 & 310 & 474 & 465 & 470 & 512 & 488 & 565 & 562 & 706 \\
\bottomrule
\end{tabular}}
\end{table}

\begin{figure}[t]
\centering
\includegraphics[width=\textwidth]{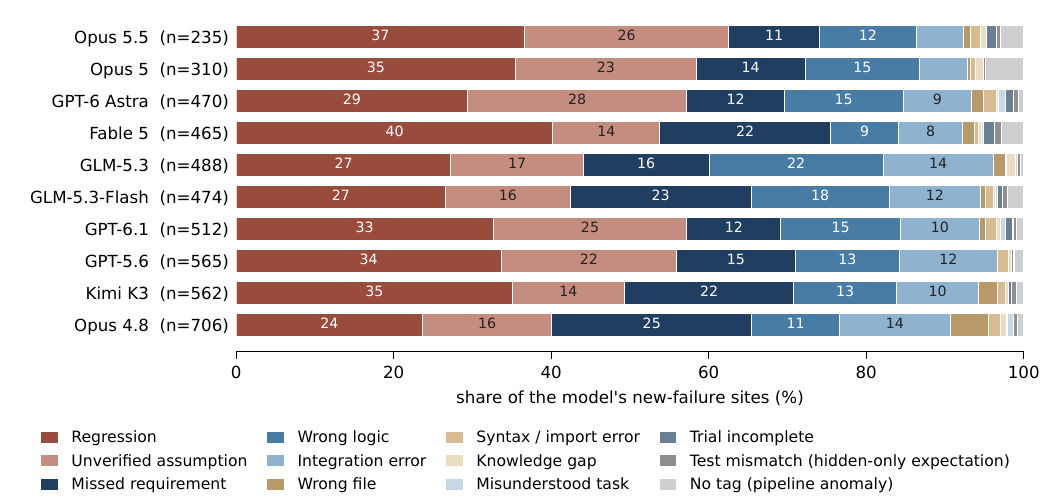}
\caption{\textbf{Breaking what already worked is the most common new failure for nine of ten models and nearly half of all first failures.} Final root-cause label (static GPT-5.6 judge, Claude Fable 5.1 sandbox judge for the test-side and withheld tags, published merge rule; LLM labels, indicative) of the 4,787 requests at which a model's own hidden tests failed or a protected test that passed before the request failed after it: first attempt of the ten models on the 280 chains evaluated to request 20 without a stopping rule (leaderboard order; bar labels in per cent). Regressions are 31\% of all new failures (48\% of the chains' first failures but 23\% of the 3,142 later ones); unverified assumptions (19\%; 22\% of later failures) and missed requirements (18\%; 21\%) follow.}
\label{fig:taxonomy-all}
\end{figure}
\begin{figure}[t]
\centering
\includegraphics[width=\textwidth]{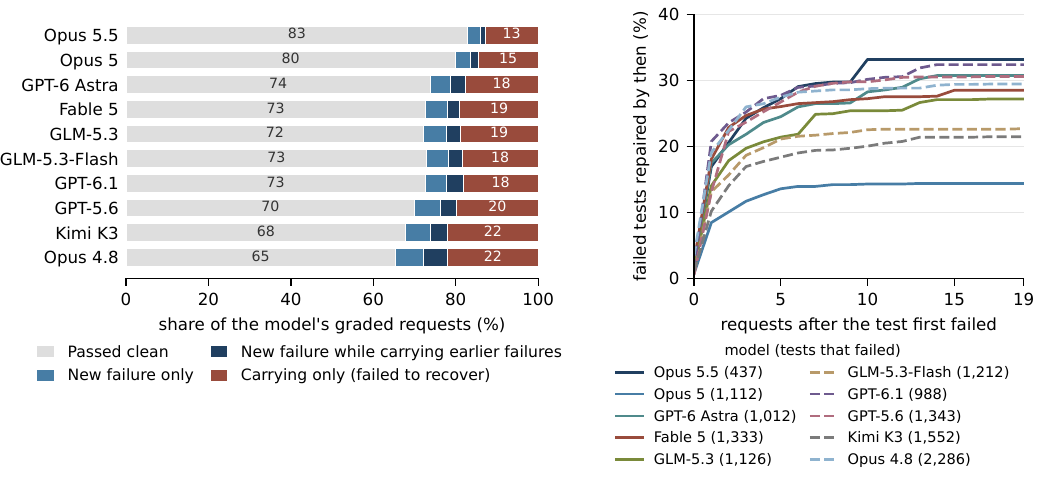}
\caption{\textbf{Failures are seldom repaired: only 14--33\% of the tests that fail during a chain ever pass again, and models carry an earlier failure at 14--28\% of their requests.} Left: outcome of each of the 55,545 graded requests (first attempt, ten models, no stopping rule): passed clean, a new failure on an otherwise clean tree, a new failure while earlier failures were still carried, or carrying only (a test from an earlier request failing before and after the request with no new failure; bar labels = carrying-only share in per cent). Right: share of the 12,401 tests that failed during a chain (misses, regressions and silent breaks; tests failing only under a later step's test overlay excluded) repaired within $d$ requests of their first failure; the repairs that happen come within 1--2 requests.}
\label{fig:failed-to-recover}
\end{figure}
\subsubsection{Regressions}
\label{app:rc-regression}
Regression is the modal class of first failures for every model (37--55\% of them): the agent
implements the request and breaks behavior the request did not ask it to change. Among later
failures its share falls to 19\%, and over all failures it is 28\% (22--39\% per model). In 1,251 of
the 1,329 regression sites a protected test fails; in the remaining 78 the new hidden test exercises
pre-existing behavior that the edit broke. The protected tests that break are, in 1,111 of the 1,251
sites, hidden tests of an earlier request of the same chain, and in 126 only tests of the original
repository: the agent breaks behavior it implemented itself, in a conversation it no longer has,
against tests it never saw. In the traced example (\hyperref[app:miss-cases]{Appendix~\ref*{app:miss-cases}}), GPT-5.6 looks in
the history for ``prior compatibility expectations'', finds that the decorators it is asked to make
stackable were ``only just added in the latest commit'' (its own request~8), concludes that there
was ``no earlier decorator contract to preserve'', and redesigns them; two tests from request~8 then
fail.

\subsubsection{Missed requirements}
\label{app:rc-missed}
One failure in seven leaves part of the specification unimplemented: the tests of the new behavior
fail and nothing else does (98\% of these sites fail only F2P nodes). The class is rarer at a chain's
first failure (11\%) than later (16\%), and 220 of its 674 sites sit on a repository that was
already carrying an earlier failure. It ranges from 8\% of GPT-6.1's failures to 21\% of Opus~4.8's,
and its share is flat along the chain (13\%, 14\% and 15\% by band).

\subsubsection{Wrong logic and integration errors}
\label{app:rc-logic}
A further 18\% of failures implement the request but wrongly: the new code computes the wrong
result (wrong logic, 11\%) or is correct but never reached, wired into a literal but not into the
constructor path the tests exercise, or into a sibling of the function the request names
(integration error, 7\%). Both are F2P-only failures in nearly every case. Integration errors are
rarest for Opus~5.5 (4\%) and most common for GLM-5.3 (10\%) and Opus~4.8 (11\%).

\subsubsection{Unverified submission}
\label{app:rc-selfverif}
The self-verification census (55,487 submitted requests of the ten models' first attempts on the 280
chains) asks what the agent did before it submitted. The agents test before almost every submission
(97\% of requests invoke a test runner; 90--100\% per model), and a green local test run barely
lowers the chance of a miss: submissions preceded by a passing run fail their request's hidden
contract 8.3\% of the time against 8.6\% for all submissions (3.8--12.2\% against 4.2--12.6\% per
model), and 59\% of the misses follow a clean local run (47--77\% per model; 96\% follow some test
run).
The agents
test what they wrote, not what was asked; a passing run of their own tests is what makes them
believe a request complete.

\subsection{Additional leaderboard details}
\label{app:leaderboard-additional}

\paragraph{Complete-chain rates.} \hyperref[tab:leaderboard-280]{Table~\ref*{tab:leaderboard-280}} gives the first attempt's pass@1,
strict pass@1 and regression rate on the 280 chains with repository-clustered 95\% bootstrap
intervals, the values that \hyperref[tab:leaderboard]{Table~\ref*{tab:leaderboard}} averages over the eight attempts, and the
number of chains that pass@1 credits although a request was missed or a passing test broken along
the way and repaired later. The converse, chains that delivered every request and never broke a
test but fail pass@1, is rarer (0 to 5 per model): a test that never passed in the verifier
environment is neither protected nor repairable and fails at request 20. pass@8 counts the chains
that at least one of eight independent attempts completes under the same criterion, an attempt
that broke a passing test being judged on its continuation to request 20.

\begin{table}[htbp]
\centering
\scriptsize
\setlength{\tabcolsep}{4pt}
\caption{\textbf{Complete-chain rates on the 280 chains (140 repositories), first attempt.} Counts
are chains; brackets are repository-clustered 95\% bootstrap intervals in percent. Healed: chains
counted by pass@1 but not by strict pass@1. pass@8: chains completed by at least one of the eight
attempts. A chain without a request-20 state counts as a failure.}
\label{tab:leaderboard-280}
\resizebox{\linewidth}{!}{%
\begin{tabular}{@{}lrlrlrlrrl@{}}
\toprule
Model & pass@1 (attempt 1) & [95\% CI] & strict pass@1 & [95\% CI] & regression rate & [95\% CI] & healed & pass@8 & [95\% CI] \\
\midrule
Claude Opus~5.5 & 187 (66.8) & [60.4, 73.0] & 156 (55.7) & [49.1, 62.2] &  59 (21.1) & [16.1, 26.4] & 36 & 235 (83.9) & [79.2, 88.4] \\
Claude Opus~5   & 170 (60.7) & [54.3, 67.0] & 143 (51.1) & [44.6, 57.4] &  77 (27.5) & [22.0, 33.2] & 30 & 232 (82.9) & [77.5, 88.0] \\
Claude Fable~5  & 145 (51.8) & [45.4, 58.0] & 102 (36.4) & [29.8, 43.0] & 118 (42.1) & [35.7, 48.9] & 44 & 202 (72.1) & [66.3, 77.6] \\
GLM-5.3-Flash   & 143 (51.1) & [44.3, 57.8] & 108 (38.6) & [31.8, 45.4] &  87 (31.1) & [24.9, 37.4] & 36 & 217 (77.5) & [71.6, 83.3] \\
GPT-6 Astra     & 143 (51.1) & [44.3, 57.8] & 101 (36.1) & [29.6, 42.6] &  98 (35.0) & [28.8, 41.5] & 42 & 179 (63.9) & [57.2, 70.5] \\
GPT-6.1-sol     & 137 (48.9) & [42.3, 55.4] & 102 (36.4) & [29.8, 43.2] & 104 (37.1) & [30.9, 43.7] & 36 & 170 (60.7) & [54.1, 67.2] \\
GLM-5.3         & 130 (46.4) & [39.7, 53.2] & 107 (38.2) & [32.1, 44.4] &  89 (31.8) & [25.6, 38.0] & 28 & 213 (76.1) & [70.0, 82.1] \\
GPT-5.6-sol     & 128 (45.7) & [38.9, 52.5] &  91 (32.5) & [26.0, 38.9] & 121 (43.2) & [36.6, 50.0] & 40 & 168 (60.0) & [53.4, 66.4] \\
Kimi~K3         & 127 (45.4) & [38.4, 52.3] &  92 (32.9) & [25.8, 40.1] & 124 (44.3) & [37.5, 51.3] & 36 & 191 (68.2) & [61.7, 74.4] \\
Claude Opus~4.8 & 107 (38.2) & [31.6, 44.9] &  77 (27.5) & [21.4, 34.0] & 115 (41.1) & [34.2, 48.3] & 32 & 168 (60.0) & [53.0, 66.8] \\
\bottomrule
\end{tabular}}
\end{table}

\paragraph{Chains without a final state.} An attempt counts as a failure in pass@1 and strict pass@1
whether or not it has a graded repository after request 20; \hyperref[tab:no-verdict]{Table~\ref*{tab:no-verdict}} lists the
first attempt's. Three kinds of chain have none
(\hyperref[tab:no-verdict]{Table~\ref*{tab:no-verdict}}). A \emph{verifier error} on a request leaves the chain without a
verdict from that request on. A \emph{budget exit} ends the chain when the two-hour tool-time
budget or the model's context window is exhausted before request 20; the chain is graded for the
request in which it happens and stops. A \emph{failed run} is one whose model requests stopped
returning (three Claude Fable~5 chains, closed after a deterministic request-timeout loop).
Counting all of these chains as failures is conservative. In the regression rate they count as
not regressed unless a protected test had already broken, and in the request completion rate they
contribute the requests they completed. Excluding them instead gives the last column of
\hyperref[tab:no-verdict]{Table~\ref*{tab:no-verdict}}.

\begin{table}[htbp]
\centering
\scriptsize
\setlength{\tabcolsep}{5pt}
\caption{\textbf{Chains without a request-20 state} on the 280 chains, by cause, and pass@1 over the
chains that have one. Verifier errors and budget exits include those of the no-stop continuation
of~a~chain.}%
\label{tab:no-verdict}
\begin{tabular}{@{}lrrrr@{}}
\toprule
Model & Verifier error & Budget exit & Failed run & pass@1 with a final state \\
\midrule
Claude Opus~5.5 & 2 & 0 & 0 & 187/278 (67.3) \\
Claude Opus~5   & 5 & 0 & 0 & 170/275 (61.8) \\
Claude Fable~5  & 1 & 0 & 3 & 145/276 (52.5) \\
GLM-5.3-Flash   & 2 & 5 & 0 & 143/273 (52.4) \\
GPT-6 Astra     & 2 & 0 & 0 & 143/278 (51.4) \\
GPT-6.1-sol     & 2 & 0 & 0 & 137/278 (49.3) \\
GLM-5.3         & 3 & 3 & 0 & 130/274 (47.4) \\
GPT-5.6-sol     & 2 & 0 & 0 & 128/278 (46.0) \\
Kimi~K3         & 3 & 3 & 0 & 127/274 (46.4) \\
Claude Opus~4.8 & 3 & 0 & 0 & 107/277 (38.6) \\
\bottomrule
\end{tabular}
\end{table}

\paragraph{Failures along the chain.} \hyperref[tab:leaderboard-survival]{Table~\ref*{tab:leaderboard-survival}} follows the chains
request by request. A chain is \emph{clean through request $k$} if every request up to $k$ passed
its own tests on its own step and no passing test broke; clean through 20 is strict pass@1. The
\emph{first-failure hazard} of a quarter is the share of requests in that quarter at which a chain
still clean picked up its first failure, a silent miss or a broken protected test, over the
requests such chains reached in the quarter. For every model the hazard of the last quarter is at
least that of the first, so the fall of the clean share over the chain is not front-loaded: it is
the compounding of a roughly constant per-request rate. The miss and the regression are both
common as a first failure: of the chains with one, 49 to 62\% first missed a request, 31 to 45\%
first broke a passing test, and 4 to 8\% did both at the same request. At request 20 the models
pass that request's own tests in 84-94\% of the chains with a final state, yet 29-55\% of those
chains carry a failing test from an earlier request or from the original repository.

\begin{table}[htbp]
\centering
\scriptsize
\setlength{\tabcolsep}{4pt}
\caption{\textbf{Clean share by request and first-failure hazard by quarter on the 280 chains.}
Clean through $k$: share of chains with no silent miss and no broken protected test at any request
up to $k$ (a chain without a verdict at some request $\le k$ counts as not clean). Hazard: first
failures at requests 1--5, 6--10, 11--15 and 16--20 over the requests reached by chains that were
still clean. Request completion: share of all graded requests that passed their tests on their
own step with no protected test broken.}
\label{tab:leaderboard-survival}
\resizebox{\linewidth}{!}{%
\begin{tabular}{@{}lrrrrrrrrrr@{}}
\toprule
 & \multicolumn{4}{c}{Clean through request $k$ (\%)} & & \multicolumn{4}{c}{First-failure hazard by quarter (\%)} & Request \\
\cmidrule(lr){2-5}\cmidrule(lr){7-10}
Model & $k{=}5$ & $k{=}10$ & $k{=}15$ & $k{=}20$ & & 1--5 & 6--10 & 11--15 & 16--20 & completion (\%) \\
\midrule
Claude Opus~5.5 & 90.7 & 75.4 & 66.4 & 55.7 & & 1.9 & 3.5 & 2.5 & 3.5 & 84.0 \\
Claude Opus~5   & 88.2 & 73.9 & 62.1 & 51.1 & & 2.5 & 3.2 & 3.4 & 3.8 & 80.8 \\
Claude Fable~5  & 80.7 & 60.4 & 46.1 & 36.4 & & 4.1 & 5.4 & 5.2 & 4.6 & 73.2 \\
GLM-5.3-Flash   & 83.6 & 63.9 & 48.9 & 38.6 & & 3.5 & 5.1 & 5.2 & 4.6 & 73.9 \\
GPT-6 Astra     & 84.3 & 67.1 & 51.1 & 36.1 & & 3.3 & 4.4 & 5.4 & 6.7 & 75.2 \\
GPT-6.1-sol     & 83.6 & 64.6 & 49.6 & 36.4 & & 3.5 & 4.9 & 5.2 & 5.9 & 74.2 \\
GLM-5.3         & 82.5 & 63.2 & 49.3 & 38.2 & & 3.8 & 5.1 & 4.7 & 4.9 & 73.0 \\
GPT-5.6-sol     & 82.1 & 60.0 & 44.6 & 32.5 & & 3.8 & 6.1 & 5.8 & 6.2 & 71.8 \\
Kimi~K3         & 76.8 & 57.5 & 43.2 & 32.9 & & 5.1 & 5.5 & 5.4 & 5.4 & 68.8 \\
Claude Opus~4.8 & 73.6 & 52.5 & 40.7 & 27.5 & & 5.8 & 6.4 & 5.0 & 7.4 & 66.2 \\
\bottomrule
\end{tabular}}
\end{table}

\begin{figure}[htbp]
\centering
\includegraphics[width=\textwidth]{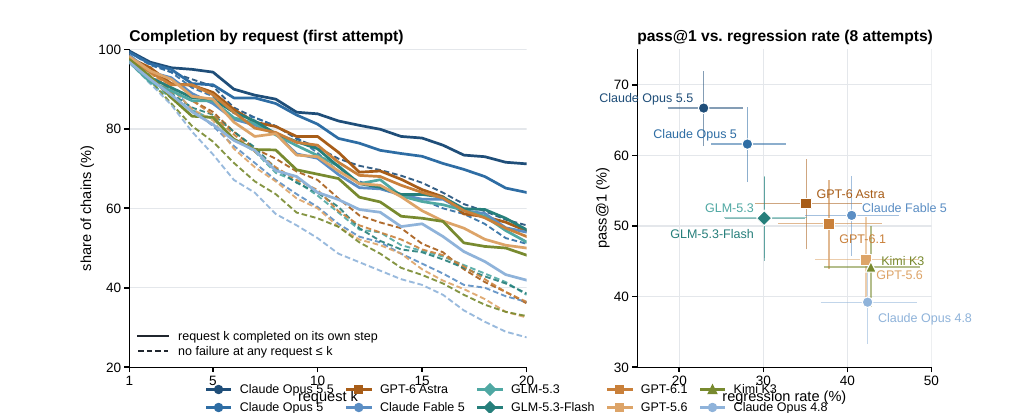}
\caption{\textbf{Completion erodes request by request, and the models that break the fewest passing tests
complete the most chains.} Left: for each model,
the share of chains whose request $k$ passed its own tests on its own step with no protected test
broken (solid) and the share with no failure at any request up to $k$ (dashed), first attempt, 280
chains. Right: pass@1 against the regression rate, both estimated from the eight attempts of each
chain, with repository-clustered 95\% intervals.}
\label{fig:chain-curves}
\end{figure}

\paragraph{Retries and the constant-rate fit.} Every chain gets eight independent attempts; pass@8
counts the chains at least one of whose attempts passes under the pass@1 criterion. Eight attempts
complete 12 to 30 points more chains than the first attempt alone (\hyperref[tab:leaderboard-280]{Table~\ref*{tab:leaderboard-280}}):
60 to 84\% of the chains, against 38 to 67\% on the first attempt and 39 to 67\% per attempt on
average (\hyperref[tab:leaderboard]{Table~\ref*{tab:leaderboard}}), and 46 to 78\% with every
request delivered on its own step and nothing broken in at least one attempt. The chains that no
model completes in any of its eight attempts number 19 (44 are completed by no model on the first
attempt). With $h$ a model's pooled
first-failure hazard, $(1-h)^{20}$ predicts a clean share at request 20 of 56.8, 53.0, 37.6, 40.1,
39.0, 38.6, 39.5, 34.3, 33.5 and 28.7\% against the first attempt's strict pass@1 of 55.7, 51.1, 36.4,
38.6, 36.1, 36.4, 38.2, 32.5, 32.9 and 27.5\%, so a constant per-request rate overestimates every
model by 0.6 to 2.9 points: the strict record is well described by a flat hazard that compounds
over twenty requests. pass@1 is not: it depends on what happens after a failure, which
\hyperref[app:rootcauses]{Appendix~\ref*{app:rootcauses}} measures request by request.

\paragraph{Shared outcomes.} On the first attempt, 44, 21, 18, 23, 22, 23, 20, 22, 15, 26 and 46
chains are completed by 0, 1, \dots, 10 models; if models failed independently at their own rates,
0.2 chains would be completed by none and 0.3 by all ten. When two models both fail somewhere in a
chain, their first failures fall on the same request in 44.6\% of cases (2{,}624 of 5{,}885
pairs), against 5.4\% if failure positions were independent. Any two models agree on pass or fail
for 63--90\% of chains (chance: 46--54\%). The 44 chains that no model completes on the first
attempt come from 35 repositories; on them the models complete a median of 7 to 10 of the 20
requests, and 19 of them resist every model's eight attempts.

\paragraph{Clustered intervals.} Chains from one repository share code, tests and history, so
every interval in \hyperref[tab:leaderboard-280]{Table~\ref*{tab:leaderboard-280}} resamples the 140 repositories with replacement
(10{,}000 draws), recomputes the statistic as a ratio of sums over the resampled chains
(passing chains over chains) and reports the 2.5th and 97.5th percentiles; they are 1 to 3 points
wider than the per-chain Wilson intervals of \hyperref[tab:leaderboard]{Table~\ref*{tab:leaderboard}}. The 140 repositories
contribute one to five chains each (34, 81, 18, 5 and 2 repositories respectively).

\section{Reproducibility and Evidence Scope}
\label{app:repro}

\paragraph{What the paper documents.}
\hyperref[app:qa]{Appendix~\ref*{app:qa}} gives the construction pipeline, the specification prompt, grader rules, and
every filtering round; \hyperref[app:eval]{Appendix~\ref*{app:eval}} gives models, settings, and budgets; the
\texttt{handoff/} directory of the release carries the scoring document, the per-model results
with their intervals and the per-chain, per-request records behind \hyperref[tab:leaderboard]{Table~\ref*{tab:leaderboard}};
\hyperref[tab:analysis-sets]{Table~\ref*{tab:analysis-sets}} gives the population of each study.

\begin{table}[htbp]
\centering\small
\caption{\textbf{Populations by study.} The unit and eligible subset depend on the experiment.}
\label{tab:analysis-sets}
\begin{tabularx}{\linewidth}{@{}lX@{}}
\toprule
Study & Population and scope\\
\midrule
Baseline & 280 chains, 140 repositories; ten models; pass@1 estimated over eight attempts per chain (2,240 per model), single-attempt values alongside; pass@8 over the same attempts.\\
Whole-chain repair & 280 chains, 140 repositories; five models, continuous-conversation first attempts; under the handoff protocol, ten models paired with their Autonomous first attempts (\hyperref[app:trust]{Appendix~\ref*{app:trust}}).\\
Task-size and dependency statistics & 280 chains, 140 repositories.\\
Verified-prefix comparison & Retained boundaries from the 140 repositories; eleven models, model-specific retention.\\
Failed-state continuation & Eligible silent misses from the continuous-conversation first attempts, with model-specific missing outcomes; gold-state control on 120 / 130 / 256 of those states (Fable~5 / GPT-6 Astra / Kimi~K3), two attempts each.\\
Blind implementation review & 109 wrong implementations per author from the Autonomous first attempts, each paired with a correct one: 2,180 items per reviewer, ten reviewers, 21,800 reviews.\\
Repair dose & Eligible misses on the 280 chains; GPT-5.6 and Kimi~K3 in the continuous conversation; ten models under the handoff protocol (\hyperref[app:trust]{Appendix~\ref*{app:trust}}).\\
Inference cost & Continuous-conversation first attempts on 280 chains, five models; usage available for 273 / 280 / 280 / 279 / 266 attempts (Fable / Astra / Kimi / GPT-5.6 / GLM). Standardized API estimates with a shared 98\% cache-read assumption (\hyperref[app:cost]{Appendix~\ref*{app:cost}}).\\
Verifier compute & Timed checks for Fable, GPT-5.6, Kimi and GLM, priced with an assumed CPU-hour rate; not measured supervision overhead.\\
False-alarm study & 100 stratified chains; sealed run with 100 uncensored attempts per level for both models.\\
Harness comparison & 58 chains, 52 repositories (a 60-chain draw made before the last filtering round, minus the two chains of a repository that round excluded); four models under three other harnesses, paired with their mini-swe-agent first attempts.\\
Prior-exposure probes & Hunk samples from the 140 repositories; all ten models.\\
Failure anatomy and self-checks & Failure anatomy: every graded request of the ten models' first attempts (55,545 requests, 4,787 failures); self-verification: every submitted request of the same attempts (\hyperref[app:rc-selfverif]{Appendix~\ref*{app:rc-selfverif}}); self-checks: 60 silent misses and 60 accepted submissions per model from the continuous-conversation first attempts (\hyperref[app:trial-records]{Appendix~\ref*{app:trial-records}}).\\
\bottomrule
\end{tabularx}
\end{table}

\paragraph{Replication.}
A replication needs the task images by digest, the grader, the specifications, and the per-attempt
verdicts, together with the model and harness versions of \hyperref[app:eval]{Appendix~\ref*{app:eval}} and, for the
leaderboard, the scoring document and the per-request records of the release's \texttt{handoff/}
directory, which define every verdict.

\paragraph{Statistical scope.}
Baseline intervals are repository-clustered bootstrap 95\% intervals for the eight-attempt
estimates and Wilson 95\% intervals per chain for the single-attempt values, over all 280 chains
of an arm; attempts without a request-20 record count as failures.
2,000 draws. Pairwise completion intervals use 20,000 draws.
These counts belong to different analyses, not inconsistent repeats of one procedure.
The prefix, review, and repair tables use the intervals and tests of each study.
We infer no paired uncertainty from marginal rates.

\section{How well do coding benchmarks separate models?}
\label{app:separation}

This appendix gives the pairwise tests behind the leaderboard (\hyperref[app:leaderboard]{Appendix~\ref*{app:leaderboard}}) and
compares the resolving power of \textsc{BOBA-Bench} with that of the other repository-evolution
leaderboards: that no two of the top five models on SWE-Milestone or SlopCodeBench can be told apart from the
counts those papers report, that SWE-Chain reports no dispersion at all, and that on
\textsc{BOBA-Bench} 26 of the 45 pairs among ten models are separated, the leader from every
other model at $p\le0.036$. The script (\texttt{separation\_analysis.py}) and its output accompany the paper.

\paragraph{Data.}
For each benchmark we take the per-model figures from the paper's main results table and the unit
the figure is a fraction of (\hyperref[tab:sep-data]{Table~\ref*{tab:sep-data}}). Where the table reports a percentage rather
than a count we recover the count as $\mathrm{round}(\text{rate}\times n)$; for SlopCodeBench the
denominator is fixed at 196 by construction, and for SWE-Milestone the rates are per-evolution-range averages, so the implied count
is approximate to within one milestone. SWE-Milestone's primary metric, Score, is a mean of
recall$\times$precision and not a fraction of trials, so we use its Resolve Rate (a milestone is
fully resolved or not). SWE-Chain's resolving rate is a mean over 12 chains of test-level
micro-averages from a single run, with no per-chain values or dispersion published; it cannot be
tested, and we show binomial bounds for orientation only. For \textsc{BOBA-Bench} we use attempt-1
outcomes on the 280-chain set of \hyperref[sec:benchmark]{Section~\ref*{sec:benchmark}}; a chain without a request-20 record (a
verifier error, a budget or context exit before request 20, or a run whose model requests stopped
returning; 2 to 7 chains per model) counts as a failure, as in the published rates.

\begin{table}[h]
\centering\scriptsize
\setlength{\tabcolsep}{4pt}
\caption{Inputs. Unit: what one trial is. Count: how the per-model count was obtained.}
\label{tab:sep-data}
\begin{tabular}{@{}lp{5.0cm}lrp{2.4cm}@{}}
\toprule
Benchmark & Metric (source) & Unit & $n$ & Count \\
\midrule
SWE-Milestone \citep{swemilestone2026} & Resolve Rate (Table 2), 15 agent--model configurations & milestone & 98 & round(rate$\times n$), $\pm1$ \\
SlopCodeBench \citep{slopcodebench2026} & strict and isolated solve (Table 1; best run per model) & checkpoint & 196 & round(rate$\times n$) \\
SWE-Chain \citep{swechain2026} & Build+Fix resolving (Table 3), single run & chain & 12 & not a count \\
\textsc{BOBA-Bench} & pass@1, attempt 1 & chain & 280 & exact \\
\bottomrule
\end{tabular}
\end{table}

\paragraph{Tests.}
Intervals are Wilson 95\% score intervals. For the other benchmarks the papers publish aggregates,
not per-task outcomes, so two models can only be compared with an unpaired test: we use the two-sided
Fisher exact test on the two counts at $\alpha=0.05$. For \textsc{BOBA-Bench} every model ran the same
280 chains, so we use the exact two-sided McNemar test on the discordant chains---the test
\citet{liu2026converged} apply to adjacent SWE-bench Verified entries---and, because the 280 chains
come from 140 repositories (one to five chains each), a repository-clustered paired bootstrap of the
difference in pass@1 (repositories resampled with replacement, all their chains entering, 20{,}000
resamples, seed 20260921). We also
report the unpaired Fisher test on our own counts so that the comparison with the other benchmarks
uses the same test. No multiplicity correction is applied on either side; a correction would only
reduce the number of separated pairs, and of the claims below it would remove the leader's separation
from the runner-up (a Holm correction over the 45 pairs keeps 17 of the 26 separated pairs).

\paragraph{Other benchmarks.}
\hyperref[tab:sep-others]{Table~\ref*{tab:sep-others}} gives, for each benchmark, how many of all model pairs, of adjacent pairs
in the ranking, and of pairs among the top five are separated. On none of them is any adjacent pair
or any pair among the top five separated. Separation exists only between a leader and clearly weaker
models: on SWE-Milestone Gemini 3 Pro at 13.4\% is separable only from MiniMax M2.5 at 1.3\%; on
SlopCodeBench GPT 5.5 at 14.8\% is first separable from Sonnet 4.6 at 7.1\% (strict) or Kimi K2.6 at 18.9\% from
28.1\% (isolated). On SWE-bench Verified, \citet{liu2026converged} find that exact McNemar tests
separate none of the 29 adjacent pairs among the top thirty entries, whose scores span 8.8 points.

\begin{table}[h]
\centering\scriptsize
\setlength{\tabcolsep}{3pt}
\caption{Pairs of models separated at $\alpha=0.05$ (two-sided Fisher exact test on reported counts). Spread: best to worst reported rate.}
\label{tab:sep-others}
\begin{tabular}{@{}llrrrrrp{3.8cm}@{}}
\toprule
Benchmark & Metric & Models & Spread (\%) & All pairs & Adjacent & Top five & Leader vs.\ runner-up (Wilson 95\%) \\
\midrule
SWE-Milestone & Resolve Rate & 15 & 13.4--1.3 & 9/105 & 0/14 & 0/10 & 13/98 [7.9, 21.4] vs.\ 12/98 [7.1, 20.2] \\
SlopCodeBench & strict solve & 15 & 14.8--2.0 & 27/105 & 0/14 & 0/10 & 29/196 [10.5, 20.4] vs.\ 22/196 [7.5, 16.4] \\
SlopCodeBench & isolated solve & 15 & 28.1--4.1 & 46/105 & 0/14 & 0/10 & 55/196 [22.2, 34.7] vs.\ 51/196 [20.4, 32.6] \\
SWE-Chain & resolving, Build+Fix & 9 & 60.8--20.2 & \multicolumn{4}{p{7.2cm}}{not testable: single run, means over 12 chains, no dispersion published; a binomial bound with 12 units separates 0/36} \\
\bottomrule
\end{tabular}
\end{table}

\paragraph{\textsc{BOBA-Bench}.}
\hyperref[tab:sep-ours]{Table~\ref*{tab:sep-ours}} gives the ten models' pass@1 with Wilson intervals and
\hyperref[tab:sep-ours-pairs]{Table~\ref*{tab:sep-ours-pairs}} every pairwise comparison. Twenty-six of the 45 pairs are separated by the McNemar test and 26 by the clustered bootstrap
(25 in common: Opus~5.5 vs.\ Opus~5 is separated by the McNemar test alone, its bootstrap interval touching zero, and Fable~5 vs.\ GPT-5.6 by the bootstrap alone, $p=0.064$), 20 by the unpaired Fisher test; three of the nine adjacent pairs are separated, and seven of the ten pairs among the top five.
Opus~5.5 is separated from every other model ($p\le0.036$: the runner-up Opus~5 at $p=0.036$, the other eight at $p\le2.1\times10^{-5}$;
$p\le0.048$ if chains without a request-20 record are excluded instead of counted as failures). Below the two Opus models, seven models from 51.8\% to 45.4\% form a group within which
only two of the 21 pairs are separated (GLM-5.3-Flash vs.\ Kimi~K3, $p=0.044$; GPT-6 Astra vs.\ GPT-5.6, $p=0.049$), and Opus~4.8 is separated from all nine
others; the ranking therefore has four tiers, Opus~5.5, then Opus~5 (the two separated by the McNemar test only), then the seven, then Opus~4.8.
The number of models differs from the other leaderboards (10 against 9--15), so
the comparable statements are about adjacent pairs, pairs among the top five, and the leader against
the field, not the raw count of separated pairs.

\begin{table}[h]
\centering\scriptsize
\setlength{\tabcolsep}{5pt}
\caption{\textsc{BOBA-Bench}, 280 chains, attempt 1: pass@1 with Wilson 95\% intervals.}
\label{tab:sep-ours}
\begin{tabular}{@{}lrl@{}}
\toprule
Model & pass@1 & Wilson 95\% \\
\midrule
Opus 5.5 & 187/280 (66.8\%) & [61.1, 72.0] \\
Opus 5 & 170/280 (60.7\%) & [54.9, 66.2] \\
Fable 5 & 145/280 (51.8\%) & [46.0, 57.6] \\
GLM-5.3-Flash & 143/280 (51.1\%) & [45.2, 56.9] \\
GPT-6 Astra & 143/280 (51.1\%) & [45.2, 56.9] \\
GPT-6.1 & 137/280 (48.9\%) & [43.1, 54.8] \\
GLM-5.3 & 130/280 (46.4\%) & [40.7, 52.3] \\
GPT-5.6 & 128/280 (45.7\%) & [40.0, 51.6] \\
Kimi K3 & 127/280 (45.4\%) & [39.6, 51.2] \\
Opus 4.8 & 107/280 (38.2\%) & [32.7, 44.0] \\
\bottomrule
\end{tabular}
\end{table}

\begin{table}[h]
\centering\scriptsize
\setlength{\tabcolsep}{5pt}
\caption{\textsc{BOBA-Bench} pairwise comparisons on the same 280 chains. Discordant: chains passed only by the first / only by the second model. McNemar: exact two-sided $p$ on the discordant chains. Fisher: unpaired two-sided $p$ on the two counts. Bootstrap: repository-clustered 95\% interval of the difference in percentage points.}
\label{tab:sep-ours-pairs}
\begin{tabular}{@{}llrrlll@{}}
\toprule
& & Diff.\ (pp) & Discordant & McNemar $p$ & Fisher $p$ & Bootstrap 95\% \\
\midrule
Opus 5.5 & Opus 5 & $+6.1$ & 38 / 21 & $0.036$ & $0.16$ & [$+0.0$, $+11.9$] \\
Opus 5.5 & Fable 5 & $+15.0$ & 69 / 27 & $2.1\times10^{-5}$ & $4.1\times10^{-4}$ & [$+8.1$, $+21.8$] \\
Opus 5.5 & GLM-5.3-Flash & $+15.7$ & 68 / 24 & $4.9\times10^{-6}$ & $2.1\times10^{-4}$ & [$+8.8$, $+22.6$] \\
Opus 5.5 & GPT-6 Astra & $+15.7$ & 59 / 15 & $2.6\times10^{-7}$ & $2.1\times10^{-4}$ & [$+9.8$, $+21.8$] \\
Opus 5.5 & GPT-6.1 & $+17.9$ & 66 / 16 & $2.3\times10^{-8}$ & $2.6\times10^{-5}$ & [$+11.7$, $+24.1$] \\
Opus 5.5 & GLM-5.3 & $+20.4$ & 75 / 18 & $1.9\times10^{-9}$ & $1.7\times10^{-6}$ & [$+13.2$, $+27.4$] \\
Opus 5.5 & GPT-5.6 & $+21.1$ & 75 / 16 & $2.7\times10^{-10}$ & $7.1\times10^{-7}$ & [$+14.0$, $+28.0$] \\
Opus 5.5 & Kimi K3 & $+21.4$ & 78 / 18 & $4.4\times10^{-10}$ & $4.6\times10^{-7}$ & [$+14.5$, $+28.4$] \\
Opus 5.5 & Opus 4.8 & $+28.6$ & 89 / 9 & $1.1\times10^{-17}$ & $1.7\times10^{-11}$ & [$+22.0$, $+35.1$] \\
Opus 5 & Fable 5 & $+8.9$ & 51 / 26 & $5.9\times10^{-3}$ & $0.041$ & [$+2.9$, $+14.9$] \\
Opus 5 & GLM-5.3-Flash & $+9.6$ & 53 / 26 & $3.2\times10^{-3}$ & $0.027$ & [$+3.1$, $+16.4$] \\
Opus 5 & GPT-6 Astra & $+9.6$ & 53 / 26 & $3.2\times10^{-3}$ & $0.027$ & [$+3.2$, $+16.1$] \\
Opus 5 & GPT-6.1 & $+11.8$ & 58 / 25 & $3.8\times10^{-4}$ & $6.5\times10^{-3}$ & [$+5.6$, $+18.2$] \\
Opus 5 & GLM-5.3 & $+14.3$ & 62 / 22 & $1.5\times10^{-5}$ & $9.3\times10^{-4}$ & [$+7.5$, $+21.0$] \\
Opus 5 & GPT-5.6 & $+15.0$ & 58 / 16 & $9.7\times10^{-7}$ & $5.0\times10^{-4}$ & [$+8.7$, $+21.3$] \\
Opus 5 & Kimi K3 & $+15.4$ & 64 / 21 & $3.3\times10^{-6}$ & $3.7\times10^{-4}$ & [$+8.8$, $+21.9$] \\
Opus 5 & Opus 4.8 & $+22.5$ & 70 / 7 & $3.5\times10^{-14}$ & $1.4\times10^{-7}$ & [$+16.8$, $+28.3$] \\
Fable 5 & GLM-5.3-Flash & $+0.7$ & 46 / 44 & $0.92$ & $0.93$ & [$-6.2$, $+7.7$] \\
Fable 5 & GPT-6 Astra & $+0.7$ & 45 / 43 & $0.92$ & $0.93$ & [$-5.5$, $+7.0$] \\
Fable 5 & GPT-6.1 & $+2.9$ & 46 / 38 & $0.45$ & $0.55$ & [$-2.9$, $+8.7$] \\
Fable 5 & GLM-5.3 & $+5.4$ & 59 / 44 & $0.17$ & $0.24$ & [$-1.8$, $+12.4$] \\
Fable 5 & GPT-5.6 & $+6.1$ & 46 / 29 & $0.064$ & $0.18$ & [$+0.4$, $+11.8$] \\
Fable 5 & Kimi K3 & $+6.4$ & 48 / 30 & $0.054$ & $0.15$ & [$+0.0$, $+13.0$] \\
Fable 5 & Opus 4.8 & $+13.6$ & 55 / 17 & $8.1\times10^{-6}$ & $1.6\times10^{-3}$ & [$+8.0$, $+19.2$] \\
GLM-5.3-Flash & GPT-6 Astra & $+0.0$ & 41 / 41 & $1.0$ & $1.0$ & [$-6.9$, $+6.9$] \\
GLM-5.3-Flash & GPT-6.1 & $+2.1$ & 42 / 36 & $0.57$ & $0.67$ & [$-4.3$, $+8.4$] \\
GLM-5.3-Flash & GLM-5.3 & $+4.6$ & 42 / 29 & $0.15$ & $0.31$ & [$-1.5$, $+10.6$] \\
GLM-5.3-Flash & GPT-5.6 & $+5.4$ & 46 / 31 & $0.11$ & $0.24$ & [$-1.1$, $+11.6$] \\
GLM-5.3-Flash & Kimi K3 & $+5.7$ & 36 / 20 & $0.044$ & $0.20$ & [$+0.4$, $+10.9$] \\
GLM-5.3-Flash & Opus 4.8 & $+12.9$ & 55 / 19 & $3.4\times10^{-5}$ & $2.9\times10^{-3}$ & [$+6.8$, $+19.0$] \\
GPT-6 Astra & GPT-6.1 & $+2.1$ & 17 / 11 & $0.34$ & $0.67$ & [$-1.7$, $+5.9$] \\
GPT-6 Astra & GLM-5.3 & $+4.6$ & 52 / 39 & $0.21$ & $0.31$ & [$-2.7$, $+12.0$] \\
GPT-6 Astra & GPT-5.6 & $+5.4$ & 33 / 18 & $0.049$ & $0.24$ & [$+0.4$, $+10.4$] \\
GPT-6 Astra & Kimi K3 & $+5.7$ & 47 / 31 & $0.089$ & $0.20$ & [$-1.1$, $+12.4$] \\
GPT-6 Astra & Opus 4.8 & $+12.9$ & 53 / 17 & $1.9\times10^{-5}$ & $2.9\times10^{-3}$ & [$+7.1$, $+18.8$] \\
GPT-6.1 & GLM-5.3 & $+2.5$ & 51 / 44 & $0.54$ & $0.61$ & [$-4.9$, $+9.7$] \\
GPT-6.1 & GPT-5.6 & $+3.2$ & 29 / 20 & $0.25$ & $0.50$ & [$-1.4$, $+7.8$] \\
GPT-6.1 & Kimi K3 & $+3.6$ & 39 / 29 & $0.27$ & $0.45$ & [$-2.7$, $+9.7$] \\
GPT-6.1 & Opus 4.8 & $+10.7$ & 51 / 21 & $5.3\times10^{-4}$ & $0.013$ & [$+5.0$, $+16.4$] \\
GLM-5.3 & GPT-5.6 & $+0.7$ & 46 / 44 & $0.92$ & $0.93$ & [$-6.6$, $+8.0$] \\
GLM-5.3 & Kimi K3 & $+1.1$ & 41 / 38 & $0.82$ & $0.87$ & [$-5.3$, $+7.5$] \\
GLM-5.3 & Opus 4.8 & $+8.2$ & 53 / 30 & $0.015$ & $0.060$ & [$+2.2$, $+14.3$] \\
GPT-5.6 & Kimi K3 & $+0.4$ & 30 / 29 & $1.0$ & $1.0$ & [$-5.6$, $+6.2$] \\
GPT-5.6 & Opus 4.8 & $+7.5$ & 40 / 19 & $8.6\times10^{-3}$ & $0.087$ & [$+2.1$, $+13.1$] \\
Kimi K3 & Opus 4.8 & $+7.1$ & 40 / 20 & $0.013$ & $0.10$ & [$+1.7$, $+12.6$] \\
\bottomrule
\end{tabular}
\end{table}

\paragraph{What the comparison does and does not establish.}
Three caveats bound the claim. First, the other benchmarks are tested unpaired because per-task
outcomes are not published; with those outcomes a paired test could separate more pairs than we
find, which is why \hyperref[sec:related]{Section~\ref*{sec:related}} says ``from their reported counts''. Second, treating
tasks, milestones or checkpoints as independent trials ignores their clustering within repositories
or problems (SWE-Milestone's 98 milestones come from seven repositories, SlopCodeBench's 196
checkpoints from 36 problems); the true intervals are therefore wider than those in \hyperref[tab:sep-others]{Table~\ref*{tab:sep-others}}, which
makes the finding of no separation conservative in that respect. Third, the leaderboards evaluate
different model sets under different scaffolds, so the comparison is about the resolving power of
each benchmark on the models its authors ran, not about any \mbox{particular model}.%
\section{Trial Records: Failure Anatomy, Self-Checks, and Traced Regressions}
\label{app:trial-records}

This appendix gives the long form of three results of \hyperref[sec:mf]{Section~\ref*{sec:mf}} (``How agents miss''). It reads the
per-request grader records and the agent conversations of the first attempts on the 280
chains; the failure anatomy covers every graded request of the handoff first attempts, and the traced
cases are first failures, the first request of a chain whose own tests fail or that breaks a protected
test. Unless
stated otherwise, intervals are repository-clustered bootstrap 95\% intervals.

\subsection{Failure anatomy}
\label{app:failure-anatomy}

\paragraph{Method.} The grader of request $k$ runs every test of the request's universe, its own
hidden tests (F2P$_k$), the pass-to-pass list of state $k$ and the accumulated tests of earlier
requests (\hyperref[app:qa-grader]{Appendix~\ref*{app:qa-grader}}), once on the tree the agent received and once on the tree it
submitted, and records each node as passed, failed, uncollectable or missing (a deleted or renamed
test keeps its frozen id and counts as missing). A request has a \emph{new failure} if one of its own
tests does not pass afterwards (a silent miss) or if a protected test, one that passed before the
request, does not pass afterwards (a regression); it is \emph{carrying} a failure if a test that was
already failing when it was issued is still failing afterwards. Every graded request of every chain,
55,545 across the ten models, therefore has one of four outcomes: passed clean, new failure only,
carrying only, or both. A new failure is \emph{own only} if only the request's own tests fail,
\emph{earlier only} if only protected tests fail, and a \emph{regression} if at least one protected
test fails. Every class comes from grader output; no model judges it. Sixty of the 4,787 new
failures have a failing node that passes when re-run in the sealed image (flaky or order-dependent
tests); they are kept as graded. Tests that fail on the received tree only because the request's own
test overlay rewrote or newly imports them are grader artefacts and are excluded from the carried
counts; so are original repository tests that never pass in the verifier environment. Every
classified failure followed the agent's own submission of the request as complete. The
\emph{first failure} of a chain is its first request with a new failure (1,645 of the 2,800
model~$\times$~chain slots; 1,079 chains passed all 20 requests, 31 ended on a verifier or budget
error, or did not finish, before any failure, and 45 chains' first graded failure was only a
carried earlier test): it is the one failure that happens on a state in which every earlier
request passed, so nothing it breaks was inherited. \hyperref[app:rootcauses]{Appendix~\ref*{app:rootcauses}} labels the same
4,787 sites with a judge model and reaches a regression share of 44.0\% of first failures against
44.4\% here, and 27.8\% of all failures against 29.3\%, by a different route: the judge places some
broken protected tests on the benchmark side and tags 78 failures of a request's own tests as
regressions of behavior the new test exercises.

\begin{table}[htbp]
\centering
\small
\setlength{\tabcolsep}{4pt}
\caption{\textbf{The first failure of a first attempt, and the failures after it}, from grader output alone
(280 chains, ten models, first attempts). Each first failure is classed by its failing tests: those of
the new request (own) and those of behavior that already existed (protected tests of the original
repository or of an earlier request). A regression fails at least one protected test. Shares are
percentages of a model's first failures ($n$); the pooled row gives repository-clustered bootstrap
95\% intervals. ``Earlier request broken'': first failures that break a test added or changed by an
earlier request of the same chain. The last two columns give the new failures after the first one and
their regression share.}
\label{tab:failure-anatomy}
\resizebox{\linewidth}{!}{%
\begin{tabular}{lrrrrrrrr}
\toprule
 & \multicolumn{6}{c}{First failure} & \multicolumn{2}{c}{Later failures} \\
\cmidrule(lr){2-7} \cmidrule(lr){8-9}
 & $n$ & Own tests only & Earlier tests only & Both & Regression & Earlier request broken & $n$ & Regression \\
\midrule
Opus~5.5      & 116 & 56.0 & 37.9 & 6.0 & 44.0 & 37.9 & 119 & 17.6 \\
Opus~5        & 130 & 53.8 & 41.5 & 4.6 & 46.2 & 41.5 & 180 & 18.3 \\
GLM-5.3-Flash & 165 & 61.2 & 31.5 & 7.3 & 38.8 & 33.3 & 309 & 17.5 \\
Fable~5       & 170 & 47.6 & 45.9 & 6.5 & 52.4 & 47.6 & 295 & 31.2 \\
GPT-6 Astra   & 170 & 55.3 & 39.4 & 5.3 & 44.7 & 37.1 & 300 & 19.3 \\
GPT-6.1       & 170 & 53.5 & 40.6 & 5.9 & 46.5 & 37.1 & 342 & 21.1 \\
GLM-5.3       & 163 & 60.1 & 35.0 & 4.9 & 39.9 & 35.0 & 325 & 16.3 \\
GPT-5.6       & 184 & 53.8 & 42.4 & 3.8 & 46.2 & 37.5 & 381 & 23.6 \\
Kimi~K3       & 180 & 52.2 & 40.0 & 7.8 & 47.8 & 40.0 & 382 & 25.9 \\
Opus~4.8      & 197 & 61.4 & 32.0 & 6.6 & 38.6 & 33.5 & 509 & 19.6 \\
\midrule
Pooled & 1645 & 55.6 {\scriptsize[50.6, 60.8]} & 38.5 {\scriptsize[33.2, 43.8]} & 5.9 {\scriptsize[3.5, 8.5]}
       & 44.4 {\scriptsize[39.4, 49.6]} & 37.9 {\scriptsize[32.8, 43.1]} & 3142 & 21.4 {\scriptsize[17.4, 26.0]} \\
\bottomrule
\end{tabular}}
\end{table}

\paragraph{Where and how far back.} The regression share is 37.8\% [29.8, 46.3] of first failures at
requests 1--5, 49.6\% at requests 6--10, 51.5\% at requests 11--15, and 38.3\% at requests 16--20.
The rise from requests 1--5 to requests 6--15 is 12.6 points [2.6, 22.8], but the share falls again
at the end of the chain, so we do not claim a monotone trend. Of the 731 first-failure regressions,
624 (85.4\% [79.3, 90.8]) break a test that an earlier request of the chain added or changed; 103
break only tests of the original repository. The nearest such request lies a median of three requests
back (interquartile range 2--6), and 35.3\% [28.7, 42.0] of these regressions break a request five or
more back. Every broken test had passed on the tree the agent received: the grader protects only
such tests, and at a first failure every earlier request had passed. In 38.5\% [33.4, 44.1] of first
failures, and in 24.1\% [20.3, 28.0] of all failures, every own test passes and only earlier
behavior breaks, so a grader of the new request's tests alone would accept the submission. The
contract checks only tested behavior. Breakage of untested behavior is not measured, so the true
regression share is likely higher.

\paragraph{After a failure.} Nothing stops a chain, so what follows a failure is observed rather than
inferred (\hyperref[tab:carried]{Table~\ref*{tab:carried}}). A new failure occurs at 8.6\% of all graded requests: at
3.6--9.6\% of the requests issued on a repository in which every earlier test passed and at
7.7--20.7\% of those issued on a repository already carrying a failure; the second group sits later in
the chain and in chains that were already failing, so the gap is descriptive, not a controlled
contrast (\hyperref[app:resume-results]{Appendix~\ref*{app:resume-results}} gives the controlled one). The failures that follow a
first failure are of a different kind: 44\% of first failures are regressions against 21\% of the
3,142 later ones, which are mostly misses of the request's own contract (79\%). What breaks mostly
stays broken. Of the 12,401 tests that fail at some request (5,919 as a miss, 5,639 as a regression,
843 broken while outside the universe of the intervening requests, so that the break is seen only
later), 3,314 (27\%;
14--33\% per model) pass again at a later request, typically one or two requests later (median per
model 1--2), and 9,283 are still failing at the chain's end. At 14--28\% of all graded requests a
model is carrying at least one earlier failure, and a model's chains spend 17--35\% of their 20
requests in that state. Of the 1,697 chains (of 2,800) that break at some
request, 360 (16--29\% per model) are fully green again at request 20; these are the chains that
pass@1 counts and strict pass@1 does not (\hyperref[sec:howfar]{Section~\ref*{sec:howfar}}).

\begin{table}[htbp]
\centering
\small
\setlength{\tabcolsep}{3pt}
\caption{\textbf{Carried failures}: every graded request of the first attempts (280 chains per model) by
its outcome, and what happens to the tests that fail. Clean: every test of the universe passes; New:
a new failure on a repository with no earlier failure; Carrying: an earlier failure is still failing
and nothing new fails; Both. ``New failure rate'' is the share of requests with a new failure among
those issued on a clean repository and among those issued on a repository carrying a failure
(descriptive; the two groups differ in position and chain). Recovered: tests that fail at some request
and pass again at a later one (median delay in requests). Healed: chains broken at some request
whose repository passes every test at request 20. Tests that fail on the received tree only because
the request's test overlay rewrote them are excluded.}
\label{tab:carried}
\resizebox{\linewidth}{!}{%
\begin{tabular}{lrrrrrrrrrrr}
\toprule
 & \multicolumn{5}{c}{Graded requests (\%)} & \multicolumn{2}{c}{New failure rate (\%)} & \multicolumn{3}{c}{Tests that fail} & Chains \\
\cmidrule(lr){2-6} \cmidrule(lr){7-8} \cmidrule(lr){9-11} \cmidrule(lr){12-12}
 & $n$ & Clean & New & Carrying & Both & clean repo. & carrying & $n$ & Recovered (\%) & Delay & Healed / broken \\
\midrule
Opus~5.5      & 5574 & 82.9 & 3.1 & 12.9 & 1.1 & 3.6 & 7.7  & 437  & 33.2 & 2 & 36 / 123 \\
Opus~5        & 5549 & 79.9 & 3.6 & 14.5 & 2.0 & 4.3 & 11.9 & 1112 & 14.4 & 1 & 30 / 134 \\
GLM-5.3-Flash & 5528 & 73.0 & 5.2 & 18.5 & 3.3 & 6.7 & 15.3 & 1212 & 22.7 & 1 & 36 / 168 \\
Fable~5       & 5525 & 72.6 & 5.4 & 19.0 & 3.0 & 6.9 & 13.8 & 1333 & 28.5 & 1 & 44 / 174 \\
GPT-6 Astra   & 5567 & 73.8 & 4.9 & 17.7 & 3.5 & 6.2 & 16.7 & 1012 & 30.7 & 1 & 42 / 177 \\
GPT-6.1       & 5571 & 72.6 & 5.0 & 18.2 & 4.2 & 6.5 & 18.6 & 988  & 32.4 & 1 & 36 / 176 \\
GLM-5.3       & 5536 & 72.3 & 5.5 & 18.9 & 3.3 & 7.1 & 14.9 & 1126 & 27.2 & 2 & 28 / 171 \\
GPT-5.6       & 5581 & 70.1 & 6.1 & 19.8 & 4.0 & 8.1 & 16.7 & 1343 & 30.6 & 2 & 40 / 188 \\
Kimi~K3       & 5545 & 67.8 & 6.0 & 22.0 & 4.2 & 8.1 & 16.0 & 1552 & 21.5 & 2 & 36 / 186 \\
Opus~4.8      & 5569 & 65.3 & 6.9 & 22.0 & 5.7 & 9.6 & 20.7 & 2286 & 29.4 & 1 & 32 / 200 \\
\midrule
All           & 55545 & 73.0 & 5.2 & 18.3 & 3.4 & 6.6 & 15.8 & 12401 & 26.7 & -- & 360 / 1697 \\
\bottomrule
\end{tabular}}
\end{table}

\subsection{Self-checks before submission}
\label{app:selfcheck}

\paragraph{Sample.} The sample comes from the first attempts of our earlier runs of the five models
of \hyperref[app:resume-results]{Appendix~\ref*{app:resume-results}}, in which one conversation ran through a chain and the chain
ended at its first failed request; the fixed rules below read a request's commands and outputs and do
not depend on that protocol. From first attempts with a finished record on the 280 chains, we drew,
per model, 60 silent misses and 60 accepted submissions uniformly at random with a fixed seed: 600
submitted and graded requests from 525 attempts in 136 repositories. A miss is the failed request
that ended its attempt, so the 300 misses come from 300 different attempts.

\paragraph{Rules.} Fixed regular-expression rules read only the sampled request's commands and tool
outputs; no model labels the reported rates. The agent \emph{ran tests} if it called a test runner
(pytest, unittest, tox, nox, or a \texttt{make} test target) or ran a Python assertion script that
writes no repository file. It \emph{ran the repository suite} if a runner call had no targets, or
only repository paths that the agent had not written in that request. \emph{Wrote test} counts
writes of test files, scratch files included. The \emph{last run} is the outcome of the last test
command. It failed if a runner summary reports failures or errors, a test is listed as failed, or a
script raises a traceback or exits with an error; it passed if a summary reports passes and no
failures, or a script exits cleanly; otherwise it is unknown. The agent \emph{claimed success} if
one of its last three non-empty messages states success (for example, that all tests pass or that
the work is verified or complete) and the final message has no hedge. \emph{Green, then claimed}
requires a test run, a passing last run, and a success claim.

\paragraph{Rule check.} We checked the rules against labels that a language model assigned from the
full transcripts, on requests not used to write the rules. On 20 such requests, the action rules
agreed on 100 of 100 labels and the last-run outcome on 18 of 20. After one revision, the
success-claim rule agreed on 18 of 20 further requests. An independent re-implementation of the
rules gives 100\% for running tests, 82.0\% for the repository suite, and 56.3\% for green, then
claimed, against 99.7\%, 85.0\%, and 61.3\% in \hyperref[tab:selfcheck]{Table~\ref*{tab:selfcheck}}.

\paragraph{Results.} Because the sample is stratified on the outcome, we compare misses with
accepted submissions through Mantel--Haenszel odds ratios across models. Neither running the
repository suite (1.44 [0.90, 2.29]) nor a passing last run (1.33 [0.91, 1.98]) is associated with
fewer silent misses. In half of the green-then-claimed misses (93 of 184), the passing last run was
the repository suite; in the rest it was the agent's own script or a targeted run. In at least 43 of
the 300 misses (14.3\%, Wilson [10.8, 18.8]), the agent's own test output named a test that the
grader later failed, and the agent's last mention of that test still reported a failure. This count
is a lower bound: agents often print only summary lines, and the records keep at most 50 failing
test identifiers per request. \hyperref[app:rc-selfverif]{Appendix~\ref*{app:rc-selfverif}} counts every submitted request of
the ten models' handoff first attempts instead of a sample and finds that 59\% of the misses
(47--77\% per model) follow a clean local run; that census applies its own test for a clean run and
requires no success claim, so its shares are not directly comparable with~\hyperref[tab:selfcheck]{Table~\ref*{tab:selfcheck}}.%

\begin{table}[htbp]
\centering
\small
\caption{\textbf{Self-checks before submission}, first attempts. Each cell gives the percentage of
sampled silent misses / accepted submissions (60 each per model) that show the behavior. ``Green,
then claimed'': the last test run passed and the closing message claimed success. ``Wrote test''
includes scratch files. Claim rates depend on whether a model writes closing text, and the
success-claim odds ratio is driven by Fable~5 (without it: 1.35, 0.71--2.70). Pooled intervals and
odds ratios (OR, Mantel--Haenszel across models, silent miss vs.\ accepted) use a
repository-clustered bootstrap.}
\label{tab:selfcheck}
\setlength{\tabcolsep}{3pt}
\resizebox{\linewidth}{!}{%
\begin{tabular}{lccccccc}
\toprule
Model & Ran tests & Repo suite & Wrote test & Saw failure & Last run green & Claimed success & Green, then claimed \\
\midrule
Fable~5     & 100.0 / 98.3 & 73.3 / 58.3 & 26.7 / 5.0 & 56.7 / 41.7 & 71.7 / 68.3 & 65.0 / 26.7 & 46.7 / 20.0 \\
GPT-6 Astra & 100.0 / 100.0 & 91.7 / 90.0 & 6.7 / 0.0 & 90.0 / 78.3 & 78.3 / 65.0 & 98.3 / 96.7 & 76.7 / 61.7 \\
Kimi~K3     & 98.3 / 98.3 & 85.0 / 83.3 & 25.0 / 15.0 & 75.0 / 65.0 & 66.7 / 65.0 & 96.7 / 98.3 & 66.7 / 65.0 \\
GPT-5.6     & 100.0 / 100.0 & 86.7 / 88.3 & 13.3 / 15.0 & 80.0 / 61.7 & 95.0 / 91.7 & 40.0 / 35.0 & 40.0 / 31.7 \\
GLM-5.2     & 100.0 / 96.7 & 88.3 / 80.0 & 36.7 / 15.0 & 81.7 / 65.0 & 78.3 / 75.0 & 96.7 / 91.7 & 76.7 / 70.0 \\
\midrule
All (300 / 300) & 99.7 / 98.7 & 85.0 / 80.0 & 21.7 / 10.0 & 76.7 / 62.3 & 78.0 / 73.0 & 79.3 / 69.7 & 61.3 / 49.7 \\
\quad 95\% CI, misses & 98.9--100 & 80.4--89.3 & 16.7--26.6 & 71.2--81.9 & 73.1--82.8 & 74.1--84.1 & 55.7--66.9 \\
\quad OR (95\% CI) & -- & 1.44 (0.90--2.29) & 2.56 (1.57--4.62) & 2.08 (1.41--3.07) & 1.33 (0.91--1.98) & 2.26 (1.40--3.71) & 1.71 (1.21--2.42) \\
\bottomrule
\end{tabular}}
\end{table}

\subsection{Two traced regressions}
\label{app:miss-cases}

We chose these two first-attempt regressions for clarity, not at random. Both meet five
conditions: the specification states the behavior; the agent tested its change and declared
success; the failing hidden test shows what was broken; that test belongs to an earlier request;
and another model or attempt passed the same request. Quotes are verbatim from the request's
conversation, square brackets mark our edits, and code is abridged.

\begin{table}[t]
\centering\small
\caption{\textbf{A regression that undoes accepted work} (\texttt{python-poetry/cleo}, GPT-5.6,
first attempt; abridged, quotes verbatim). Request~9 is silent on a lone decorator, and the agent's
conversation holds nothing of request~8. A grader of request~9's new test alone would accept this
code. Chosen for clarity, not at random (\hyperref[app:miss-cases]{Appendix~\ref*{app:miss-cases}}).}
\label{tab:miss-example}
\begin{tabularx}{\linewidth}{@{}>{\bfseries}l>{\raggedright\arraybackslash}X@{}}
\toprule
Request 8 & Add \texttt{@app.command}, \texttt{@app.argument} and \texttt{@app.option}; a lone
\texttt{@app.argument} or \texttt{@app.option} registers a command under the function's name.
GPT-5.6's request-8 code does this, and the request passes.\\
\midrule
Request 9 & Let the three decorators stack on one function. In a fresh conversation, the agent
reads the history, finds the decorators ``only just added in the latest commit'' and concludes:
``There was no earlier decorator contract to preserve''. It makes \texttt{argument} and
\texttt{option} store definitions only, checks the stacked form (``all 41 application tests
pass'') and submits.\\
\midrule
Hidden tests & Request~9's new test passes. Two request-8 tests fail: \texttt{application.has(...)}
is \texttt{False} after a lone decorator. Opus~5.5, Opus~5, Fable~5, GLM-5.3 and GLM-5.3-Flash
pass request~9 on their first attempts.\\
\bottomrule
\end{tabularx}
\end{table}

\paragraph{Case 1: \texttt{python-poetry/cleo}, GPT-5.6, request 9 (\hyperref[tab:miss-example]{Table~\ref*{tab:miss-example}}).}
Request~8 adds three decorators to \texttt{Application}. Each registers a command. Used alone,
\texttt{@app.argument} and \texttt{@app.option} register the command under the decorated function's
name. The agent's request-8 code does exactly that, each decorator building a \texttt{Command} and
calling \texttt{self.add}, and the request passes. Request~9, in a new conversation that holds
nothing of request~8, asks that \texttt{@command}, \texttt{@argument} and \texttt{@option} stack on
one function, with \texttt{@command} applied last. It states: ``The \texttt{argument} and
\texttt{option} decorators contribute definition entries; the \texttt{command} decorator finalizes
registration.'' It says nothing about a decorator used alone. The agent reads the three methods
(``each currently creates and registers its own independent command'') and, before editing, looks
for ``any prior compatibility expectations (such as standalone \texttt{argument}/\texttt{option}
behavior), since preserving those where possible will reduce regressions''. The history shows the
methods added by the previous commit, its own request~8, and it concludes: ``There was no earlier
decorator contract to preserve---the methods were only just added in the latest commit---so I'll
implement the specified finalize-on-\texttt{command} behavior directly.'' It rewrites
\texttt{argument} and \texttt{option} so that they only store a definition on the function for
\texttt{@command} to collect. Its checks use the stacked form; the repository's visible tests, which
do not contain request~8's hidden tests, pass (``all 41 application tests pass, and the broader
suite passes 152 tests when excluding the two pre-existing unrelated progress-rendering failures'').
After 15 model calls it submits: ``Implementation and validation are complete.'' The one new test of
request~9 passes. Two of the 197 protected tests fail: \texttt{test\_argument\_decorator} and
\texttt{test\_option\_decorator}, both from request~8, at
\texttt{self.assertTrue(application.has('decorated\_foo\_code'))} with \texttt{AssertionError: False
is not true}. A correct change keeps the request-8 behavior for a decorator used alone and adds the
stacked behavior. Opus~5.5, Opus~5, Fable~5, GLM-5.3 and GLM-5.3-Flash pass request~9 on their
first attempts; GPT-6 Astra, GPT-6.1, Kimi~K3 and Opus~4.8 fail the same two tests, Opus~4.8 in all
eight of its attempts and Opus~5.5 in none. Across all graded handoff attempts, 12 of 24 pass
request~9. The two tests stay broken for the rest of GPT-5.6's chain: every later request is issued
on a repository carrying them, request~16 adds a miss of its own, and the final state fails.

\paragraph{Case 2: \texttt{joshtemple/lkml}, Fable~5, request 4.}
Request~2 adds \texttt{serialize\_dict}; when the dictionary has a \texttt{name} key, ``The output
begins with the value of \texttt{"name"}, followed by a space, then \texttt{\{}''. Request~3 extends
it to nested dictionaries and repeats the form: ``use \texttt{<name\_value> \{}, with no key or
colon before the name value''. Both pass. Request~4 asks only for a \texttt{serialize\_list} method
and says nothing about \texttt{serialize\_dict}. The agent, in a new conversation, adds the method,
checks the two examples of the specification, and runs the repository's test suite. Two of the
repository's own serializer tests fail: they predate the chain and still expect
\texttt{dimension: \{}, with a colon; the hidden test patch of request~2 rewrote them, but hidden
patches are applied only in the verifier. The agent confirms with \texttt{git stash} that they also
fail on the tree it received: ``Pre-existing failures. These are in \texttt{serialize\_dict}
(missing colon), outside my segment but let me check if fixing it breaks anything else''; then: ``The
dict tests fail pre-existing because \texttt{serialize\_dict} yields \texttt{name \{} instead of
\texttt{name: \{}. Since the serializer isn't used elsewhere, I'll fix that too so all tests pass''.
It adds the colon, reports ``All 87 tests pass'' and submits after 15 model calls. Both new tests of
request~4 pass. Two of the 89 protected tests fail: \texttt{test\_serialize\_dict\_with\_name} from
request~2 and \texttt{test\_serialize\_nested\_dict} from request~3, each expecting
\texttt{<name> \{} and receiving \texttt{<name>: \{}. A correct change adds
\texttt{serialize\_list} and leaves \texttt{serialize\_dict} alone. The nine other models pass
request~4 on their first attempts, and Fable~5 passes it in five of its eight. Across all graded
handoff attempts, 43 of 47 pass request~4. Here the chain heals: the two tests are still failing after
requests~5 and~6 and pass again after request~7, and Fable~5's final state passes every test.

\section{Supporting Details for Related Work}
\label{app:related-details}

This appendix retains quantitative and construction details omitted from
\hyperref[sec:related]{Section~\ref*{sec:related}}. These details describe the evaluation setting and its controls;
they do not establish that tasks across benchmarks have equal difficulty or that a larger
benchmark necessarily distinguishes every model pair.

\paragraph{Counting requests consistently.}
We count one separately delivered specification document as one request, without splitting
its internal requirements into additional tasks. SWE-Milestone contains 98 graded
specifications and three additional, ungraded specifications that agents still implement
\citep{swemilestone2026}. SWE-Chain contains 155 version-transition specifications, each
bundling upgrade requirements \citep{swechain2026}. SlopCodeBench contains 196 checkpoint
specifications across 36 problems \citep{slopcodebench2026}. Including the ungraded
SWE-Milestone requests gives 452 specifications in total. These counts were checked against
SWE-Milestone dataset revision \texttt{3500478e}, SWE-Chain dataset revision
\texttt{7080d652}, and the SlopCodeBench problems release \texttt{v1.0}
(\texttt{4d38d300}). The corresponding source snapshots and counting scripts accompany our
local research artifacts.

\textsc{BOBA-Bench} contains 280 chains of 20 requests across 140 repositories, yielding
5,600 request occurrences but 4,404 distinct specifications. The final window of a retained
run overlaps the preceding window when needed to include its tail, producing 1,196 repeated
occurrences at different chain positions. The distinct-request ratio is therefore
$4{,}404/452=9.74$, not the occurrence ratio $5{,}600/452=12.39$. Repeated requests have
different prefixes but are not independent observations for repository-level inference.
These are counts of specification documents, not comparable amounts of implementation work.

\paragraph{Task-quality evidence.}
OpenAI's SWE-bench Verified audit establishes a known-defect floor above 16\% of the full
benchmark, based on an audit that targeted frequently failed tasks rather than a random
sample \citep{openai2026swebenchverified}. Its SWE-bench Pro analysis reports defects in
roughly 30\% of the public split, with the exact proportion depending on the audit procedure
\citep{openai2026signalnoise}. These figures concern different auditing procedures and should
not be interpreted as directly comparable prevalence estimates. Other studies identify
solution leakage, weak tests, and opportunities for exploitation
\citep{aleithan2024swebenchplus,yu2025utboost,zhong2025impossiblebench}.

SWE-Milestone generates requirements from reference patches, aligns them with tests, and
uses expert review to check solvability and implementation leakage \citep{swemilestone2026}.
BOBA specification generation receives test changes and development descriptions, but not
the source patch. This removes direct access to the reference implementation at generation
time, rather than proving that every resulting specification is complete or free of hints.
\hyperref[app:qa-specs]{Appendix~\ref*{app:qa-specs}} describes generation and reference-code validation.

The isolated-success figure formerly quoted in Related Work concerns sampled requests,
not every request in every chain. For Claude Fable~5, 1,089 of 1,120 scheduled attempts on
one sampled request per repository passed, or 97.23\%, counting the 28 ungraded attempts
as failures. Among graded attempts the fraction was 1,089/1,092. These denominators must
not be confused with distinct-request coverage or the probability of at least one success
across repeated attempts. The count comes from the filtered per-target isolated-attempt
summary, not from the earlier candidate-certification campaign.

\paragraph{Model comparison and uncertainty.}
The manuscript's earlier Related Work combined illustrative binomial interval widths with
claims about ranking models. Counts of repositories, chains, and checkpoints are not
interchangeable statistical units. Within-repository dependence must be accounted for,
and an aggregate rate cannot in general be multiplied by a nominal task count to reconstruct
valid binomial outcomes. In particular, a macro-averaged milestone rate is not a binomial
proportion over all milestones. The absence of published uncertainty does not demonstrate
that models are indistinguishable. We therefore do not retain the earlier numerical
resolution argument or the blanket non-separation claim in \hyperref[sec:related]{Section~\ref*{sec:related}}.
The predecessor calculations in \hyperref[app:separation]{Appendix~\ref*{app:separation}} require revision before they
can support comparative significance claims. BOBA comparisons should use the observed
paired outcomes and repository-clustered uncertainty, without treating a nonsignificant
difference as evidence of equivalence.
\section{Trust, but Verify: Protocols and Estimates}
\label{app:trust}

This appendix gives, for each arm of \hyperref[sec:trust]{Section~\ref*{sec:trust}}, what the agent was told, which
runs were used, and the per-model estimates with intervals. The mechanism studies of
\hyperref[app:trust-arms]{Appendices~\ref*{app:trust-arms}}--\ref{app:trust-noisy} ran in the continuous conversation: one
conversation per chain, continued after the miss so that the verifier's message arrives as the
observation to the agent's submit command. There, completion means that every request passes on
its own step, and the baseline is the first attempt under that conversation, not the pass@1 of
\hyperref[tab:leaderboard]{Table~\ref*{tab:leaderboard}}, which starts every request in a fresh conversation and grades the
final state; the two are not interchangeable. \hyperref[app:trust-handoff]{Appendix~\ref*{app:trust-handoff}} delivers the same
message under the benchmark's protocol and reports single-attempt pass@1, and \hyperref[app:trust-scaffold]{Appendix~\ref*{app:trust-scaffold}}
compares harnesses under it. Intervals are Wilson 95\% for rates, repository-clustered bootstrap
95\% for contrasts (2{,}000 replicates in the mechanism studies, 10{,}000 in the handoff
sections), unless stated.

\hyperref[tab:trust]{Table~\ref*{tab:trust}} summarizes the section's estimates; the subsections below give counts,
intervals and~\mbox{protocols}.%

\begin{table}[h]
\centering
\scriptsize
\setlength{\tabcolsep}{4.5pt}
\renewcommand{\arraystretch}{1.05}
\begin{tabular}{@{}lcccccc@{}}
\toprule
 & \multicolumn{4}{c}{Reject (once, at the miss)} & \multicolumn{2}{c}{Gate (every request): completion (\%)} \\
\cmidrule(lr){2-5}\cmidrule(lr){6-7}
 & Repair & \multicolumn{3}{c}{Next request fails (\%) on a state that is} & No & Perfect \\
Model & rate (\%) & repaired & unrepaired & certified & supervisor & supervisor \\
\midrule
Fable~5      & 44.6 & 5.7  & 22.7 & 3.8  & 65.7 & 77.5 \\
GPT-6 Astra  & 45.4 & 5.1  & 28.2 & 4.3  & 52.1 & 68.2 \\
Kimi~K3      & 29.4 & 3.4  & 25.0 & 9.6  & 48.2 & 56.1$^{\dagger}$ \\
GPT-5.6      & 30.0 & 12.4 & 30.0 & 16.9 & 43.9 & 59.3 \\
GLM-5.2      & 37.4 & 8.5  & 31.1 & 11.9 & 27.9 & 41.4 \\
\bottomrule
\end{tabular}

\smallskip
\begin{tabular}{@{}lccccccc@{}}
\toprule
\multicolumn{8}{@{}l}{\textit{Noisy supervisor: completion (\%), 100-chain subset per model}} \\
\addlinespace[2pt]
Model & No supervisor & $p{=}0$ & $p{=}0.1$ & $p{=}0.2$ & $p{=}0.3$ & $p{=}0.4$ & $p{=}0.5$ \\
\midrule
GPT-6 Astra & 49 & 64 & 64 & 58 & 61 & 53 & 53 \\
GPT-5.6 & 44 & 60 & 58 & 54 & 46 & 49 & 47 \\
\bottomrule
\end{tabular}
\caption{\textbf{What one bit from outside recovers: summary of \hyperref[sec:trust]{Section~\ref*{sec:trust}}} (280
chains; one conversation per chain, continued after the miss; completion = every request passes
on its own step). \emph{Reject}: the message
once, at a silent miss, with one repair round; \emph{Gate}: the same after every failed
submission of a chain. The certified column is the reference of a state that passed its
contract. The lower block reports the separate 100-chain \emph{Noisy} study: $p$ is the
false-alarm rate and $p{=}0$ a perfect supervisor. $^{\dagger}$Lower bound
(\hyperref[app:trust]{Appendix~\ref*{app:trust}}).}
\label{tab:trust}
\end{table}

\subsection{Arms and messages}
\label{app:trust-arms}

\paragraph{Starting points.} The Reject, Name, Insist and fresh-conversation arms start
from the silent misses of \hyperref[sec:compound]{\S\ref*{sec:compound}}, taken from the continuous-conversation first
attempts: every request $k\le19$ that a model submitted as
complete and that failed its hidden contract. The failing state is restored and the same
conversation is continued. Counts per model are in \hyperref[tab:resume-results]{Table~\ref*{tab:resume-results}}.

\paragraph{The message.} In every arm the supervisor's message is the same string, sent as the
observation to the agent's submit command:
\begin{quote}\small\itshape
External verification of your previous submission failed: the repository does not satisfy the
previous specification. Fix the repository so that the previous specification is satisfied,
then submit again with the same command. Do not modify or create test files.
\end{quote}
The message is an oracle bit, and it bounds what any verifier can deliver: the checks an agent
can run without hidden tests---its own test run or test suite, another model's review---fall
far short of it (\hyperref[sec:compound]{\S\ref*{sec:compound}}, \hyperref[app:cv]{Appendix~\ref*{app:cv}}, \hyperref[sec:cv]{\S\ref*{sec:cv}}).
\emph{Reject} (arm E) sends it once and allows one repair round before request $k{+}1$ is
issued. \emph{Name} (arm E1-names) appends the identifiers of the failing test nodes.
\emph{Insist} (arm E8) re-sends it after each failed repair, for up to eight rounds.
\emph{Fresh conversation} (arm F) restores the same broken state into a new conversation with no
message, as a control for the effect of a restart. The next request is graded on its own tests
(own-work failure), so a failure inherited from the open miss is not counted against it; the
reference rate is the same model's own-work failure after a certified pass of request $k$,
standardised to the arm's distribution of $k$ (\hyperref[tab:resume-results]{Table~\ref*{tab:resume-results}}). The repaired and
unrepaired subsets are selected on outcome; their rates describe, they do not estimate a causal
effect.

\paragraph{Gate.} The continuous-conversation first attempts are audited from the exact state at
which they first failed (identity-checked restore): one repair round is allowed at that request
and at every later request that fails, including the twentieth, and a chain counts as complete
only if all twenty requests pass. Kimi~K3's continuations that reached the 12-hour wall while
still passing are counted as failures (18 of 145 audited chains), so its completion is a lower
bound; every other model's audit is fully resolved on the 280 chains.

\paragraph{Noisy supervisor.} Fresh chains are run from the first request on 100 chains of the
280-chain set (100 distinct repositories; 44 that GPT-5.6 completed on its first attempt, 30
that it failed by request 10 and 26 that it failed later), with GPT-5.6-sol and GPT-6 Astra in
the continuous conversation, with the budgets of the Gate audit. After every submission the hidden grader runs: after a failure the
message above is sent (a true alarm); after a pass it is sent with probability $p$ (a false
alarm), otherwise the usual acceptance. One uniform draw per (chain, request), fixed before
launch, decides the false alarms at every level $p\in\{0.1,0.2,0.3,0.4,0.5\}$, so the levels
are nested. Each alarm allows one repair round, after which the verdict is final: a false alarm
that makes the agent break a correct submission ends the chain. The references are the same 100
chains without a supervisor (first attempt in the continuous conversation) and under the Gate audit above
($p{=}0$). A first run of this arm wrote its bookkeeping---the level, the chain's alarm
schedule and the kind of each alarm---into a directory that the harness mounts into the agent's
sandbox; GPT-5.6 read at least one of these files in 13--19 of 100 attempts per level and GPT-6
Astra in at most one. The arm was re-run with the bookkeeping sealed outside the sandbox and a
per-attempt check that nothing of it was visible. \hyperref[sec:false-alarm]{Section~\ref*{sec:false-alarm}} reports the
sealed run; the first run is reported below as a sensitivity analysis.

\subsection{Reject, one round}
\label{app:trust-onebit}

\hyperref[tab:resume-results]{Table~\ref*{tab:resume-results}} gives the per-model estimates on the 280-chain set: starting
points, repair rate with Wilson intervals, own-work failure at $k{+}1$ on repaired, unrepaired
and certified states, and the paired contrast between the fresh-conversation and silent arms
($F-D$: Fable~5 $+4.1$\,pp [0.7, 8.2] on 465 pairs; Kimi~K3 $+0.4$ [$-1.9$, 2.7] on 773;
GPT-5.6 $-1.2$ [$-5.1$, 2.5] on 939; GPT-6 Astra and GLM-5.2 have no arm F).

\subsection{Name and Insist}
\label{app:trust-names}

Both variants were run for GPT-5.6 and Kimi~K3 on the same starting points as their Reject arm
and are reported on the 280 chains, recomputed from the continuation records;
\hyperref[tab:trust-names]{Table~\ref*{tab:trust-names}} gives the
estimates. Contrasts are paired on starting points and clustered by repository. Insist is
intention-to-treat: a continuation censored by the tool-time cap or the wall before a pass counts
as unrepaired at every round, and the arms' denominators differ slightly for Kimi~K3 for that
reason.

\begin{table}[h]
\centering
\scriptsize
\setlength{\tabcolsep}{2.5pt}
\newcommand{\lab}[1]{\parbox[t]{1.9in}{\raggedright\strut #1\strut}}
\caption{\textbf{Name and Insist} (280 chains). Repair rate of the starting-point miss
after $r$ rounds; increments per round; calls until the miss passes among repaired misses and
calls spent among misses never repaired (Insist arm). Wilson intervals for rates,
repository-clustered bootstrap for the paired contrast.}
\label{tab:trust-names}
\begin{tabular}{@{}lcc@{}}
\toprule
 & GPT-5.6 & Kimi~K3 \\
\midrule
\lab{Starting points, Reject / Name / Insist (repositories)} & 939 / 939 / 939 (100) & 834 / 827 / 821 (96) \\
\lab{Reject, repair rate after one round (\%)} & 30.0 [27.2, 33.0] & 29.5 [26.5, 32.7] \\
\lab{Name, repair rate after one round (\%)} & 46.4 [43.3, 49.6] & 46.8 [43.4, 50.2] \\
\lab{Name $-$ Reject, paired (pp)} & 16.4 [11.0, 22.0] & 17.1 [11.2, 22.8] \\
\addlinespace
\lab{Insist, repair rate after round 1, 2, \dots, 8 (\%)} & 30.0, 38.1, 44.3, 47.6, 49.8, 52.2, 53.0, 53.9 & 31.1, 42.3, 48.2, 52.6, 54.1, 56.0, 57.6, 58.7 \\
\lab{Increment of rounds 2, \dots, 8 (pp)} & 8.1, 6.2, 3.3, 2.2, 2.3, 0.9, 0.9 & 11.2, 6.0, 4.4, 1.5, 2.0, 1.6, 1.1 \\
\lab{Round 8, Wilson interval (\%)} & 53.9 [50.7, 57.1] & 58.7 [55.3, 62.0] \\
\addlinespace
\lab{Calls until pass, repaired misses, median} & 16.5 & 19 \\
\lab{Calls spent, misses never repaired, median (mean)} & 78 (87.8) & 73 (75.5) \\
\bottomrule
\end{tabular}
\end{table}

\subsection{Gate}
\label{app:trust-project}

\hyperref[tab:trust-project]{Table~\ref*{tab:trust-project}} gives the audit on the 280 chains: every failed first attempt is
resumed from the exact state at its failure with the rejection and one repair round, and
the chain continues under the same rule until it completes or a repair round fails.

\begin{table}[h]
\centering
\scriptsize
\setlength{\tabcolsep}{2.5pt}
\newcommand{\lab}[1]{\parbox[t]{1.85in}{\raggedright\strut #1\strut}}
\caption{\textbf{Gate} (280 chains). Completion with no supervisor (first attempt in the
continuous conversation) and
with a perfect supervisor (one repair round at every failed request); Wilson intervals over
280 chains.}
\label{tab:trust-project}
\begin{tabular}{@{}lccccc@{}}
\toprule
 & Fable~5 & GPT-6 Astra & Kimi~K3 & GPT-5.6 & GLM-5.2 \\
\midrule
\lab{No supervisor: first attempts completing 20 (of 280)} & 184 & 146 & 135 & 123 & 78 \\
\lab{Failed first attempts audited} & 96 & 134 & 145 & 157 & 202 \\
\lab{Rescued by verification and repair} & 33 & 45 & 22 & 43 & 38 \\
\lab{Perfect supervisor: completion (\%)} & 77.5 [72.3, 82.0] & 68.2 [62.5, 73.4] & 56.1 [50.2, 61.8] & 59.3 [53.4, 64.9] & 41.4 [35.8, 47.3] \\
\lab{Gain (pp; relative)} & 11.8 (18\%) & 16.1 (31\%) & 7.9 (16\%) & 15.4 (35\%) & 13.6 (49\%) \\
\lab{Unresolved continuations (counted as failures)} & 0 & 0 & 18 & 0 & 0 \\
\lab{Upper bound if all unresolved had completed (\%)} & --- & --- & 62.5 & --- & --- \\
\bottomrule
\end{tabular}
\end{table}

\subsection{Noisy supervisor}
\label{app:trust-noisy}

\hyperref[tab:trust-noisy]{Table~\ref*{tab:trust-noisy}} gives the sealed run (its bands are drawn in \hyperref[fig:false-alarm-ci]{Figure~\ref*{fig:false-alarm-ci}}): ten arms of 100 chains, 1{,}000 attempts, none
censored, every model call at the benchmark's reasoning setting, and no harness file visible in
any sandbox. Completion intervals are clustered by chain (each chain is its own repository in
this sample). We also report, per model, the outcome of every phantom round---unchanged
resubmit (identical tree), changed and still passing, or broken---the calls it required, its share
of all model calls, and the repair rate of true alarms, which is comparable in
protocol but not in population to the Reject arm (here the misses arise inside fresh attempts).
The call shares are pooled within each arm, not averaged across chains. We report calls rather
than dollar shares because the cache and pricing uncertainties of \hyperref[app:cost]{Appendix~\ref*{app:cost}}
also affect the monetary attribution of phantom~rounds.%

\begin{table}[h]
\centering
\scriptsize
\setlength{\tabcolsep}{2.5pt}
\newcommand{\lab}[1]{\parbox[t]{2.05in}{\raggedright\strut #1\strut}}
\caption{\textbf{Noisy supervisor} (sealed run; 100 chains per level, 0 censored). Completion
against the false-alarm rate $p$, its paired difference from the perfect supervisor on the same
chains, and the fate of the phantom rounds. Brackets: repository-clustered bootstrap 95\%
intervals (2{,}000 replicates).}
\label{tab:trust-noisy}
\begin{tabular}{@{}lccccc@{}}
\toprule
 & $p{=}0.1$ & $p{=}0.2$ & $p{=}0.3$ & $p{=}0.4$ & $p{=}0.5$ \\
\midrule
\multicolumn{6}{@{}l}{\textit{GPT-5.6-sol} --- no supervisor 44 [35, 54]; perfect supervisor 60 [50, 70]} \\
\lab{Completion (\%)} & 58 [49, 67] & 54 [44, 64] & 46 [37, 56] & 49 [39, 59] & 47 [37, 57] \\
\lab{Difference from perfect supervisor, paired (pp)} & $-2$ [$-9$, 5] & $-6$ [$-13$, 1] & $-14$ [$-23$, $-5$] & $-11$ [$-19$, $-2$] & $-13$ [$-21$, $-5$] \\
\lab{Chains ended by a false alarm (\%)} & 3 & 2 & 8 & 7 & 10 \\
\lab{Phantom rounds ($n$); broken (\%); unchanged (\%)} & 147; 2.0; 2.0 & 298; 0.7; 1.7 & 416; 1.9; 0.2 & 581; 1.2; 1.2 & 701; 1.4; 2.0 \\
\lab{Calls per phantom round, median} & 8 & 8 & 7.5 & 8 & 8 \\
\lab{Phantom rounds' share of all model calls (\%)} & 10 & 19 & 23 & 28 & 34 \\
\lab{True alarms repaired (\%)} & 33 (19/58) & 34 (23/67) & 27 (17/63) & 24 (14/58) & 25 (14/57) \\
\addlinespace
\multicolumn{6}{@{}l}{\textit{GPT-6 Astra} --- no supervisor 49 [40, 59]; perfect supervisor 64 [54, 73]} \\
\lab{Completion (\%)} & 64 [55, 73] & 58 [48, 68] & 61 [51, 71] & 53 [44, 63] & 53 [42, 62] \\
\lab{Difference from perfect supervisor, paired (pp)} & 0 [$-7$, 7] & $-6$ [$-13$, 1] & $-3$ [$-10$, 4] & $-11$ [$-19$, $-3$] & $-11$ [$-19$, $-2$] \\
\lab{Chains ended by a false alarm (\%)} & 6 & 6 & 7 & 8 & 12 \\
\lab{Phantom rounds ($n$); broken (\%); unchanged (\%)} & 157; 3.8; 0.0 & 311; 1.9; 0.3 & 463; 1.5; 0.9 & 593; 1.3; 0.8 & 771; 1.6; 0.6 \\
\lab{Calls per phantom round, median} & 6 & 6 & 6 & 5.5 & 6 \\
\lab{Phantom rounds' share of all model calls (\%)} & 9 & 16 & 21 & 27 & 31 \\
\lab{True alarms repaired (\%)} & 47 (27/57) & 36 (20/56) & 40 (21/53) & 33 (19/58) & 35 (19/54) \\
\bottomrule
\end{tabular}
\end{table}

Pooled over the five levels, GPT-5.6's 2{,}143 phantom rounds ended in an unchanged resubmit
1.4\% of the time [0.6, 2.4], in a change that still passed 97.2\%, and in a broken submission
1.4\% [0.8, 2.1]; GPT-6 Astra's 2{,}295 in 0.7\% [0.0, 1.8], 97.6\% and 1.7\% [0.9, 2.7].
A phantom round cost a median of 8 (GPT-5.6) and 6 (GPT-6 Astra) model calls and about 6 and
3\,s of tool execution. True alarms were repaired in 28.7\% [22.4, 34.8] and 38.1\% [27.9, 48.4]
of cases, against 30.0\% and 45.4\% in the Reject arm's population.

\paragraph{Sensitivity: the first run.} The first run, in which the alarm schedule was
readable, gives completion within 7 points of the sealed run at every level for both models
(GPT-5.6: 55, 61, 47, 42, 44\%; GPT-6 Astra: 63, 60, 57, 55, 57\%) and the same broken rate per
phantom round (1.5\% for both). Where it differs is in how GPT-5.6 answered a false alarm: 11.3\%
[7.8, 15.2] of its 2{,}145 phantom rounds were unchanged resubmits, against 1.4\% once sealed;
restricted to the attempts whose model never read a bookkeeping file, the first run's rate is
1.2\%, so the unchanged resubmits were the exposed attempts. GPT-6 Astra, which read the files
in at most one attempt per level, shows no difference (0.8\% against 0.7\%). One first-run cell,
GPT-5.6's true-alarm repair rate of 10\% at $p{=}0.5$ (5 of 50), did not recur in the sealed run
(25\%, 14 of 57).

\subsection{Verification under the handoff protocol}
\label{app:trust-handoff}

The arms above continue the conversation in which the miss was made. Under the benchmark's
protocol (\hyperref[sec:settings]{\S\ref*{sec:settings}}) that conversation is gone when the verdict arrives, so the bit is
delivered as follows: whenever request $k\le19$ leaves a test of requests $1..k$ failing, the agent
receives, before request $k{+}1$, one repair round in a fresh conversation whose task is the
specification of request $k$ followed by the message of \hyperref[app:trust-arms]{Appendix~\ref*{app:trust-arms}}; request
$k{+}1$ then starts in its own fresh conversation. The repair is graded together with request
$k{+}1$, so strict pass@1 never credits it; a failing original test of the repository does not
trigger the message; nothing stops the chain, and a failed request 20 gets no repair. Harness,
budgets and settings are those of the \textsc{\BobaMark{autonomous}Autonomous} run, and every
contrast is paired with its first attempt on the same 280 chains (\hyperref[tab:trust-handoff]{Table~\ref*{tab:trust-handoff}}):
the Verified arm has one attempt, so these are single-attempt comparisons, not the eight-attempt
estimate of \hyperref[tab:leaderboard]{Table~\ref*{tab:leaderboard}}, and the Autonomous columns below are attempt-1 values. Eight of the
2{,}800 Verified chains were still running when the table was built and count as failures (three
for Fable~5, two each for GPT-5.6 and GLM-5.3-Flash, one for Kimi~K3); no number can move by more
than 1.1 points.

\begin{table}[h]
\centering
\scriptsize
\setlength{\tabcolsep}{2pt}
\renewcommand{\arraystretch}{1.05}
\caption{\textbf{Verified under the handoff protocol} (ten models, 280 chains each, one attempt per
arm; Wilson 95\%
intervals per arm, repository-clustered bootstrap 95\% intervals for the paired differences
Verified $-$ Autonomous; A = Autonomous (attempt 1), V = Verified). \emph{Tests repaired} is the share of tests that fail for the
first time after a request and pass after the next one: after the repair round in the Verified arm,
by the model's own later work in the Autonomous arm. Calls per chain are medians and include the
repair rounds.}
\label{tab:trust-handoff}
\begin{tabular}{@{}lccccccc@{}}
\toprule
 & \multicolumn{2}{c}{pass@1 (\%)} & \multicolumn{2}{c}{Regression rate (\%)} & Strict (\%) & Tests repaired (\%) & Calls \\
\cmidrule(lr){2-3}\cmidrule(lr){4-5}
Model & A / V & $\Delta$ (pp) & A / V & $\Delta$ (pp) & A / V & A / V & A / V \\
\midrule
Opus~5.5 & 66.8 / 73.9 & $+7.1$ [2.5, 12.0] & 21.1 / 25.7 & $+4.6$ [0.7, 8.6] & 55.7 / 57.9 & 30 / 60 & 217 / 234 \\
Opus~5 & 60.7 / 69.6 & $+8.9$ [3.6, 14.3] & 27.5 / 32.5 & $+5.0$ [$-0.7$, 10.6] & 51.1 / 50.0 & 26 / 43 & 401 / 486 \\
GPT-6 Astra & 51.1 / 53.9 & $+2.9$ [$-2.4$, 8.0] & 35.0 / 42.5 & $+7.5$ [3.0, 12.2] & 36.1 / 36.4 & 12 / 39 & 182 / 209 \\
Fable~5 & 51.8 / 63.9 & $+12.1$ [6.8, 17.5] & 42.1 / 46.1 & $+3.9$ [$-1.4$, 9.4] & 36.4 / 38.9 & 19 / 38 & 293 / 350 \\
GLM-5.3 & 46.4 / 68.6 & $+22.1$ [15.8, 28.4] & 31.8 / 35.0 & $+3.2$ [$-2.9$, 9.8] & 38.2 / 40.7 & 16 / 49 & 560 / 684 \\
GLM-5.3-Flash & 51.1 / 67.1 & $+16.1$ [10.2, 21.9] & 31.1 / 34.3 & $+3.2$ [$-2.5$, 9.1] & 38.6 / 40.7 & 16 / 49 & 739 / 902 \\
GPT-6.1 Sol & 48.9 / 52.9 & $+3.9$ [$-0.4$, 8.1] & 37.1 / 45.7 & $+8.6$ [4.0, 13.5] & 36.4 / 35.0 & 17 / 37 & 190 / 222 \\
GPT-5.6 Sol & 45.7 / 46.8 & $+1.1$ [$-2.9$, 4.9] & 43.2 / 47.9 & $+4.6$ [0.0, 9.4] & 32.5 / 30.0 & 12 / 31 & 299 / 355 \\
Kimi~K3 & 45.4 / 52.5 & $+7.1$ [1.5, 12.6] & 44.3 / 47.9 & $+3.6$ [$-2.1$, 9.0] & 32.9 / 35.4 & 8 / 28 & 389 / 497 \\
Opus~4.8 & 38.2 / 44.6 & $+6.4$ [1.7, 11.1] & 41.1 / 50.0 & $+8.9$ [2.6, 15.1] & 27.5 / 27.9 & 22 / 36 & 367 / 496 \\
\bottomrule
\end{tabular}
\end{table}

\begin{figure}[htbp]
\centering
\includegraphics[width=\textwidth]{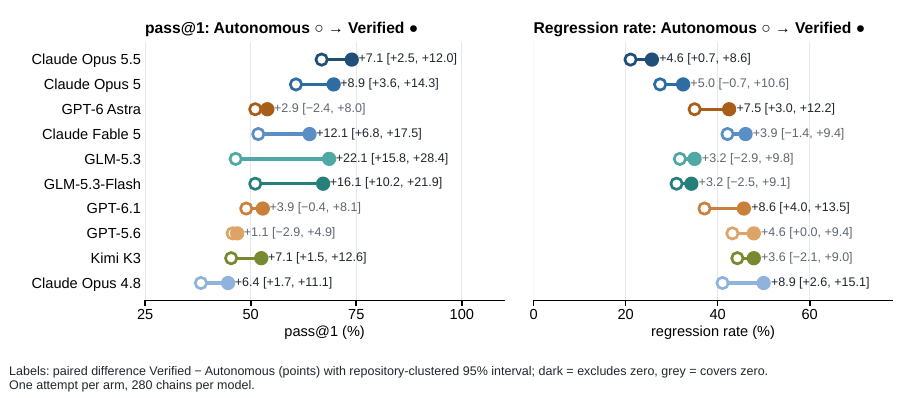}
\caption{\textbf{One ``verification failed'' message raises pass@1 for every model, and raises the
regression rate too.} Single-attempt pass@1 (left) and regression
rate (right) of the \textsc{\BobaMark{autonomous}Autonomous} run (hollow) and of the Verified arm
(filled) for ten models on the same 280 chains; labels give the paired difference with its
repository-clustered 95\% interval, dark where the interval excludes zero (seven of ten for pass@1,
four of ten for the regression rate).}
\label{fig:trust-handoff}
\end{figure}

The bit repairs misses: the share of new failing tests that pass after the next request rises
from 8--30\% to 28--60\%, and pass@1 rises by 1.1 to 22.1 points, with seven of ten intervals
excluding zero (\hyperref[fig:trust-handoff]{Figure~\ref*{fig:trust-handoff}}). It also breaks things: the regression rate rises for every model, by 3.2 to 8.9
points (four intervals exclude zero), because the repair round starts without the conversation in
which the work was written and has to rediscover it before it can mend it. Strict pass@1, which
grades each request on its own step, cannot see a repair and moves by $-2.5$ to $+2.5$ points; the
repair rounds add 47 to 294 calls per chain (12 to 98 per round).

\paragraph{Name and Insist under handoff.} Both variants start from every repair slot of the
Verified arm---a request $k\le19$ after which fixable tests were failing---restored to the exact
repository state after $k$ with the arm's budget counters, and send the message in a fresh
conversation. \emph{Name} appends the failing tests' identifiers; \emph{Insist} re-sends the bare
message after each failed round, up to four rounds (rounds 5--8 of \hyperref[app:trust-names]{Appendix~\ref*{app:trust-names}}
added under 2 points each), its first round being Reject once in isolation, the paired control for
Name on identical slots (\hyperref[tab:trust-handoff-variants]{Table~\ref*{tab:trust-handoff-variants}}).

\begin{table}[h]
\centering
\scriptsize
\setlength{\tabcolsep}{2.2pt}
\renewcommand{\arraystretch}{1.05}
\caption{\textbf{Name and Insist under the handoff protocol.} Repair rate of the starting-point
miss after one round with the failing tests named (Name) and with the bare message (Reject once =
Insist round 1), their paired difference on identical slots (repository-clustered bootstrap 95\%),
the bare message after four rounds, the Insist rounds needed to reach Name's one round, and the
share of rounds that broke a protected test. Wilson 95\% intervals in the source; `--': slots still
running when the table was built.}
\label{tab:trust-handoff-variants}
\begin{tabular}{@{}lccccccc@{}}
\toprule
 & Slots & Name, & Reject once, & Name $-$ Reject once, & Insist, & Insist rounds & Round broke a test (\%), \\
Model & (Name / Insist) & 1 round (\%) & 1 round (\%) & paired (pp) & 4 rounds (\%) & to match Name & Name / Reject once \\
\midrule
Claude Opus~5.5 & 517 / 517 & 44.5 & 16.8 & $+27.7$ [18.5, 37.3] & 28.8 & $>4$ & 12.0 / 7.2 \\
Claude Opus~5 & 632 / 632 & -- & -- & -- & -- & -- & -- \\
GPT-6 Astra & 1090 / 1090 & 35.6 & -- & -- & -- & -- & 12.2 / -- \\
Claude Fable~5 & 875 / 875 & -- & -- & -- & -- & -- & -- \\
GLM-5.3 & 808 / 808 & 27.2 & -- & -- & -- & -- & 9.8 / -- \\
GLM-5.3-Flash & 762 / 762 & -- & -- & -- & -- & -- & -- \\
GPT-6.1 Sol & 1155 / 1155 & 37.1 & 12.6 & $+24.5$ [18.6, 30.9] & 14.4 & $>4$ & 12.6 / 4.9 \\
GPT-5.6 Sol & 1304 / 1304 & 28.2 & -- & -- & -- & -- & 12.9 / -- \\
Kimi~K3 & 1212 / 1212 & 32.5 & -- & -- & -- & -- & 12.4 / -- \\
Claude Opus~4.8 & 1430 / 1430 & -- & -- & -- & -- & -- & -- \\
\bottomrule
\end{tabular}
\end{table}

For Claude Opus~5.5 and GPT-6.1 Sol, naming the failing tests repairs 44.5\% and 37.1\% of the
misses in one round against 16.8\% and 12.6\% for the bare message, which four rounds of insisting
(28.8\% and 14.4\%) do not reach; the named round breaks a protected test more often than the bare
one (12.0\% and 12.6\% against 7.2\% and 4.9\% of rounds), because it changes more.

\paragraph{Noisy watcher under handoff.} Fresh chains on the 100-chain subset of
\hyperref[app:trust-arms]{Appendix~\ref*{app:trust-arms}}, at $p\in\{0.1,0.5\}$ with the same alarm schedule, under the
handoff protocol: a failing request gets the repair round of the Verified arm (a true alarm), a
passing one gets it with probability $p$ (a phantom alarm), one round per alarm, nothing stops
the chain, and the references are the \textsc{\BobaMark{autonomous}Autonomous} run and the
Verified arm ($p{=}0$) on the same chains (\hyperref[tab:trust-handoff-noisy]{Table~\ref*{tab:trust-handoff-noisy}}). Because a
phantom round is a fresh conversation that is told to fix work that is not broken, its cost shows
in what it breaks rather than in chains it ends.

\begin{table}[h]
\centering
\scriptsize
\setlength{\tabcolsep}{2.5pt}
\renewcommand{\arraystretch}{1.05}
\caption{\textbf{Noisy watcher under the handoff protocol} (100 chains per cell, one attempt per
arm; paired
differences with repository-clustered bootstrap 95\% intervals). Single-attempt pass@1 of the noisy arm next to
the \textsc{Autonomous} run (attempt 1) and the Verified arm ($p{=}0$) on the same chains; regression rate
of the three; the fate of the phantom rounds (share that broke a passing test) and their share of
all model calls. `--': chains still running when the table was built.}
\label{tab:trust-handoff-noisy}
\begin{tabular}{@{}llcccccc@{}}
\toprule
 & & pass@1 (\%) & $\Delta$ vs Autonomous & $\Delta$ vs $p{=}0$ & Regression (\%) & Phantom rounds & Phantom share \\
Model & $p$ & noisy / Auton.\ / $p{=}0$ & (pp) & (pp) & noisy / Auton.\ / $p{=}0$ & ($n$; broke, \%) & of calls (\%) \\
\midrule
Claude Opus~5.5 & 0.1 & 71 / 69 / 76 & $+2$ [$-6$, 10] & $-5$ [$-13$, 2] & 41 / 20 / 28 & 169; 9.5 & 11 \\
 & 0.5 & 67 / 69 / 76 & $-2$ [$-11$, 7] & $-9$ [$-18$, $-1$] & 44 / 20 / 28 & 769; 4.5 & 33 \\
Claude Opus~5 & 0.1 & -- & -- & -- & -- & -- & -- \\
 & 0.5 & -- & -- & -- & -- & -- & -- \\
GPT-6 Astra & 0.1 & 55 / 52 / 53 & $+3$ [$-5$, 11] & $+2$ [$-6$, 10] & 48 / 41 / 47 & 143; 3.5 & 6 \\
 & 0.5 & 51 / 52 / 53 & $-1$ [$-9$, 7] & $-2$ [$-11$, 7] & 52 / 41 / 47 & 719; 4.0 & 24 \\
Claude Fable~5 & 0.1 & -- & -- & -- & -- & -- & -- \\
 & 0.5 & -- & -- & -- & -- & -- & -- \\
GLM-5.3 & 0.1 & -- & -- & -- & -- & -- & -- \\
 & 0.5 & -- & -- & -- & -- & -- & -- \\
GLM-5.3-Flash & 0.1 & -- & -- & -- & -- & -- & -- \\
 & 0.5 & -- & -- & -- & -- & -- & -- \\
GPT-6.1 Sol & 0.1 & 57 / 51 / 53 & $+6$ [$-2$, 14] & $+4$ [$-5$, 13] & 46 / 41 / 49 & 149; 5.4 & 6 \\
 & 0.5 & 49 / 51 / 53 & $-2$ [$-10$, 6] & $-4$ [$-11$, 3] & 56 / 41 / 49 & 716; 4.2 & 24 \\
GPT-5.6 Sol & 0.1 & 41 / 49 / 45 & $-8$ [$-16$, 0] & $-4$ [$-12$, 4] & 59 / 47 / 55 & 142; 7.0 & 6 \\
 & 0.5 & -- & -- & -- & -- & -- & -- \\
Kimi~K3 & 0.1 & 49 / 48 / 56 & $+1$ [$-10$, 12] & $-7$ [$-17$, 4] & 51 / 46 / 46 & 140; 11.4 & 7 \\
 & 0.5 & -- & -- & -- & -- & -- & -- \\
Claude Opus~4.8 & 0.1 & -- & -- & -- & -- & -- & -- \\
 & 0.5 & -- & -- & -- & -- & -- & -- \\
\bottomrule
\end{tabular}
\end{table}

Where the cells are final, a phantom round breaks a passing test in 3.5--11.4\% of cases at
$p{=}0.1$ and in 4.0--4.5\% at $p{=}0.5$, against 1.4--1.7\% inside a continuous conversation
(\hyperref[tab:trust-noisy]{Table~\ref*{tab:trust-noisy}}); the regression rate of the noisy chains exceeds the Verified
arm's by up to 16 points (Claude Opus~5.5 at $p{=}0.5$, [6, 26]). On pass@1 one false alarm in
ten is within the interval of a perfect supervisor for every final model ($-7$ to $+4$ points), and
at $p{=}0.5$ Claude Opus~5.5 gives back the whole gain ($-9$ [$-18$, $-1$] against $p{=}0$) and
sits where the \textsc{Autonomous} run sits ($-2$ [$-11$, 7]).

\subsection{Scaffold sensitivity}
\label{app:trust-scaffold}

\begin{figure}[htbp]
\centering
\includegraphics[width=\textwidth]{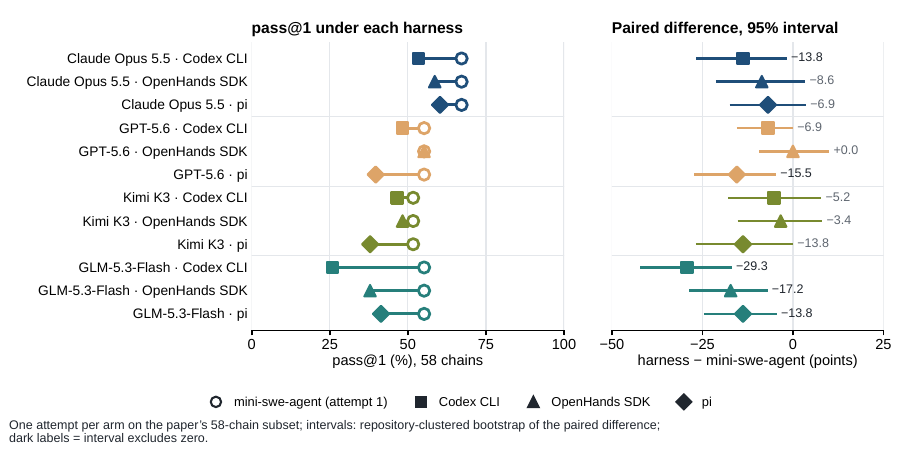}
\caption{\textbf{A different harness does not help: none raises pass@1 for any model.} Single-attempt pass@1 of four models
under Codex CLI, the OpenHands SDK and pi next to their mini-swe-agent first attempt on the same 58
chains (left), and the paired difference harness $-$ mini-swe-agent with its repository-clustered
95\% interval (right; dark labels where the interval excludes zero): five
of the twelve intervals lie entirely below zero and none above it.}
\label{fig:scaffold-handoff}
\end{figure}

\hyperref[tab:scaffold]{Table~\ref*{tab:scaffold}} gives the harness comparison used in \hyperref[sec:trust-benchmarks]{Section~\ref*{sec:trust-benchmarks}}
(and referenced in \hyperref[sec:howfar]{Section~\ref*{sec:howfar}}): on a stratified 58-chain subset (52 repositories; a
60-chain draw made before the last filtering round, minus the two chains of a repository that round
excluded), four models run under Codex CLI 0.146.1, the OpenHands SDK 1.46.0 and pi 0.84.2, each
installed inside the task container and invoked afresh for every request on the persisting
repository, with the hidden grader, the no-stop rule and the model settings of the
\textsc{\BobaMark{autonomous}Autonomous} run; every arm is paired with the same model's
mini-swe-agent first attempt on the same chains. No harness raises pass@1 for any model: the paired
difference runs from $0.0$ (GPT-5.6, OpenHands SDK) to $-29.3$ points (GLM-5.3-Flash, Codex CLI),
seven of the twelve intervals cover zero and five lie below it (\hyperref[fig:scaffold-handoff]{Figure~\ref*{fig:scaffold-handoff}}). Between 67 and 90\% of the chains
get the same verdict under a harness as under mini-swe-agent, and when both fail, 139 of 264 pairs
fail first at the same request. GLM-5.3-Flash is the exception, and it is a property of the model:
under every harness it breaks protected tests far more often than under mini-swe-agent (regression
rate 36--62\% against 26\%). pi has no per-command timeout; four chains stalled in one shell call
for more than two hours and were closed under the benchmark's tool-time rule, and eight chains ended
in a verifier error; both count as failures, as in the \textsc{\BobaMark{autonomous}Autonomous} run.
The cumulative tool-time budget cannot be enforced inside the harnesses; a per-call limit stands in
for it, and the OpenHands SDK, the one harness that reports its tool time, stays far below the
budget.

\begin{table}[h]
\centering
\scriptsize
\setlength{\tabcolsep}{3.5pt}
\renewcommand{\arraystretch}{1.05}
\caption{\textbf{Scaffold sensitivity} (58 chains, 52 repositories; handoff protocol; one attempt per
arm). Single-attempt pass@1
of each model under three harnesses next to its mini-swe-agent run (attempt 1) on the same chains; Wilson 95\%
intervals per arm, repository-clustered bootstrap 95\% intervals for the paired difference; strict
pass@1 and regression rate as chain counts; chains with the same pass@1 verdict under the harness
and under mini-swe-agent.}
\label{tab:scaffold}
\begin{tabular}{@{}llccccc@{}}
\toprule
Model & Harness & pass@1 (of 58) & Paired $\Delta$ (pp) & Strict & Regressed & Same verdict \\
\midrule
Claude Opus~5.5 & mini-swe-agent & 39 [54.4, 77.9] & --- & 29 & 15 & --- \\
                & Codex CLI      & 31 [40.8, 65.7] & $-13.8$ [$-26.7$, $-1.7$] & 27 & 16 & 44 \\
                & OpenHands SDK  & 34 [45.8, 70.4] & $-8.6$ [$-21.3$, 3.4] & 30 & 17 & 47 \\
                & pi             & 35 [47.5, 71.9] & $-6.9$ [$-17.5$, 3.6] & 29 & 12 & 48 \\
\addlinespace
GPT-5.6-sol     & mini-swe-agent & 32 [42.5, 67.3] & --- & 20 & 27 & --- \\
                & Codex CLI      & 28 [35.9, 60.8] & $-6.9$ [$-15.5$, 0.0] & 21 & 25 & 52 \\
                & OpenHands SDK  & 32 [42.5, 67.3] & $0.0$ [$-9.3$, 10.0] & 21 & 23 & 52 \\
                & pi             & 23 [28.1, 52.5] & $-15.5$ [$-27.3$, $-4.7$] & 13 & 27 & 45 \\
\addlinespace
Kimi~K3         & mini-swe-agent & 30 [39.2, 64.1] & --- & 19 & 26 & --- \\
                & Codex CLI      & 27 [34.3, 59.2] & $-5.2$ [$-17.9$, 7.8] & 20 & 27 & 43 \\
                & OpenHands SDK  & 28 [35.9, 60.8] & $-3.4$ [$-15.3$, 8.1] & 20 & 29 & 46 \\
                & pi             & 22 [26.6, 50.8] & $-13.8$ [$-26.8$, 0.0] & 16 & 26 & 42 \\
\addlinespace
GLM-5.3-Flash   & mini-swe-agent & 32 [42.5, 67.3] & --- & 24 & 15 & --- \\
                & Codex CLI      & 15 [16.3, 38.4] & $-29.3$ [$-42.1$, $-16.9$] & 11 & 36 & 39 \\
                & OpenHands SDK  & 22 [26.6, 50.8] & $-17.2$ [$-28.6$, $-7.0$] & 15 & 27 & 46 \\
                & pi             & 24 [29.6, 54.2] & $-13.8$ [$-24.6$, $-4.5$] & 20 & 21 & 48 \\
\bottomrule
\end{tabular}
\end{table}

\end{document}